\documentclass[a4paper,fleqn,usenatbib]{mnras}

\usepackage[T1]{fontenc}
\usepackage{ae,aecompl}

\usepackage{graphicx}	
\usepackage{amsmath}	
\usepackage{amssymb}	
\usepackage{txfonts}
\usepackage{color, colortbl}
\usepackage{enumitem}

\newcommand{\be}{\begin{equation}}
\newcommand{\ee}{\end{equation}}

\def\rp{r^\prime}

\def\rmd{{\rm d}}
\def\vir{{\rm vir}}

\newcommand*\percent{\protect\scalebox{1}{\%}}

\title[A model for core formation by outflow episodes]{A model for core formation in dark matter haloes and ultra diffuse galaxies by outflow episodes}

\author[J. Freundlich et al.]{Jonathan Freundlich,$^{1}$\thanks{E-mail: jonathan.freundlich@mail.huji.ac.il}
Avishai Dekel,$^{1,2}$ 
Fangzhou Jiang,$^{1}$
Guy Ishai,$^{1}$
\newauthor 
Nicolas Cornuault,$^{1}$
Sharon Lapiner,$^{1}$
Aaron A. Dutton,$^{3}$
and Andrea V. Macci\`o$^{3,4}$
\\
$^1$Centre for Astrophysics and Planetary Science, Racah Institute of Physics, The Hebrew University, Jerusalem 91904, Israel\\
$^2$Santa Cruz Institute for Particle Physics, University of California, Santa Cruz, CA 95064, USA\\
$^3$New York University Abu Dhabi, PO Box 129188, Saadiyat Island, Abu Dhabi, United Arab Emirates\\
$^4$Max Planck Institute f\"ur Astronomie, K\"onigstuhl 17, 69117 Heidelberg, Germany\\
}

\date{Accepted XXX. Received YYY; in original form ZZZ}

\pubyear{2019}

\begin{document}
\label{firstpage}
\pagerange{\pageref{firstpage}--\pageref{lastpage}}
\maketitle

\begin{abstract}
%

We present a simple model for the response of a dissipationless spherical system to an instantaneous mass change at its center, describing the formation of flat cores in dark matter haloes and ultra-diffuse galaxies (UDGs) from feedback-driven outflow episodes in a specific mass range. This model generalizes an earlier simplified analysis of an isolated shell into a system with continuous density, velocity and potential profiles. The response is divided into an instantaneous change of potential at constant velocities due to a given mass loss or gain, followed by energy-conserving relaxation to a new Jeans equilibrium. The halo profile is modeled by a two-parameter function with a variable inner slope and an analytic potential profile (Dekel et al. 2017), which enables determining the associated kinetic energy at equilibrium. The model is tested against NIHAO cosmological zoom-in simulations, where it successfully predicts the evolution of the inner dark-matter profile between successive snapshots in about 75\% of the cases, failing mainly in merger situations. This model provides a simple understanding of the formation of dark-matter halo cores and UDGs by supernova-driven outflows, and a useful analytic tool for studying such processes.

\end{abstract}




\section{Introduction}
\label{section:introduction}

\subsection{The cusp-core discrepancy}

The Lambda cold dark matter (LCDM) model of structure formation is extremely successful at describing the evolution of the Universe, from the nearly homogeneous state shown by the cosmic microwave background to the present day clustered distribution of matter with galaxies, clusters of galaxies and voids between them. But while it agrees with observations on the large-scale structure of the Universe, it has faced different challenges at galactic scales. 
In particular, while LCDM numerical simulations predict steep, `cuspy' density profiles for dark matter halos, observations of dark matter dominated dwarf, low-surface-brightness and dwarf satellite galaxies as well as clusters favor shallower `cores' \citep[e.g., ][]{Flores1994,Moore1994,McGaugh1998a,vandenBosch2001, deBlok2008,deBlok2010, Kuzio2011,Oh2011,Oh2015,Newman2013a,Newman2013b,Adams2014}.
Proposed solutions to this `cusp-core' discrepancy and to potentially related challenges of LCDM cosmology such as the `too big to fail' problem\footnote{LCDM simulations predict a population of subhaloes around Milky Way-sized galaxies that have higher central surface densities than inferred from observed satellites of the Milky Way and its neighbours \citep[e.g.,][]{Boylan-Kolchin2011}} can be broadly divided into solutions considering fundamental changes in the physics of the model and solutions focusing on the baryonic processes at stake during galaxy formation and evolution. The first category comprises alternatives to LCDM such as warm dark matter, self-interacting dark matter and models that fundamentally change the gravitational law \citep[e.g., ][]{Milgrom1983a,Burkert2000, Colin2000, Goodman2000, Hu2000,
Spergel2000, Bode2001,
Gentile2011,Famaey2012,Maccio2012a, 
Destri2013, 
Peter2013, 
Marsh2014}.
Solutions invoking baryonic processes within the LCDM framework are motivated by the fact that the discrepancies between model and observations precisely occur at the scale at which baryons start to play an important role, notably through powerful stellar feedback processes that generate significant movements of the gas. Moreover, hydrodynamical simulations with different feedback implementations are able to reproduce dark matter cores \citep[e.g., ][]{Governato2010,Governato2012, Maccio2012b, Martizzi2013,Teyssier2013,DiCintio2014,Chan2015, Peirani2017}. 

Baryons mostly affect dark matter halos through their gravitational potential. When baryons cool and contract, they accumulate at the center of the halo, which steepens the potential and causes the dark matter to contract as well through adiabatic contraction \citep{Blumenthal1986}. 
When a massive object such as a satellite galaxy or a clump of gas moves with respect to the dark matter {\color{black} background, it can transfer part of its kinetic energy to the background through dynamical friction, as the concentration of particles increases in its wake and generates a drag force} \citep{Chandrasekhar1943}. 
Dynamical friction can thus `heat' the dark matter halo and contribute to remove the central cusp \citep{El-Zant2001,El-Zant2004,Tonini2006,RomanoDiaz2008,Goerdt2010,Cole2011,Nipoti2015}.
Alternatively, repeated gravitational potential fluctuations induced by stellar winds, supernova explosions and active galactic nuclei (AGNs) can also dynamically heat the dark matter and lead to the formation of a core \citep{Dekel1986, Read2005, Mashchenko2006, Mashchenko2008, Pontzen2012, Pontzen2014, Governato2012, Zolotov2012, Martizzi2013, Teyssier2013, Madau2014, Dutton2016, El-Zant2016, Peirani2017}. In this case, variations in the baryonic mass distribution induce violent potential fluctuations which progressively disperse dark matter particles away from the center of the halo.

\subsection{Ultra Diffuse Galaxies}

Deep imaging observations reveal the existence of a population of low central surface brightness ($\rm \mu_{g,0}>24~mag.arcsec^{-2}$) galaxies with stellar masses of dwarfs and effective radii comparable to that of the Milky Way ($\rm r_{eff}> 1.5\rm ~kpc$). 
These Ultra Diffuse Galaxies (UDGs) appear to be ubiquitous in groups and clusters \citep[e.g., ][]{VanDokkum2015,vanDokkum2015b,Janowiecki2015,Koda2015,Munoz2015,Mihos2015, Mihos2017, Merritt2016,vanderBurg2016,Yagi2016,Janssens2017, Shi2017} but are also observed in the field \citep[e.g.,][]{Roman2017,Leisman2017}. 
Possible formation scenarii include them being (i) failed Milky Way-like galaxies that lost their gas after forming their first stars \citep{VanDokkum2015, vanDokkum2015b, vanDokkum2016, Yozin2015, Peng2016, Lim2018}, possibly due to the denser environment in groups and clusters, (ii) the high-spin tail of the dwarf galaxy population \citep{Amorisco2016, Rong2017,Shi2017}, (iii) tidal debris from mergers or tidally disrupted dwarfs \citep{Beasley2016b,Greco2018,Jiang2018,Carleton2019} or (iv) galaxies whose spatial extend is due to episodes of inflows and outflows from stellar feedback \citep{DiCintio2017,Chan2018, Jiang2018}.

From deep images and stacks, \cite{vanDokkum2015b} and \cite{Mowla2017} find no evidence for the majority of Coma clusters UDGs to be tidally disrupted. 
Although the abundance of globular clusters in UDGs compared to dwarf galaxies of similar masses displays a significant scatter, its often large value may indicate intense star formation episodes in the past \citep{Beasley2016,Peng2016,vanDokkum2016,vanDokkum2017,Lim2018,Amorisco2018}. Some UDGs also seem to harbour signs of recent or ongoing star formation \citep{Martinez-Delgado2016,Trujillo2017,Shi2017,Pandya2018} while both \cite{Janowiecki2015} and \cite{Shi2017} report group UDGs rich in atomic HI gas.
\cite{Martin-Navarro2019} further interpret the high [Mg/Fe] abundance ratio of a field UDG with an extended star formation history as a possible result of supernova-driven gas outflows.
These observations would fit with UDGs being formed by outflows resulting from stellar feedback.
Indeed, the feedback-induced gravitational potential fluctuations leading to dark matter core formation can also cause the expansion of the stellar component \citep{Teyssier2013, Dutton2016, El-Badry2016} and this process is precisely expected to be most efficient in the mass range of UDGs \citep{DiCintio2014,DiCintio2014b,Tollet2016, Dutton2016}. In the NIHAO simulations \citep{Wang2015} studied by \cite{DiCintio2017}, prolonged and persistent bursty star formation histories lead to the simultaneous expansion of both the dark matter and stellar distributions of UDGs.

\subsection{Motivation for this work}

Although hydrodynamical simulations are able to reproduce the observed cores of low-mass dark matter haloes and a UDG population by adding prescriptions for baryonic processes such as stellar winds, stellar radiation, and supernova feedback, they do not describe nor quantify the exact mechanisms through which baryons affect the dark matter and stellar collisionless spatial distributions. 
\cite{Dekel1986} show that a significant supernova-driven gas loss can explain the low surface brightness of dwarf ellipticals, 
\cite{Pontzen2012, Pontzen2014} that a fluctuating gravitational potential can irreversibly transfer energy to collisionless particles. 
Attempting to isolate further the physical mechanism at stake during dark matter core formation, \cite{El-Zant2016} propose a theoretical model comparable to two-body relaxation in stellar systems. In this model, core formation occurs from dark matter particles experiencing successive `kicks' resulting from feedback-induced stochastic potential fluctuations, which cumulatively lead the particles to deviate from their trajectories as in a diffusion process. 
In the context of a fuzzy dark matter halo, \cite{Bar-Or2018} similarly describe the effect on stars and black holes of stochastic density fluctuations due to the large de Broglie wavelength of the dark matter particles they assume while \cite{Marsh2018} and \cite{El-Zant2019} use the \cite{El-Zant2016} model to place constraints on the fuzzy dark matter particle mass.
In yet another context, \cite{Penoyre2018} study how particles on Keplerian orbits respond to a sudden drop in the central mass, notably following how the density of the system varies as they move to higher-energy orbits.

In this article, we present and test a simple theoretical model in which core formation results from cycles of inflows and outflows. In particular, we test how an instantaneous inflow or outflow episode followed by relaxation to a new equilibrium can affect the density profile of a spherical dissipationless mass distribution. 
The model we develop builds upon that presented by \cite{Dutton2016} in their section 4.3 and recalled in Section~\ref{section:isolated}, which focuses on an isolated spherical shell of collisionless matter, neglects the full halo mass distribution and approximates the shell kinetic energy by its expression derived from the equation of virial equilibrium. As shown in Section~\ref{section:isolated}, these assumptions are not sufficient to describe the whole structure of a collisionless halo and its evolution. We provide here a consistent analytical framework to understand and predict the evolution of a collisionless spherical matter distribution when mass is added or removed at different radii and we test this model with the NIHAO simulations. We show that it accurately predicts the evolution of the dark matter density profile between two successive simulation snapshots in a large majority of cases, failing mostly in violent perturbed situations of major mergers.

This paper unfolds progressively as follows.  
In section~\ref{section:isolated}, we recall the isolated shell model proposed by \cite{Dutton2016} and highlight its main limitations.
In section~\ref{section:halo}, we present our new spherical collisionless halo response model for an instant mass change, which enables retrieving the evolution of the halo mass distribution using accurate analytic parametrizations of the potential and kinetic energy stemming from the \cite{Dekel2017} density profile. 
In section~\ref{section:test}, we successfully test this model on NIHAO \citep{Wang2015} cosmological zoom-in simulations. 
In section~\ref{section:discussion}, we discuss the reasons for the model's failures and its ability to predict the evolution of the halo mass distribution over multiple episodes of inflows and outflows.
In section~\ref{section:conclusion}, we summarize our results and conclude.

\section{Isolated shell model and its limits}
\label{section:isolated}

\subsection{The isolated shell model of Dutton et al. (2016b)}
\label{section:Dutton2016}

In their Section 4.3, \cite{Dutton2016} consider the evolution of a spherical halo shell enclosing a dark matter mass $M$ initially at radius $r_i$ when a baryonic mass $m$ is instantaneously added (or removed, in which case $m<0$) at the shell center. 
They assume that the halo is spherical, that orbits are circular, that shells do not cross and that energy is conserved between shells enclosing a given collisionless mass before and after the mass inflow (outflow).
However, they ignore variations in the mass outside the shell and assume that the expression of the kinetic energy stemming from the virial theorem holds for an isolated shell.

Since the mass variation occurs on a sufficiently short time scale compared to the dynamical time of the system, \cite{Dutton2016} assume an intermediate transitional state in which the gravitational potential adapts instaneously while the velocities remain frozen at their initial values, following the impulse approximation. They write the initial energy associated with the shell enclosing an isolated dark matter mass $M$ at radius $r_i$
\begin{equation}
\label{eq:Dutton-Ei}
\begin{array}{@{}l}
\displaystyle E_i = U_i +K_i\\
\displaystyle \text{with }~ U_i = -\frac{GM}{r_i} ~\text{ and }~ K_i = \frac{1}{2} \frac{GM}{r_i}, 
\end{array}
\end{equation}
which becomes 
\begin{equation}
\label{eq:Dutton-Et}
\begin{array}{@{}l}
\displaystyle E_t = U_t +K_t\\
\displaystyle \text{with }~ U_t= U_i-\frac{Gm}{r_i} ~\text{ and }~ K_t = K_i
\end{array}
\end{equation}
right after the mass variation in this intermediate transitional state. The potential energy per unit mass $U$ corresponds to an isolated spherical system of mass $M$, while the expression for the kinetic energy $K$ is assumed to be that stemming from the virial theorem, even if it is applied to a single shell and not to the whole halo. 
The system described by Eq.~(\ref{eq:Dutton-Et}) right after the mass variation is out of equilibrium but the shell eventually settles in a new equilibrium state at radius $r_f$, where the kinetic energy is again given by the virial theorem. The final energy is then written
\begin{equation}
\label{eq:Dutton-Ef}
\begin{array}{@{}l}
\displaystyle E_f = U_f +K_f\\
\displaystyle\text{with }~ U_f= -\frac{G(M+m)}{r_f} ~\text{ and }~ K_f = \frac{G(M+m)}{2 r_f}. 
\end{array}
\end{equation}
Assuming energy conservation for the shell after the mass variation, i.e., $E_f = E_t$, they derive an expression for the ratio $r_f/r_i$ as a function of the ratio $f=m/M$ (their Eq.~(16)). 

\cite{Dutton2016} further consider the case in which a slow inflow of mass $m$ is followed by an instantaneous outflow of mass $\eta m$ as well as a succession of multiple such episodes. 
Expressing the conservation of gas mass by the bathtub model \citep{Bouche2010, Lilly2013, Dekel2014} and assuming that all baryons accreted at the virial radius penetrate through the halo into the central galaxy, they relate $\eta$ to the integrated star formation efficiency $\epsilon_{\rm SF} = M_\star/f_{\rm b} M_{\rm vir}$ with $\eta = 1-\epsilon_{\rm SF}$,
where $M_\star$ is the stellar mass, $f_{\rm b}$ the average baryon fraction in the Universe and $M_{\rm vir}$ the virial mass.
Adjusting the number of cycles $N$ to match the simulations given $\epsilon_{\rm SF}$, they obtain simulated halo responses comparable to those expected from the model (cf. their Fig.~14).  
This agreement supports the general idea according to which the cusp-core transformation occurs through cycles of inflows and outflows and further justifies trying to develop a more precise and more predictive model. 

\begin{figure}
	\centering
	\includegraphics[width=1\linewidth,trim={0cm 4.4cm 0.4cm 1.4cm},clip]{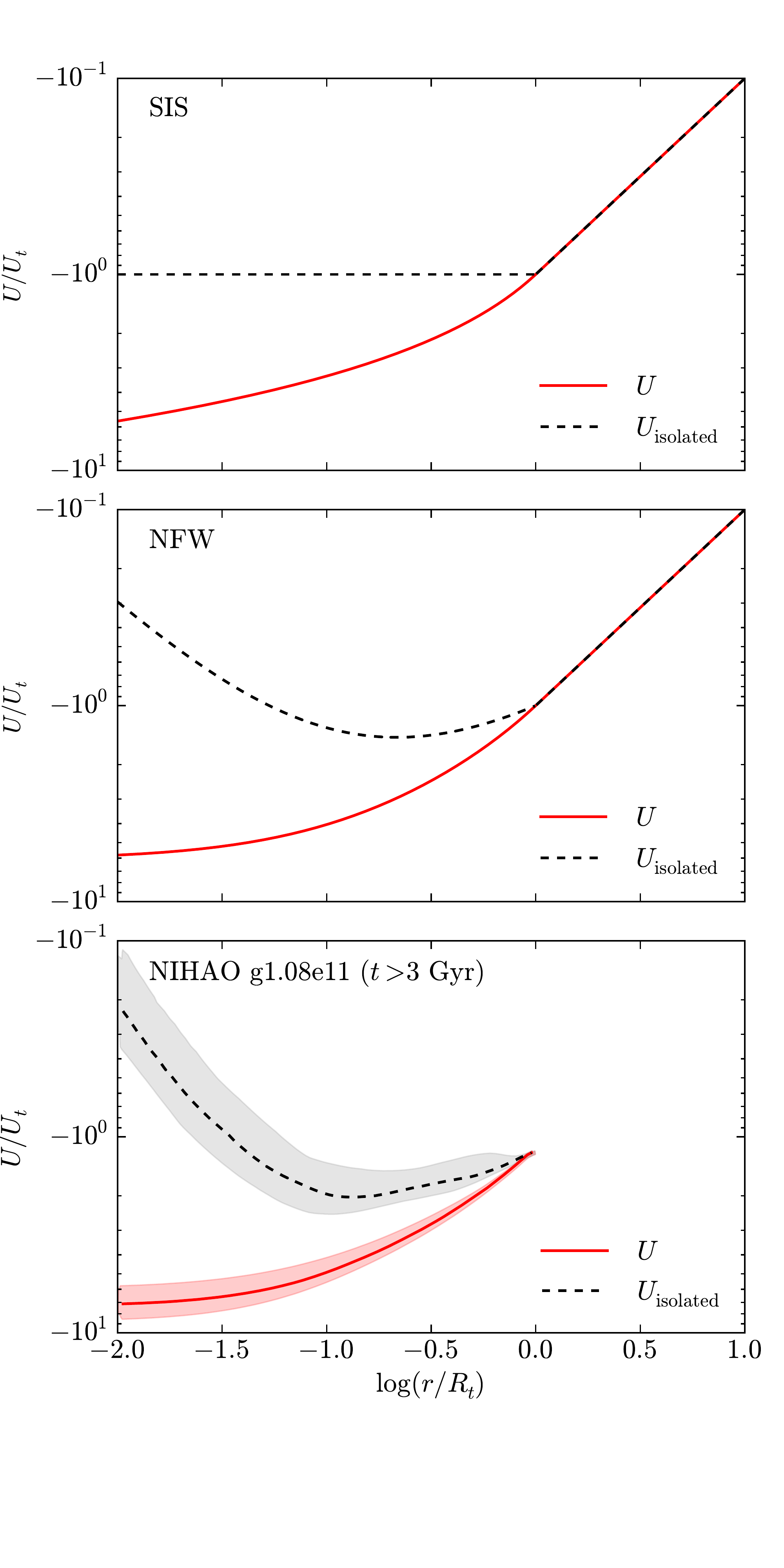}
	\vspace{-0.4cm}
	\caption{Components of the gravitational potential energy of test particles at different radii in a singular isothermal sphere (top), in an NFW halo of concentration $c=10$ (middle) and in all outputs after 3 Gyr of the NIHAO simulation \texttt{g1.08e11} (bottom), expressed in units of $U_t= GM_t/R_t$. The solid red line corresponds to the potential energy derived from Eq.~(\ref{eq:Urb}), integrated analytically for the top and middle panels, numerically for the bottom one. The dashed line corresponds to $U_{\rm isolated}(r)= -GM(r)/r$, i.e., to the potential when neglecting the outer layers. 
		For the bottom panel, we take $R_t = R_{\vir}$ the virial radius, $M_t = M_{\vir}$ the virial mass and only show the profiles within $R_{\vir}$. 
		In this case, lines correspond to the median curves while the shaded areas highlight the ranges. The discrepancy between solid red and dashed black lines highlights the need for taking the halo outer layers into account when evaluating the gravitational potential.}
	\vspace{-0.4cm}
	\label{fig:U-all}
\end{figure}

\subsection{Gravitational potential of a non-isolated shell}
\label{section:wholehalo}

\cite{Dutton2016} only take into account the enclosed mass in the potential energy although the mass distribution in the outer shells matters because its response to the central mass change affects the potential at the shell in question. 
Indeed, the potential energy per unit mass $U(r)$ is minus the energy needed to extract a unit mass from $r$ to infinity, or equivalently the work done by the gravitational force between $r$ and infinity: 
\begin{equation}
\label{eq:Ur}
\displaystyle U(r)  = \int_r^{+\infty} F(y)dy
       = -\int_r^{R_t} \frac{G M(y)}{y^2} dy - \frac{GM_t}{R_t},
\end{equation}
assuming that the halo is truncated at a certain radius $R_t$ enclosing a mass $M_t$. 
Integrating by parts leads to
\begin{equation}
\label{eq:Urb}
U(r) = - \frac{G M(r)}{r} -G \int_r^{R_t} \frac{1}{y}\frac{dM}{dy} dy = - \frac{G M(r)}{r} - \int_r^{R_t} 4\pi y G \rho(y) dy;  
\end{equation}
hence $U(r)$ is the sum of a contribution corresponding to the shells inside $r$, $U_{\rm isolated}=-GM(r)/r$, and of a contribution of the shells outside $r$ given by the remaining integral.

Except when applied to the outer layer at $R_t$, the gravitational potential is not equal to $U_{\rm isolated}$ as was assumed by \cite{Dutton2016}, and the contribution of the outer shells is always negative: adding the outer shells leads to a deeper potential as it becomes more difficult to extract a particle from any radius owing to the additional mass. 
As an example, Fig.~\ref{fig:U-all} shows how taking into account the outer layers affects the potential energy per unit mass of test particles at different radii in the cases of a singular isothermal sphere (SIS), of an NFW profile and in a simulated galaxy from the NIHAO sample \citep[\texttt{g1.08e11}; ][]{Wang2015}. 
The figure focuses on simulated outputs after 3 Gyr in order to put aside the more erratic first outputs corresponding to the building up of the halo. As can be seen in the figure, neglecting the outer layers induces a shallower gravitational potential. 
More importantly, changes in the outer mass distribution change the potential at the radius in question. In fact, the bottom panel of Fig.~\ref{fig:U-all} also shows that most of the changes in the simulated shell potential $U(r)$ come from changes in the outer mass distribution. Indeed, $U_{\rm isolated}(r)$ is more than an order of magnitude smaller than $U(r)$ in absolute value below $0.1 R_{\rm vir}$, where changes in the potential are the most significant.
In Section~\ref{section:density}, we indicate the expression of the potential derived from Eq.~(\ref{eq:Ur}) for a density profile parametrized following \cite{Dekel2017}.

\subsection{Shell kinetic energy and virial theorem}
\label{section:virial}

\begin{figure}
	\centering
	\includegraphics[width=1\linewidth,height=0.615\linewidth,trim={0.6cm .6cm 0.4cm .1cm},clip]{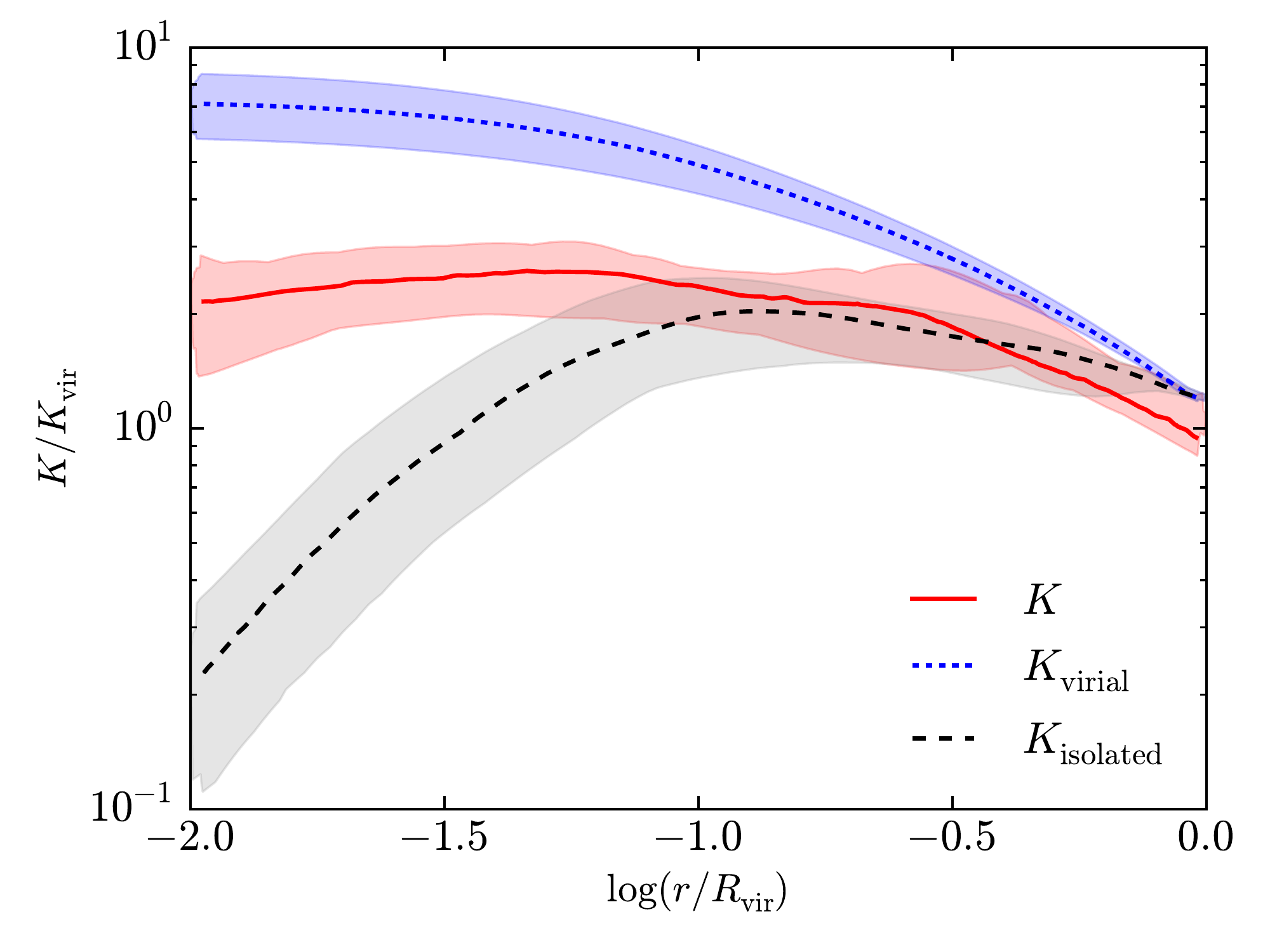}
	\vspace{-0.4cm}
	\caption{Comparison between the simulated local kinetic energy per unit mass $K$ in the different outputs of NIHAO galaxy \texttt{g1.08e11} after 3 Gyr (red), that derived from the equation of virial equilibrium $K_{\rm virial}=-U/2$ where $U$ is determined from Eq.~(\ref{eq:Urb}) (blue) and $K_{\rm isolated}=GM/2r$ (dashed black). Since the additional term corresponding to the outer layers in Eq.~(\ref{eq:Urb}) is always negative, $K_{\rm virial}$ is always above $K_{\rm isolated}$. The kinetic energy is in units of $K_{\rm vir} = G M_{\rm vir}/2R_{\rm vir}$. Lines correspond to the median curves while the shaded areas highlight the ranges.
		The discrepancy between the different curves shows that the local kinetic energy of a shell within a halo can not be derived from the equation of virial equilibrium, except near the virial radius.}
	\label{fig:Tvirial}
\end{figure}

The scalar virial theorem \citep[e.g.,][section 4.8.3]{BinneyTremaine2008} states that the total kinetic energy of a collisionless system in a steady equilibrium state without surface pressure is half its total potential energy. 
\cite{Dutton2016} assume that this is still valid for a shell at radius $r$, which leads them to write the kinetic energy per unit mass of that shell as
\begin{equation}
K_{\rm isolated}(r) = \frac{1}{2} \frac{GM}{r}
\end{equation}
where $M$ is the total mass enclosed within $r$ and where they neglect the contribution of the outer shells when expressing the potential.
Fig.~\ref{fig:Tvirial} compares the local kinetic energy per unit mass in the simulated NIHAO galaxy \texttt{g1.08e11} with $K_{\rm isolated}$ and $K_{\rm virial}=U/2$, where the potential energy per unit mass $U$ stems from a numerical integration of Eq.~(\ref{eq:Urb}) as in Fig.~\ref{fig:U-all}. 
The local kinetic energy per unit mass associated with a shell at radius $r$ was computed as
\be
\label{eq:Kdef}
K(r) = \frac{d{K(<r)}}{dM} = \frac{d{K(<r)}}{dr} \frac{dR}{dr}
\ee
where $K(<r)$ is the enclosed kinetic energy and $M(r)$ the enclosed mass\footnote{
	More precisely, $K$ is determined using Eq.~(\ref{eq:Kdef}) with 
	${K(<r)} =\sum m_i v_i^2/2 $ and 
	$M(r) =\sum m_i$
	summing over all particles within radius $r$; 
	the derivatives $dK(<r)/dr$ and $dM/dr$ are numerically obtained with a Savitzky-Golay smoothing filter \citep[][implemented as \texttt{scipy.signal.savgol\_filter} in \texttt{scipy}]{SavitzkyGolay1964} with a polynomial order $n=3$ and a window size $w=11$ (optimized to smooth out the radial oscillations of the profile and to capture its average shape).  
}.
As can be seen in the figure, neither $K_{\rm isolated}$ nor $K_{\rm virial}$ follow the simulated kinetic energy $K$. The discrepancy between  $K$ and $K_{\rm virial}$ shows that the expression of the kinetic energy derived from the equation of virial equilibrium is not valid for shells embedded in a larger halo, while the discrepancy between $K$ and $K_{\rm isolated}$ shows the limits of this latter expression, in particular towards the center of the halo at $r<0.05 R_\vir$, where we explore the cusp-core evolution. 
The discrepancy between $K$ and $K_{\rm virial}$ can not be explained by non-isotropic velocities, since the anisotropy parameter is close to zero at most radii and in particular towards the centre (cf. Appendix~\ref{appendix:symmetry}, Fig.~\ref{fig:kinetic_beta}).
In section~\ref{section:kinetic}, we carefully adress the parametrization of the kinetic energy, notably by obtaining an analytical expression deriving from the Jeans equation for a density profile parametrized following \cite{Dekel2017}.

\section{Halo response model}
\label{section:halo}

\subsection{Instant mass change}
\label{subsection:model}

We aim at describing the response of a spherical collisionless halo with an initial mass profile $M_i(r)$ when mass is added (or removed) instantaneously about the center with a profile $m(r)$, and we wish to determine the final collisionless halo mass profile $M_f(r)$. Positive values of $m$ correspond to inflows, negative values to outflows. As in \cite{Dutton2016}, we consider the evolution of a spherical halo shell enclosing a given collisionless mass $M$, initially at radius $r_i$ and eventually at radius $r_f$ so that $M = M_i(r_i) = M_f(r_f)$. 

We assume that the shell is initially at equilibrium (\textit{stage~1}), its total  energy per unit mass being 
\begin{equation}
\label{eq:Ei}
E_i(r_i) = U_i(r_i) +K_i(r_i),  
\end{equation}
where $U_i$ and $K_i$ are respectively the gravitational potential and kinetic energy per unit mass.
Contrarily to \cite{Dutton2016}, we neither neglect the outer shells in the expression of the potential nor assume that the kinetic energy is that derived from the equation of halo virial equilibrium. We keep functional forms for $U_i$ and $K_i$ for the moment, which will become explicit in the following sections~\ref{section:density} and \ref{section:kinetic}. The potential $U_i(r_i)$ can be derived from the  mass profile $M_i(r)$ while the kinetic energy $K_i(r)$ is set by Jeans equilibrium.

When a given mass $m$ is instantaneously added (or removed, if $m<0$) inside $r_i$, we assume an intermediate transitional state (\textit{stage~2}) in which the potential immediately adapts to the mass variation while the velocities are frozen to their initial values and the shell still at radius $r_i$. 
This stage notably corresponds to what happens to particles on Keplerian orbits subjected to a sudden change in the central mass, as in \cite{Penoyre2018}.
The total energy of the shell right after the mass variation is 
\begin{equation}
\label{eq:Et}
\begin{array}{@{}l}
\displaystyle E_t(r_i) = U_t(r_i) +K_i(r_i)\\
\text{with }~ \displaystyle U_t(r_i)= U_{i}(r_i)-\frac{Gm}{r_i}, 
\end{array}
\end{equation}
where we further assume that the mass inflow (or outflow) is spherically symmetric and limited to within $r_i$ so that its contribution to the potential is simply $-Gm/r_i$. This expression for the potential is rigorously valid when mass is added (or removed) directly at the center of the halo, but it can also describe an inflow/outflow mass profile $m(r_i)$ provided that the shell of interest is outside the main body of the mass variations, i.e., that the contribution of the mass added (removed) at $r>r_i$ to the potential is small compared to $Gm/r_i$. 
In Section~\ref{sub:Km}, we will use the same assumption to express the contribution of the added (removed) mass to the local kinetic energy.

The system at this stage is out of equilibrium, and we assume that it subsequently relaxes to a new equilibrium state (\textit{stage~3}) where the total energy per unit mass of the shell, now at $r_f$, is
\begin{equation}
\label{eq:Ef}
\begin{array}{@{}l}
\displaystyle E_f(r_f) = U_f(r_f) +K_f(r_f,m)\\
\text{with }~ \displaystyle U_f(r_f) = U_{f}(r_f)-\frac{Gm}{r_f}.
\end{array}
\end{equation}
In this expression, the contribution to the potential of the additional mass has been singled out and the local kinetic energy $K_f$ is set by the new equilibrium state.

We now postulate that the total energy is the same for corresponding shells before and after the relaxation, $E_f(r_f) = E_t(r_i)$.
If there is no shell crossing following the sudden mass change (e.g., as expected in the case of contraction due to added mass), we can have $r_f$ and $r_i$ refer to the same Lagrangian shell, and then indeed expect conservation of energy based on the energy conservation in the whole halo and the approximate self-similarity of the halo density profile over a range of radii. When there is shell crossing, 
we make the ansatz that the equal energy equation is still valid but for shells  encompassing the same mass before and after the relaxation even when they do not represent the same Lagrangian shell. 
This approximation is not formally justified a priori -- its validity will be verified by the success or failure of the model in reproducing the response in simulations.
We note that in \cite{Penoyre2018}, shell crossing manifests itself by the presence of caustics in the perturbed radial density profile where shells overlap, which we do not take into account in our parametrisation of the density profile.
Combining Eqs.~(\ref{eq:Et}) and (\ref{eq:Ef}), the energy equality yields 
\begin{equation}
\label{eq:Et-Ef}
U_{i}(r_i)-\frac{Gm}{r_i} + K_i(r_i) = U_{f}(r_f)-\frac{Gm}{r_f} + K_f(r_f,m).
\end{equation}

Given functional forms $U(r;p)$ and $K(r;p,m)$ with parameters $p$, where $K$ stems from Jeans equilibrium, it is possible to solve numerically this equation in order to obtain the final radius $r_f$ and the final parameters $p_f$ entering in the expressions of $U(r;p)$ and $K(r;p,m)$ from given initial conditions ($r_i$, $p_i$) and the mass change $m(r_i)$. Note that the expression of the kinetic energy needs to account for the mass that is added to (or removed from) the system, as we would otherwise be left with the trivial solution $r_f = r_i$ and $p_f = p_i$. 
One can apply Eq.~(\ref{eq:Et-Ef}) to a set of $N$ values of $r_i$ in a certain range of interest (for example towards the center of the halo to describe the cusp-core transition) and thus obtain $N$ constraints that should be approximately fullfilled. By finding the best-fit values for the final parameters $p_f$ that minimize the residuals between the two sides of Eq.~(\ref{eq:Et-Ef}) across all the $N$ radii, the toy model presented here makes a prediction for the evolution of the collisionless density profile when mass is added (or removed) to the halo. 
The functional forms for the gravitational potential and the kinetic energy can be derived under certain assumptions from the density profile or its  parametrisation and from the Jeans equation. In the following section~\ref{section:density}, we describe one such parametrisation and the resulting expression for the gravitational potential. In section~\ref{section:kinetic}, we derive the corresponding kinetic energy from the Jeans equation describing the steady-state equilibrium of a collisionless system, assuming complete spherical symmetry and an isotropic velocity dispersion.

To summarize, the key ingredients and main assumptions of the model are: 
(1) the system is spheri-symmetric with a given profile that is described by a parametric function, and the velocities are isotropic; 
(2) the initial halo is in Jeans equilibrium; 
(3) the mass change is instantaneous, with a mass profile $m(r)$ about the halo center;  
(4) the change puts the halo system in an intermediate out-of-equilibrium state in which the gravitational potential is changed instantaneously while the velocities are frozed at their initial values; 
(5) the halo relaxes to a new Jeans equilibrium with a final profile that is described by the same function as initially but with new parameters; 
(6) the energy is the same before and after the relaxation process for shells that enclose a given halo mass. %

\begin{figure}
	\centering
	\includegraphics[width=1\linewidth,trim={0.6cm 4.4cm 0.2cm 3.7cm},clip]{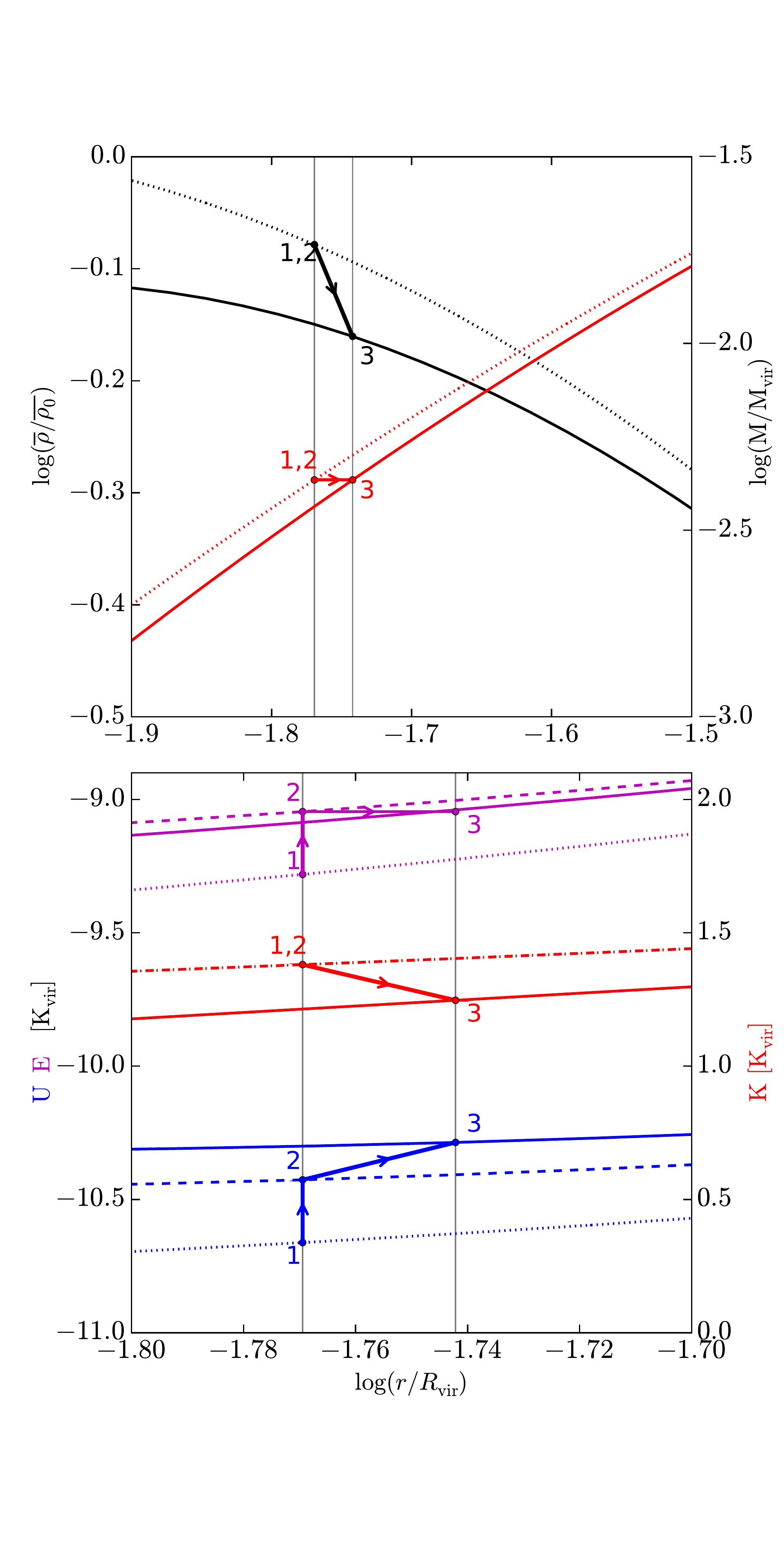}
	\vspace{-0.4cm}
	\caption{
		{\it Top}: Evolution of the density and mass profiles when a mass equal to 2\% of the virial mass is removed at the center of an ideal fiducial collisionless halo whose density profile is given by Eq.~(\ref{eq:rho32}) with parameters taken from a fit to the $t=6.7~\rm Gyr$ output of NIHAO galaxy \texttt{g1.08e11}. The average density is normalized by its initial value at $0.01 R_\vir$, $\overline{\rho}_0$.
		\protect\linebreak\mbox{\it Bottom}: Potential, kinetic and total energy in the different stages assumed by the model: {\it stage 1} (dotted line) corresponds to the initial equilibrium state of the halo described by Eq.~(\ref{eq:Ei}), {\it stage 2} (dashed line) to the intermediate out-of-equilibrium state described by Eq.~(\ref{eq:Et}), where the potential has adapted to the mass change while the velocities remain frozen to their initial values, and {\it stage 3} (plain line) to the final equilibrium state of the halo described by Eq.~(\ref{eq:Ef}). The gravitational potential is given by Eq.~(\ref{eq:U32}) and the kinetic energy by Eq.~(\ref{eq:sigma_dekel}) given the profile parameters. The energies are in unit of the virial kinetic energy $K_{\vir} = G M_{\vir}/2R_{\vir}$. 
	}
	\label{fig:steps}
\end{figure}

\begin{figure*}
	\centering
	\includegraphics[width=0.86\textwidth,trim={0cm 0cm 0cm 0.8cm},clip]{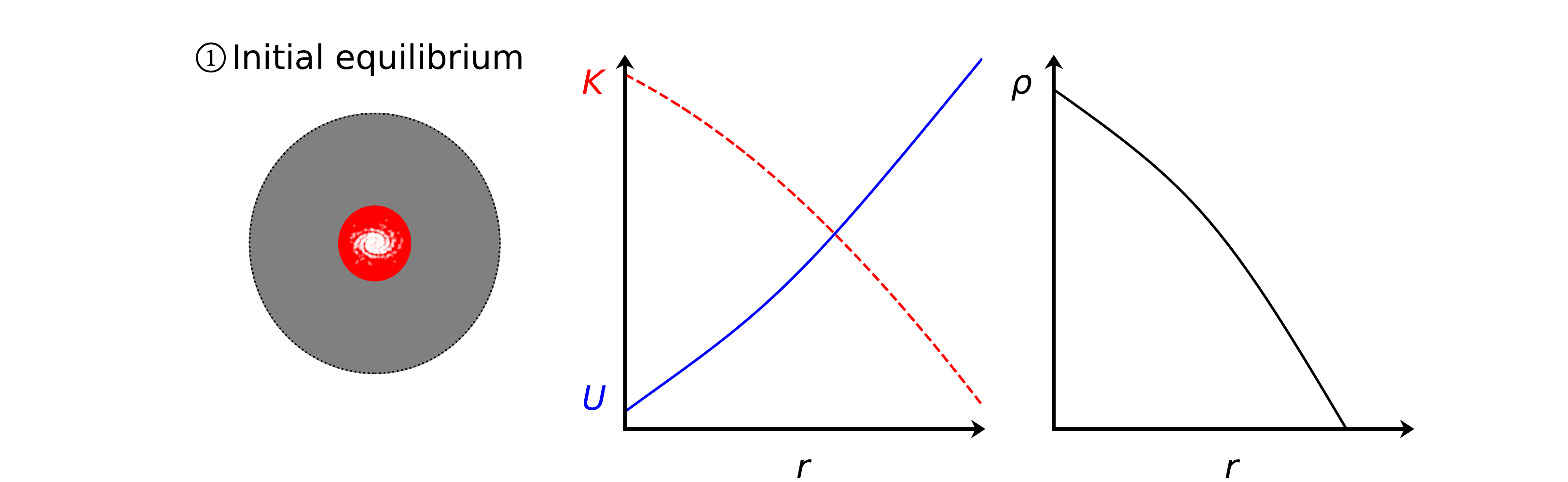}\vspace{0.2cm}      
	\includegraphics[width=0.86\textwidth,trim={0cm 0cm 0cm 0.8cm},clip]{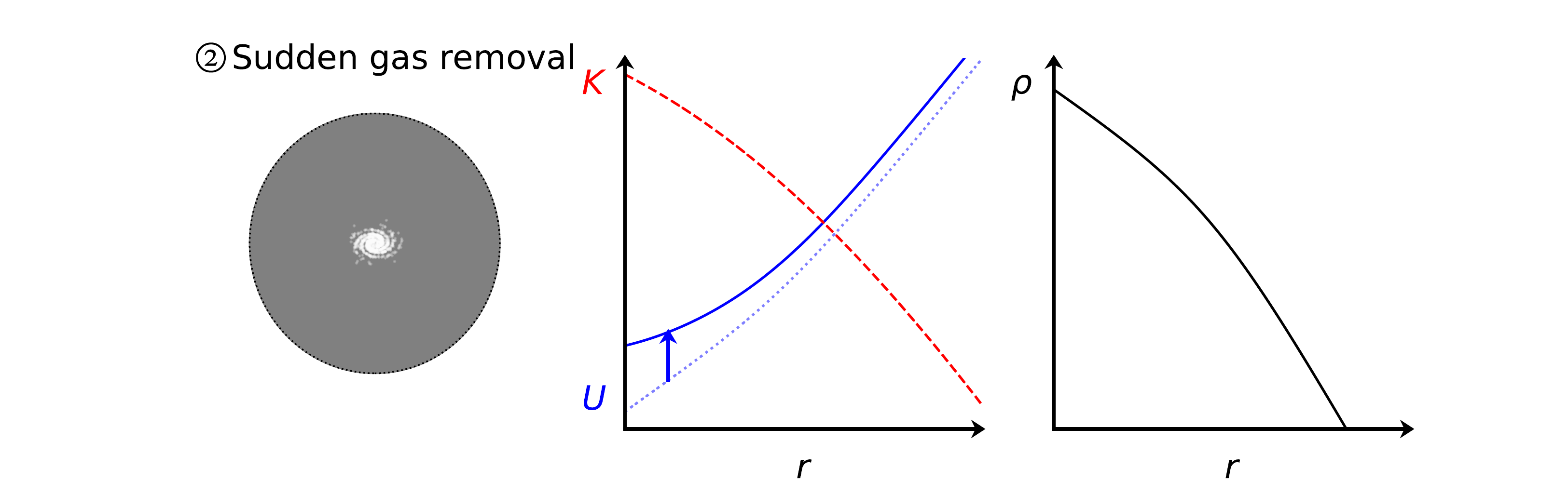}\vspace{0.2cm}       
	\includegraphics[width=0.86\textwidth,trim={0cm 0cm 0cm 0.8cm},clip]{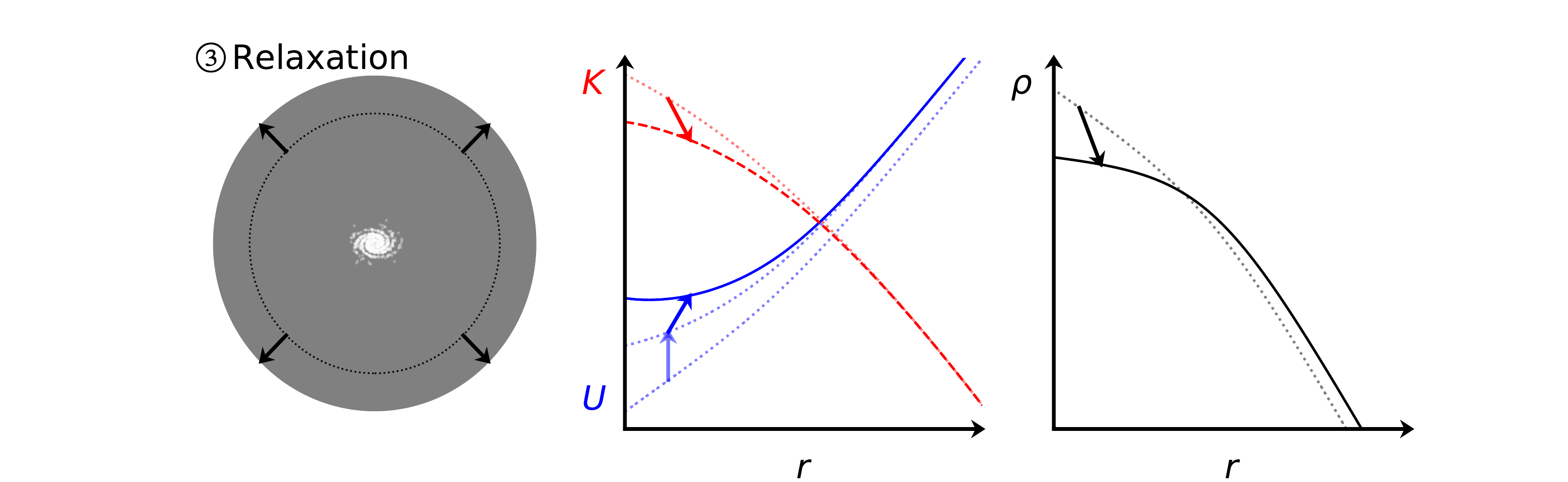}  
	\caption{Schematic representation of the different stages assumed by the model for a gas outflow episode affecting a dark matter halo: (1) the dark matter halo (in gray in the left panel), which hosts gas (in red) and stars (in white) at its center, is initially at equilibrium so its kinetic energy $K$ (dashed red line in the middle panel) stems from the Jeans equation while its gravitational potential energy $U$ (plain blue line) can be deduced from its density profile $\rho$ (right panel); (2) after a sudden gas mass loss, the potential adjusts instantly while the velocities and the kinetic energy remain frozen to their initial values; (3) the halo relaxes to a new equilibrium at constant energy $U+K$, leading to the expansion of the dark matter distribution.}
	\label{fig:cartoon}
\end{figure*}

Fig.~\ref{fig:steps} illustrates the assumed steps of this model and the evolution of the density profile for a fiducial choice of the initial collisionless mass distribution and a mass $m$ equal to 0.2\% of the virial mass directly removed at the center. The initial density profile corresponds to a fit to output 32 ($t=7.14~\rm Gyr$) of the NIHAO galaxy \texttt{g1.08e11} according to Eq.~(\ref{eq:rho32}) below.  The gravitational potential is given by Eq.~(\ref{eq:U32}) and the kinetic energy by Eq.~(\ref{eq:sigma_dekel}).  
As can be seen in the figure, the density profile becomes shallower when the mass is removed: the inner slope at $0.01 R_\vir$ drops from 0.37 to 0.25.
A given collisionless mass $M$ initially enclosed within radius $r_i$ is enclosed within $r_f>r_i$ after the relaxation, while the density decreases.
Following the model, the equilibrium state of the initial shell enclosing $M$ at $r_i$ is described by Eq.~(\ref{eq:Ei}) ({\it stage~1}); the counterpart of this shell during the transitional out-of-equilibrium state verifies Eq.~(\ref{eq:Et}) ({\it stage~2}), where the potential energy has increased because of the mass removal but the kinetic energy and radius remain those of the initial shell; and the equibrium state of final shell enclosing $M$ at $r_f$ is characterized by Eq.~(\ref{eq:Ef}) ({\it stage~3}). The potential energy associated to the final stage is higher than the initial one, but the kinetic energy lower. 
Minimizing Eq.~(\ref{eq:Et-Ef}) for shells at different initial radii enables to retrieve both the parameters of the final density profile and the final characteristics of each shell. 
Fig.~\ref{fig:cartoon} proposes a schematic representation of the different stages assumed by the model in the case of a sudden gas outflow towards the center of a dark matter halo. The cartoon is inspired by Fig.~\ref{fig:steps} but includes a stellar component for illustration.

\subsection{Parametrisation of the density profile}
\label{section:density}

\cite{Dekel2017} propose a parametrisation of the density profile of dark matter haloes with a free inner slope and an analytic potential, developed to describe the transition from cusps to cores and alterations of the profile due to environmental effects. 
This parametrisation relies on expressing the \emph{mean} density profile within a sphere of radius $r$ as:
\begin{equation}
\label{eq:rhob}
\overline{\rho}(r) = \frac{\overline{\rho_c}}{x^a (1+x^{1/b})^{b(g-a)}}
\end{equation}
where $\overline{\rho_c}$ is a characteristic density, $x=r/r_c$ with $r_c$ an intermediate characteristic radius, $a$ and $g$ the inner and outer asymptotic slopes and $b$ a middle shape parameter. As the virial radius $R_{\vir}$ is set by cosmology for a given halo mass, this functional form effectively depends on four shape parameters: $a$, $b$, $g$ and $c=R_{\vir}/r_c$, to be reduced to two free parameters in our analysis. 
Indeed, the normalisation factor $\overline{\rho_c}$ can be expressed as
$\overline{\rho_c}= c^3 \mu \overline{\rho_\vir}$, 
with 
$\mu = c^{a-3} (1+c^{1/b})^{b(g-a)}$, 
and $\overline{\rho_{\rm vir}}$ the mean mass density within $R_{\vir}$.
When $g=3$ and $b=2$, the local density profile is
\begin{equation}
\label{eq:rho32}
\rho(r)  =  \frac{3-a}{3} \frac{c^3 \mu \overline{\rho_\vir}}{x^a (1+x^{1/2})^{2(3.5-a)}} 
\end{equation}
with two remaining free parameters ($a$ and $c$).
The associated potential energy stemming from Eq.~(\ref{eq:Urb}) is 
\begin{equation}
\label{eq:U32}
\medmuskip=0.5mu
\thinmuskip=0.5mu
\thickmuskip=0.5mu
\displaystyle  U_{\rm DM}(r) =-V_\vir^2~\Bigg(1 + 2c\mu \left[ \frac{\chi_\vir^{2(2-a)}-\chi^{2(2-a)}}{2(2-a)} - \frac{\chi_\vir^{2(2-a)+1}-\chi^{2(2-a)+1}}{2(2-a)+1} \right]\Bigg) 
\end{equation}
when $a\neq 2$ or $5/2$, with 
$\chi = x^{1/2}(1+x^{1/2})^{-1}$, 
$\displaystyle \chi_\vir = \chi(c)$ 
and $V_\vir=GM_\vir/R_\vir$, as detailed in Appendix \ref{appendix:dekel}. 

We caution that parameter $a$ is not the slope at the resolution limit ($0.01 R_\vir$ in the case of the NIHAO simulations) and that $c$ does not necessarily reflect the actual concentration of the halo as for an NFW \citep{NFW1996,NFW1997} profile. In fact, these two parameters can differ significantly from the concentration and the inner slope. The logarithmic slope of the density profile when $g=3$ and $b=2$ is indeed %
\begin{equation}
\label{eq:s}
s(r) = - \frac{d\log{\rho}}{d\log r} = \frac{a+3.5 x^{1/2}}{1+x^{1/2}} 
\end{equation}
so negative values of parameter $a$ can be compatible with a positive logarithmic slope at the resolution limit, in particular for large values of $c$ since $x=c r/R_\vir$. 
For the NIHAO simulations, we define 
\vspace{-0.4cm}
\begin{equation}
\label{eq:s0}
s_0 = s(0.01 R_\vir)
\end{equation}
the inner logarithmic slope at the resolution limit.
We tried enforcing a positive value for $s_0$, but it did not improve the fits so we subsequently prefered to leave both parameters $a$ and $c$ free.
Given the expressions for the mass and velocity profiles that can be derived from Eq.~(\ref{eq:rhob}) (cf. Appendix \ref{appendix:dekel}), the velocity peaks at $x_{\rm max} = (2-a)^2$ when $g=3$ and $b=2$ such that 
\be
\label{eq:cmax}
c_{\rm max}=\frac{c}{(2-a)^2}
\ee
defines a more physical concentration parameter than $c$, coinciding with the latter only when $a=1$. General expressions for $s$ and $c_{\rm max}$ are given in \cite{Dekel2017} and recalled in Appendix~\ref{appendix:dekel}. 
At fixed $b$, $g$, $R_\vir$ and $M_\vir$, there is a bijection between the couple $(a,c)$ and $(s_{\rm 0}, c_{\rm max})$.

\begin{figure}
	\centering
	\includegraphics[width=1\linewidth,height=0.695\linewidth,trim={0.3cm .3cm 0.2cm .1cm},clip]{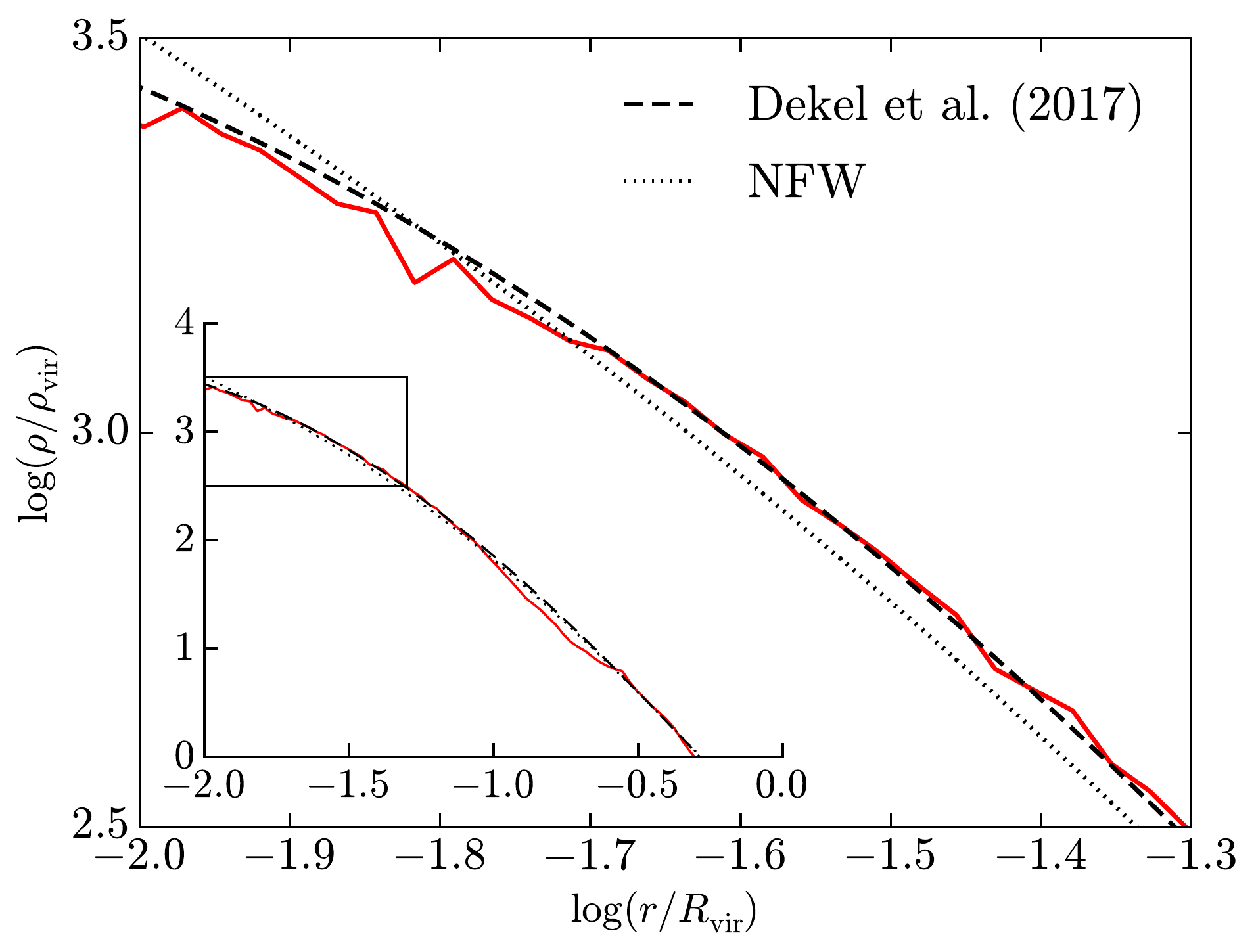}
	\vspace{-0.4cm}
	\caption{
		The dark matter density profile at $t=6.7~\rm Gyr$ for the NIHAO galaxy \texttt{g1.08e11} (plain red line) together with its fit according to the \protect\cite{Dekel2017} parametrisation of Eq.~(\ref{eq:rho32}) (dashed line). The dotted line correspond to an NFW fit. All fits were carried out within the whole range between $0.01 R_{\vir}$ and $R_{\vir}$. 
		The \protect\cite{Dekel2017} fit better accounts for the inner part of the density profile than the NFW fit.
	}
	\label{fig:dekel_nfw}
\end{figure}

\begin{figure}
	\centering
	\includegraphics[width=1\linewidth,trim={0.cm 2.8cm .6cm 1.cm},clip]{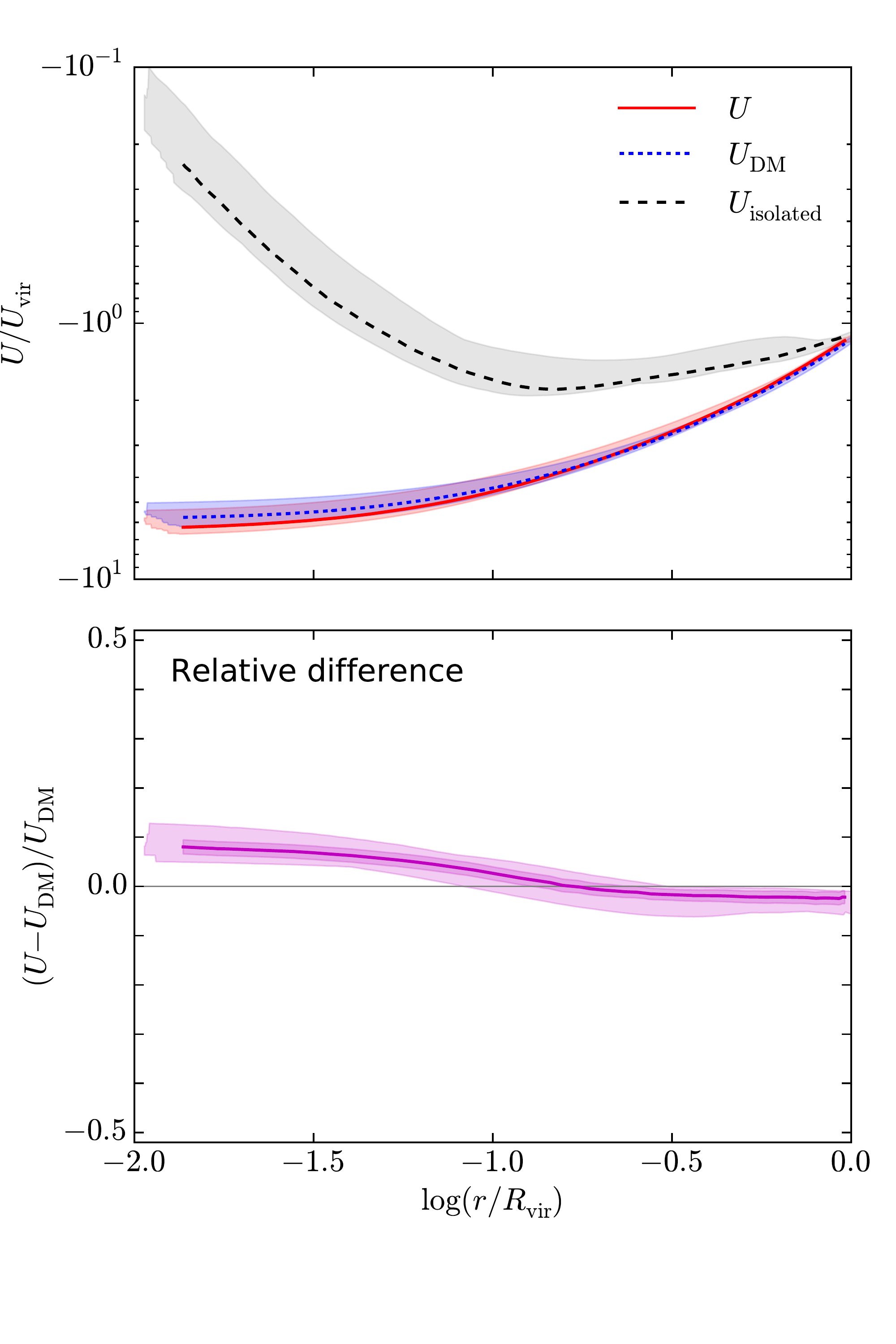}
	\vspace{-0.4cm}
	\caption{
		Comparison between the simulated gravitational potential energy $U(r)$ (plain red), obtained by numerically integrating Eq.~(\ref{eq:Urb}), and $U_{DM}$ derived from the \protect\cite{Dekel2017} parametrisation using Eq.~(\ref{eq:U32}) (dotted blue)  in the different outputs of NIHAO galaxy \texttt{g1.08e11} after 3 Gyr (top panel) together with the relative difference $(U-U_{DM})/U_{DM}$ (bottom panel). The mean RMS relative difference is 4.9\%. The potential is in units of $U_{\rm vir}=GM_{\rm vir}/R_{\rm vir}$ the potential energy at the virial radius. 
		The black dashed line recalls $U_{\rm isolated}(r)=-GM(r)/r$ assumed by \protect\cite{Dutton2016}, i.e., the potential obtained when neglecting the outer layers -- cf. Appendix~\ref{section:wholehalo}. Lines correspond to the median curves while the shaded areas highlight the ranges. In the lower panel, an additional shade shows the standard deviation from the median.
		The \protect\cite{Dekel2017} parametrisation fit to the average density profile yields the correct gravitational potential with a mean relative error $<5\%$.
	}
	\label{fig:Udekel}
\end{figure}

The functional form proposed by \cite{Dekel2017} enables to describe different inner and outer slopes for the density profile, contrarily to the usual NFW and \cite{Einasto1965} profiles, and allows analytical expressions for the density profile, the integrated mass profile as well as for the gravitational potential and the velocity dispersion when $g=3$ and $b$ a natural number. 
\cite{Dekel2017} show that this functional form yields excellent fits for haloes in simulations with and without baryons, ranging from steep cusps to flat cores. 
In particular, they show that the parametrisation with $g=3$ and $b=2$ matches simulated profiles at least as well as the usual NFW and Einasto profiles but captures cores better in addition to providing fully analytical expressions for the density, the mass and the gravitational potential. 
This motivates us to adopt the Dekel profile with $g=3$ and $b=2$ to follow the transition from cusps to cores.
As an example, Fig.~\ref{fig:dekel_nfw} compares this density profile with the NFW profile for one output of NIHAO galaxy \texttt{g1.08e11}, illustrating how it enables to better account for the inner part of the density profile.
In Fig.~\ref{fig:Udekel} we show that the expression of the gravitational potential derived from the \cite{Dekel2017} parametrisation of the dark matter density profile (Eq.~(\ref{eq:U32})) reproduces the simulated quantity. There is a small systematic trend for it to underestimate the simulated gravitational potential towards the center due to the presence of stars and gas, but the relative error remains below 10\% and its mean RMS below 5\%. 

\subsection{Parametrisation of the kinetic energy}
\label{section:kinetic}

The model presented in section~\ref{subsection:model} relies on analytic functional forms both for the gravitational potential and for the local kinetic energy $K(r)$ under Jeans equilibrium. 
Assuming complete spherical symmetry and isotropy without global rotation, we derive the local kinetic energy $K_{\rm DM}$ for a single-component halo described by a \cite{Dekel2017} density profile with $b=2$ and $g=3$ in steady-state Jeans equilibrium.
Using a simple parametrisation of the total mass profile, we further account for the multiple components (dark matter, stars and gas) of simulated galaxies and derive the corresponding local kinetic energy $K_{\rm multi}$, which is found to be in agreement with the simulations.
Additional terms can potentially be added to account for the presence of other components, for example a central mass or uniformly spread baryons.

\subsubsection{Spherical symmetry and isotropy} 
\label{sub:spherical}

The \emph{local} kinetic energy per unit mass of a spherical shell at radius $r$ within a spherical collisionless system, as defined in Eq.~(\ref{eq:Kdef}), can be written as  
\be
\label{eq:Ksim}
K(r) = \frac{3-2\beta}{2}\sigma_r^2,
\ee
where $\sigma_r$ is the radial component of the velocity dispersion and $\beta$ the anisotropy parameter \citep[e.g., ][Section 4.8.3]{BinneyTremaine2008}. This expression assumes complete spherical symmetry, i.e., that not only the potential and density profiles are spherically symmetric but also the distribution function describing the collisionless system such that there is no global rotation of the halo. 
It yields $\displaystyle K(r) = 1.5 \sigma_r^2$ when $\beta=0$,  which is what we assume in the following. In Appendix~\ref{appendix:symmetry}, we indeed show that assuming complete spherical symmetry and isotropy are valid assumptions for dark matter haloes in the NIHAO simulations.

\subsubsection{Jeans equilibrium} 
\label{sub:jeans}

The equilibrium of a spherical collisionless system can be described by the 
spherical Jeans equation stemming from the Boltzmann equation \citep[][Eq.~4.215]{BinneyTremaine2008}, which yields 
\be
\label{eq:sigmar2-0}
\sigma_r^2 (r)
= \frac{G}{\rho(r)}\int_{r}^{\infty}\rho(\rp)M(\rp)r^{-2}\rmd\rp
\ee 
when $\beta=0$ and the boundary condition is $\lim_{r\rightarrow +\infty} \sigma_r^2 = 0$.  
We show in Appendix~\ref{appendix:jeans} that the Jeans equation describing the steady-state equilibrium of a spherical collisionless system is valid in the NIHAO simulations within a mean RMS difference of 13\% for $r$ between $0.02 R_\vir$ and $R_\vir$. 
For a single-component spherical halo whose density profile is given by Eq.~(\ref{eq:rho32}), Eq.~(\ref{eq:sigmar2-0}) yields a local kinetic energy per unit mass
\be
\label{eq:sigma_dekel}
K_{\rm DM}(r)= c\mu \frac{GM_{\vir}}{R_{\vir}} \frac{(3-a)\overline{\rho_c}}{\rho(r)}
\Big[ \mathcal{B}(4-4a,9,\zeta)\Big]_\chi^1
\ee
where $\mathcal{B}(a,b,x) = \int_0^x t^{a-1}(1-t)^{b-1} dt$ is the incomplete beta function and the outer brackets refer to the difference of the enclosed function between 1 and $\chi$, i.e. $\displaystyle \left[ f(\zeta)\right]_\chi^1\equiv f(1)-f(\chi)$ for any function $f$.
Note that although the incomplete beta function is formally only defined when its parameters are positive, we extend the notation to negative parameters since Eq.~(\ref{eq:sigma_dekel}) arises from a finite integral that ensures the difference to be finite too. In fact, Eq.~(\ref{eq:sigma_dekel}) can also be expressed in terms of different finite series, whose expressions can be found in Appendix~\ref{appendix:Ksingle}.

\begin{figure}
	\centering
	\includegraphics[width=1\linewidth,height=0.714\linewidth,trim={1.1cm .6cm 0.6cm .15cm},clip]{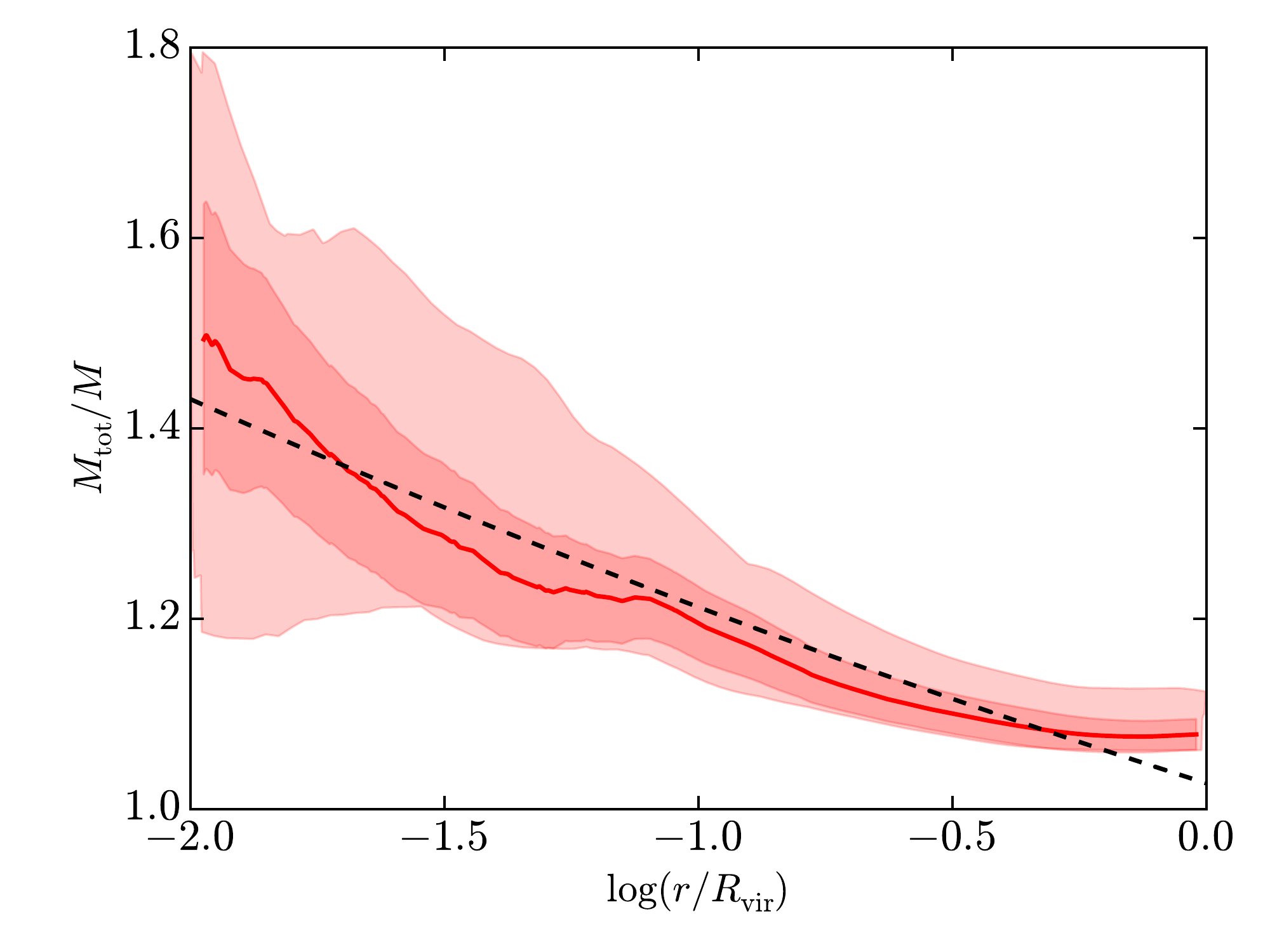}
	\vspace{-0.4cm}
	\caption{
		Ratio between the total mass $M_{\rm tot}$ and the dark matter mass $M$ as a function of radius in the outputs of NIHAO galaxy \texttt{g1.08e11} after 3 Gyr. The plain line corresponds to the median curve, the shaded areas to the standard deviation and the range, and the dashed black line to a power law fit to it as in Eq.~(\ref{eq:Mratio}). As a first approximation, the ratio $M_{\rm tot}/M$ can be approximated by a power law. The best-fit parameters to the median curve yield $X_M = 1.03$ and $n=0.07$. 
	}
	\label{fig:Mratio}
\end{figure}

\begin{figure}
	\centering
	\includegraphics[width=1\linewidth,trim={0.6cm 2.8cm 1.6cm 2.5cm},clip]{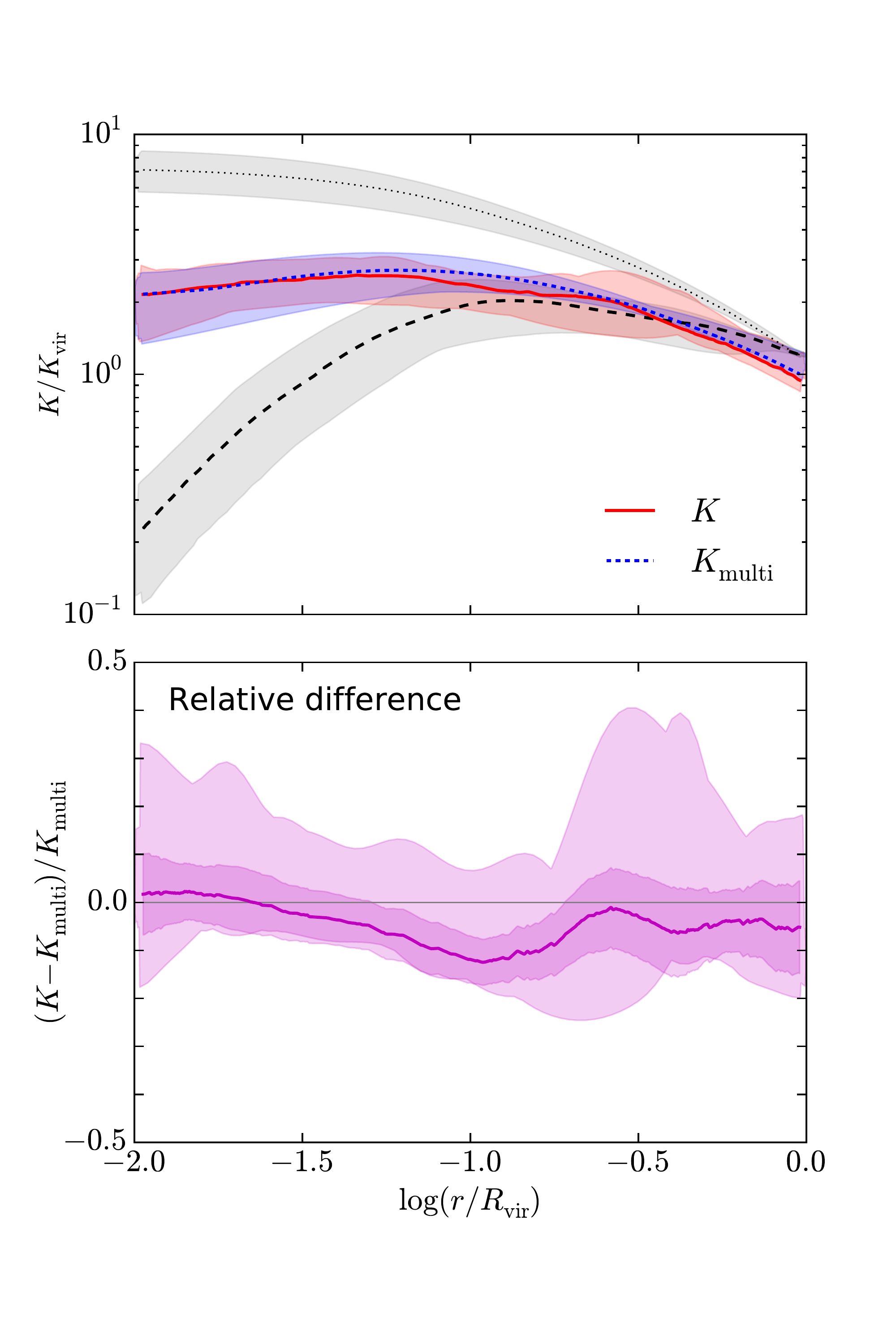}
	\vspace{-0.4cm}
	\caption{
		Comparison between the simulated kinetic energy per unit mass $K$ (red) and that derived from Eq.~(\ref{eq:Kmulti}), $K_{\rm multi}$ (blue dotted), in the different outputs of NIHAO galaxy \texttt{g1.08e11} after 3 Gyr (top panel) together with the relative difference $(K-K_{\rm multi})/K_{\rm multi}$ (bottom panel).
		The mean RMS relative difference is 8.0\%. The kinetic energy is in units of $K_{\rm vir} = G M_{\rm vir}/2R_{\rm vir}$. 
		The black dotted and dashed lines respectively recall $K_{\rm virial}=-U/2$ and $K_{\rm isolated}=GM(r)/2r$, the first one corresponding to the kinetic energy per unit mass expected from a direct application of the virial theorem and the second to that assumed by \protect\cite{Dutton2016} -- cf. Appendix~\ref{section:virial}. 
		Lines correspond to the median curves while the shaded areas highlight the ranges. In the lower panel, an additional shade shows the standard deviation from the median.
		The \protect\cite{Dekel2017} parametrisation fit to the average density profile yields the correct local kinetic energy per unit mass with a mean relative error $<$8\%.
	}
	\label{fig:Kmulti}
\end{figure}

In the case of a multi-component halo as in the simulations, the enclosed mass entering the gravitational term in the right-hand side of Eq.~(\ref{eq:sigmar2-0}) should include all components (dark matter, stars and gas). In Fig.~\ref{fig:Mratio}, we show how the total enclosed mass $M_{\rm tot}$ differs from that of the dark matter component $M$ in the different outputs of NIHAO galaxy \texttt{g1.08e11} after 3 Gyr. 
To account for the difference between $M_{\rm tot}$ and $M$, we model their ratio as a power law 
\be
\label{eq:Mratio}
\frac{M_{\rm tot}}{M} = X_M \left(\frac{r}{R_\vir} \right)^{-n}
\ee 
where $X_M$ and $n$ are ajustable parameters, and we replace $M$ by $M_{\rm tot}$ in the integral of Eq.~(\ref{eq:sigmar2-0}). In the case of the galaxy shown in Fig.~\ref{fig:Mratio}, the best-fit parameters to the median ratio are $X_M = 1.03$ and $n=0.07$.
Other approaches would be possible (e.g., solving Eq.~(\ref{eq:sigma_dekel}) numerically or running the model on the three halo components simultaneously), but this parametrisation keeps the model simple, analytic and cost-effective in terms of computational time. 
With this parametrisation, the local kinetic energy per unit mass becomes 
\be
\label{eq:Kmulti}
K_{\rm multi}(r)= \mu X_M c^{n+1} 
\frac{GM_{\vir}}{R_{\vir}} \frac{(3-a)\overline{\rho_c}}{\rho(r)}
\Big[ \mathcal{B}(4-4a-2n,9+2n,\zeta)\Big]_\chi^1
\ee
(cf. Appendix~\ref{appendix:Kmulti}). 
Fig.~\ref{fig:Kmulti} shows that this parametric expression of the local kinetic energy reproduces the simulated local kinetic energy $K$ with a relative error lower than 10\%, the agreement being particularly good towards the center. 
The small negative deviation in the relative difference around $0.1 R_\vir$ comes from an accumulation of gas at the outskirts of the stellar disk in this galaxy, which results in a relative decrease of the dark matter fraction and hence in a slight overestimation of the total mass and subsequently also of the local kinetic energy.

\subsubsection{Presence of an additional mass}
\label{sub:Km}

The key element of our model is the introduction of an additional mass $m$ (negative if mass is removed) to the system, which results in an additional term to the velocity dispersion $\sigma_r^2$ of Eq.~(\ref{eq:sigmar2-0}), 
\be
\label{eq:sigma_m}
\sigma_{m}^2 (r) =
\frac{G}{\rho(r)}\int_{r}^{\infty}\rho(\rp)m(\rp){\rp}^{-2}\rmd\rp. 
\ee
This term depends on how $m$ is spread over the halo. Assuming that it is dominated at each radius $r$ by the contribution of the mass entering (outflowing) at $r$ itself, i.e., that 
\be
\label{eq:m(r)}
\int_{r}^{\infty} \rho(\rp) m(\rp) {\rp}^{-2}\rmd\rp = m(r) \int_{r}^{\infty} \rho(\rp) {\rp}^{-2}\rmd\rp,
\ee 
the corresponding contribution to the local kinetic energy per unit mass is
\be
\label{eq:Km}
K_{m} (r) 
= 1.5 \sigma_m^2 (r)
= \frac{Gmc}{R_{\rm vir}} \frac{(3-a)\overline{\rho_c}}{{\rho(r)}} 
\Big[ \mathcal{B}(-2-2a,9,\zeta)\Big]_\chi^1
\ee
(cf. Appendix~\ref{appendix:Km}). 
The assumption explicited by Eq.~(\ref{eq:m(r)}) is valid outside the main body of $m(r)$ and in particular for the introduction (removal) of a point mass. If mass is added (removed) only up to a certain radius $r_m$, Eq.~(\ref{eq:m(r)}) is indeed rigourously valid for $r\geq r_m$. If $m(r)$ further does not oscillate between negative and positive values between the center and $r_m$, i.e. if the mass variation is a pure inflow (outflow), the right-hand side of Eq.~(\ref{eq:m(r)}) provides an upper limit to the left-hand integral and hence also to the contribution of the additional (removed) mass to the local kinetic energy. 
The assumption of Eq.~(\ref{eq:m(r)}) is motivated by a desire to obtain a simple analytic formulation. Alternative assumptions that also achieve this goal include assuming that the mass change is uniformly distributed up to a radius $r_m$, that it follows a singular isothermal sphere profile, or that if follows a \cite{Dekel2017} profile as the collisionless halo.
Expressions for the corresponding contributions to the local kinetic energy $K_m$ are given in Appendices~\ref{appendix:expressions}. 
Another possibility would be to integrate Eq.~(\ref{eq:sigma_m}) numerically  at the expense of the analytical formulation, which we avoid in the present work. 

In brief, the local kinetic energy per unit mass for a single-component spherical collisionless halo described by a \cite{Dekel2017} density profile in Jeans equilibrium in the presence of an additional mass $m(r)$ can be parametrized analytically as $K = K_{\rm DM} + K_m$; for a multi-component halo where $M_{\rm tot}/M$ is modeled as a power law in $r/R_\vir$, it can be parametrized as $K = K_{\rm multi} + K_m$. 
The different expressions for $K_m$ considered in Appendix~\ref{appendix:expressions} further enable to describe other mass distributions combining central masses, \cite{Dekel2017} haloes, singular isothermal spheres and uniform spheres.

\section{Testing the model with simulations}
\label{section:test}
\subsection{Description of the simulation suite}

We test our theoretical model using simulated galaxies taken from the Numerical Investigation of a Hundred Astrophysical Objects project \citep[NIHAO;][]{Wang2015}, which provides a set of about 90 cosmological zoom-in hydrodynamical simulations ran with the improved Smoothed Particle Hydrodynamics (SPH) code \texttt{gasoline2} \citep{Wadsley2004, Keller2014,Wadsley2017}. For each hydrodynamical simulation, a corresponding dark matter only simulation was run at the same resolution to study how baryonic processes affect the structure of dark matter haloes.
The simulations assume a flat $\Lambda$CDM cosmology with \cite{Planck2014} parameters, namely $\Omega_m = 0.3175$, $\Omega_r=0.00008$, $\Omega_\Lambda = 1-\Omega_m-\Omega_r = 0.6824$, $\Omega_b = 0.0490$, $H_0 = 67.1~\rm km.s^{-1}.Mpc^{-1}$, $\sigma_8 = 0.8344$ and $n=0.9624$. 
They include a subgrid model to account for the turbulent mixing of metals and thermal energy \citep{Wadsley2008}, cooling via hydrogen, helium and other metal lines in a uniform ultraviolet ionizing and heating background \citep{Shen2010} and star formation 
according to the Kennicutt-Schmidt relation when the temperature is below $15 000~\rm K$ and the density above $10.3 ~\rm cm^{-3}$ \citep{Stinson2013}. 
Stars inject energy back to their surrounding interstellar medium (ISM) both through ionizing feedback from massive stars \citep{Stinson2013} and through supernova feedback ejecting both energy and metals into the ISM \citep{Stinson2006}. 
The pre-supernova feedback consists in ejecting 13\% of the total stellar luminosity, which is typically $2.10^{50}~\rm erg$ per $\rm M_\odot$ of the entire stellar population over the 4 Myr preceding the explosion of high-mass stars as supernovae, into the surrounding gas. The relatively high 13\% efficiency is set to match the \cite{Behroozi2013} abundance matching results despite the increased mixing of \texttt{gasoline2}.
During the second supernova feedback phase, stars whose mass is comprised between 8 and 40 $\rm M_{\odot}$ eject 4 Myr after their formation both an energy $E_{\rm SN} = 10^{51}~\rm erg$ and metals into their surrounding ISM according to the blast-wave formalism described in \cite{Stinson2006}. 
In this formalism, the total feedback energy is distributed amongst gas particles within the maximum blastwave radius from \cite{Chevalier1977} and \cite{McKee1977}. 
In order to prevent the energy from supernova feedback to be radiated away, cooling is delayed for 30 Myr inside the blast region. Without cooling, the added supernova energy  heats the surrounding gas, which both prevents star formation and models the high pressure of the blastwave.
The NIHAO sample comprises isolated haloes chosen from dissipationless cosmological simulations \citep{Dutton2014} with halo masses between $\rm \log(M_{\vir}/M_\odot)=9.5-12.3$ but without considering their halo merging histories, concentrations and spin parameters.
Most importantly, the particle masses and force softening lengths are chosen to resolve the dark matter mass profile below 1\% of the virial radius at all masses in order to resolve the half-light radius of the galaxies. 
Stellar masses range from $5.10^4$ to $2.10^{11}~\rm M_\odot$, i.e., from dwarfs to Milky Way sized galaxies, with morphologies, colors and sizes that correspond well with observations \citep[e.g.,][]{Wang2015,Stinson2015,Dutton2016a}.

We test our model on the simulated NIHAO galaxies whose stellar mass at $z=0$ lies between $5\times 10^{7}$ and $5\times 10^9~\rm M_\odot$, which is the range where we expect the most important changes in the dark matter density profile according to \cite{Tollet2016} and \cite{Dutton2016}. 
Amongst these galaxies, we further use \texttt{g1.08e11} and \texttt{g6.12e10} as a fiducial cases given their intermediate stellar and halo masses at $z=0$, $M_\star = 8.47\times 10^8~\rm M_\odot$ and $M_\vir = 1.20 \times 10^{11}~\rm M_\odot$ for the former and  $M_\star = 9.13\times 10^7~\rm M_\odot$ and $M_\vir = 5.50 \times 10^{10}~\rm M_\odot$ for the latter, their relatively quiet merging history, and the fact that they belong to the UDGs sample studied by \cite{DiCintio2017} and \cite{Jiang2018}. Their stellar half light radius and V-band central surface brightness at $z=0$ are indeed respectively $r_e = 4.4 ~\rm kpc $ and $\mu_{0,V}=24.1~\rm mag.arcsec^{-2}$ for \texttt{g1.08e11} and $r_e = 2.6 ~\rm kpc $ and $\mu_{0,V}=24.03~\rm mag.arcsec^{-2}$ for \texttt{g6.12e10}. As they are UDGs, we expect both the dark matter and the stellar systems to expand and develop a core.

\subsection{Implementation of the model in the simulations}
\label{subsection:implementation}

For each simulation output, we first derive global properties of the central galaxy identified with the Amiga Halo Finder \citep[AHF;][]{Gill2004,Knollmann2009}. 
The virial radius $R_\vir$ is defined as the radius within which the average total density is $\Delta_c$ times the critical density of the Universe, where $\Delta_c$ is defined according to \cite{Bryan1998}. The virial mass $M_\vir$ is the total mass enclosed within $R_\vir$. 
The stellar mass $M_\star$ is calculated within $0.15 R_\vir$; the stellar half-light radius $r_{1/2}$ corresponds to the sphere enclosing half of $M_\star$.
The central V-band magnitude $\mu_{V,0}$ is computed within $0.25 R_{1/2}$ using Padova Simple stellar populations \citep{Marigo2008, Girardi2010} as implemented in \texttt{pynbody} \citep{Pontzen2013}. Galaxies with $r_{1/2}>1.5 ~\rm kpc$ and $\mu_{V,0}>24 ~\rm mag.arcsec^{-2}$ are considered as UDGs as in \cite{Jiang2018}. 
The timestep between two successive simulation outputs is 216 Myr.

We further derive radial profiles of mass ($M$), density ($\rho$), mean density ($\overline{\rho}$), radial and tangential velocity dispersions ($\sigma_r$ and $\sigma_t$) and anisotropy parameter ($\beta=1-\sigma_t^2/2\sigma_r^2$) using the \texttt{pynbody.analysis.profile} module for the different components constituting the halo, i.e., gas, stars and dark matter.
The profile radii $r$ are spaced logarithmically with $\sim$100 radii between $0.01 R_\vir$ and $R_\vir$. 
For each radius $r$, we compute the total mass including all components flowing in ($m_{\rm in}$) and out ($m_{\rm out}$) between all successive simulation outputs as well as the corresponding net mass change $m=m_{\rm in}-m_{\rm out}$, which is the main input of our theoretical model. 
We also define $f=m/M$ the fraction between this mass change and the enclosed mass of each component. 
We fit the average dark matter density profiles between $0.01R_\vir$ and $R_\vir$ according to the \cite{Dekel2017} parametrization (Eq.~(\ref{eq:rhob})) through a least-square minimization with \texttt{python}'s \texttt{lmfit} package, imposing $R_\vir$, $M_\vir$, $b=2$ and $g=3$ while leaving parameters $a$ and $c$ free. The inner slope at the resolution limit $s_0=s(0.01 R_\vir)$ and the effective concentration parameter $c_{\rm max}$ can be derived from Eqs. (\ref{eq:s}) and (\ref{eq:cmax}) for each set of parameters.

We implement the instant mass change theoretical model presented in section~\ref{subsection:model} in all successive simulation outputs excluding the first 3 Gyr of evolution, which are relatively perturbed. 
As mergers and fly-bys are expected to break the assumed spherical symmetry and to affect the halo matter distribution through processes that are not accounted for in the model, we define a merger-free subsample of successive outputs
by excluding cases where the ratio between the total mass change including all components (dark matter, stars and gas) and the total enclosed mass at $0.15 R_\vir$,
\be
f_{\rm merger}= \frac{m}{M_{\rm tot}} \Bigg|_{~ r~=~0.15R_\vir}, 
\ee
is above $f_{\rm merger, min}=10\%$ during the timestep at stake.
Amongst the processes potentially induced by mergers and fly-bys, dynamical friction can heat the background particles \citep[e.g.][]{El-Zant2001,El-Zant2004,Tonini2006,
Goerdt2010}
and gravitational torques can drive large amounts of gas towards the center in a non-spherical way \citep{Barnes1991,Mihos1996,Fensch2017}.

As presented in section~\ref{section:kinetic}, we account for the difference between the total enclosed mass $M_{\rm tot}$ and the collisionless component mass $M$ through a power-law radial dependence according to Eq.~(\ref{eq:Mratio}). We derive the best-fit parameters $X_M$ and $n$ at each output and assume that this parametrization holds until the next output time.
The principle of our model was outlined in section~\ref{subsection:model}: for a given best-fit parametrization of the collisionless density profile, a power-law parametrization of $M_{\rm tot}/M$ and the net mass change $m$ from the simulation, we make a prediction for the new collisionless density profile through a least square minimization of the energy equation $E_f(r_f) = E_t(r_i)$ explicited in Eq.~(\ref{eq:Et-Ef}). The minimization is carried out simultaneously at all radii between $0.02 R_\vir$ and $R_\vir$ (or $\log(r/R_\vir)$ between $-1.75$ and $0$), since this is the domain where the Jeans equation is valid, as shown in Appendix~\ref{appendix:jeans}.  
The final density profile is parametrized according to the \cite{Dekel2017} functional form.
As for the profile fits, its parameters $a$ and $c$ are allowed to vary while $g=3$ and $b=2$. $R_\vir$ and $M_\vir$ are fixed to their initial values. 

\begin{figure}
	\centering
	\includegraphics[width=1\linewidth,trim={0.2cm 0.2cm 0.3cm 0.2cm},clip]{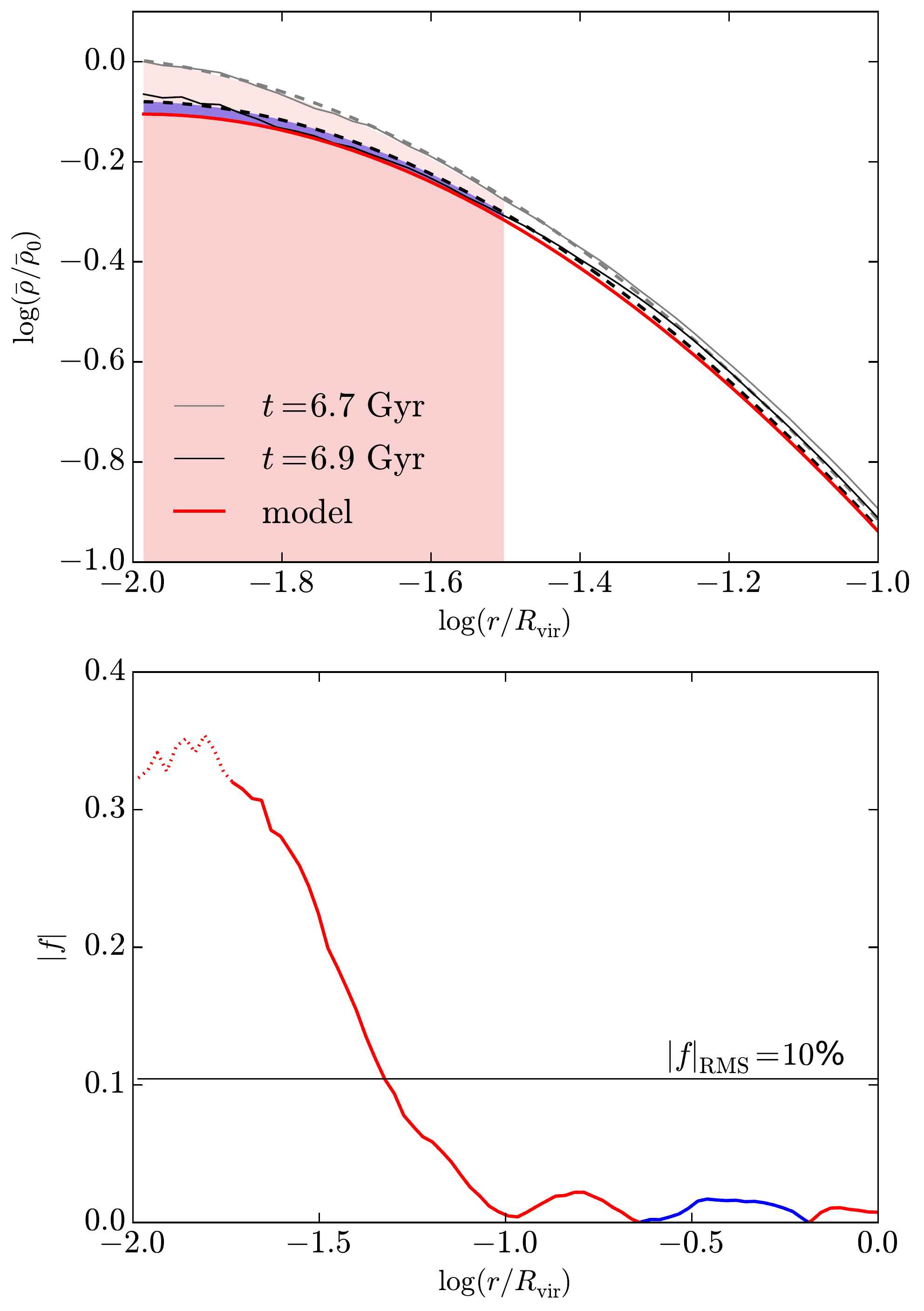}
	\vspace{-0.4cm}
	\caption{
		\textit{Top:} Evolution of the inner part of the average dark matter density profile between two successive outputs of NIHAO \texttt{g1.08e11} compared to the instant mass change model prediction. The plain gray and black lines respectively show the initial ($t=6.7$ Gyr) and final ($t=6.9$ Gyr) density profiles while the dashed curves correspond to \protect\cite{Dekel2017} fits with $g=3$ and $b=2$. The plain red line corresponds to the model prediction given the initial conditions and the net mass change at each radius. The average density is normalized by its initial value at $0.01 R_\vir$, $\overline{\rho}_0$. The shaded regions highlight the areas used to define $\delta$, $\delta_0$ and $\delta_{\rm sim}$ (Section~\ref{subsection:success}). In particular, the purple region corresponds to $|\mathcal{A}_{\rm model}-\mathcal{A}_{k+1}|$, the red shaded region underneath to $\mathcal{A}_{\rm model}$ and the sum of the two to $\mathcal{A}_{k+1}$. In the example shown here, $\delta= 2\%$, $\delta_0 = 8\%$ and $\delta_{\rm sim}=5\%$. 
		\textit{Bottom:} Radial profile of the mass ratio $f=m/M$ for the dark matter component between the two successive outputs used to derive the model prediction. Red correspond to outflows ($m>0$), blue to inflows ($m<0$). The innermost part below $\log(r/R_\vir)=-1.75$ (dotted line) was not taken into account since the Jeans equilibrium equation is not verified in this domain (Appendix~\ref{appendix:jeans}). The RMS value $|f|_{\rm RMS}=10\%$ is shown as a horizontal black line. 
		The figure shows a successful model prediction for the evolution of the average density profile, with $\delta<\delta_{\rm max}$. 
	}
	\label{fig:profile30}
\end{figure}

Fig.~\ref{fig:profile30} shows an example evolution of the dark matter average density profile between two successive outputs of the NIHAO \texttt{g1.08e11} simulation compared to the instant mass change model prediction.
The bottom panel further shows the radial profile of the mass ratio $f=m/M$ for the dark matter component stemming from the simulation that is implemented in the model. 
As can be seen in the figure, the dark matter average density profile evolves towards a flattening of its central part. This evolution agrees very well with the model prediction (red line in the upper panel) and is consistent with a mass change dominated by outflows (red portions of the curve in the bottom panel).

\subsection{Assessing the success of the model}
\label{subsection:success}

To quantitatively assess the success of the model in retrieving the inner density profile, we define $\mathcal{A}_{k}$, $\mathcal{A}_{k+1}$ and $\mathcal{A}_{\rm model}$ the areas between the horizontal line $\log(\overline{\rho}/\overline{\rho}_0)=-1$ and the \cite{Dekel2017} fit to the average density profiles $\log(\overline{\rho}/\overline{\rho}_0)$ corresponding respectively to the initial output ($t=6.7$ Gyr in the example shown in Fig.~\ref{fig:profile30}), the final output ($t=6.9$ Gyr) and to the model, $\log(r/R_\vir)$ being between $-2$ and $-1.5$. The shaded areas of Fig.~\ref{fig:profile30} help visualizing these areas, with $\mathcal{A}_{k}$ from the bottom to the gray dashed line, $\mathcal{A}_{k+1}$ from the bottom to the black dashed line and $\mathcal{A}_{\rm model}$ from the bottom to the red plain line. 
The absolute ratio 
\be
\delta = 2 \left|\frac{\mathcal{A}_{\rm model}-\mathcal{A}_{k+1}}{\mathcal{A}_{\rm model}+\mathcal{A}_{k+1}}\right|
\ee
between $\mathcal{A}_{\rm model}-\mathcal{A}_{k+1}$ and  $(\mathcal{A}_{\rm model}+\mathcal{A}_{k+1})/2$ provides a relative measurement of the difference between the model prediction and the final output, which we use to assess the success of the model. 
Although rather arbitrary, the choice of the horizontal line $\log(\overline{\rho}/\overline{\rho}_0)=-1$ corresponds well with our visual assessment of the successes.

We similarly define $\delta_0$ to measure the difference between $\mathcal{A}_{\rm model}$ and $\mathcal{A}_{k}$ (instead of $\mathcal{A}_{k+1}$), i.e. between the model prediction and the initial output, and $\delta_{\rm sim}$ between $\mathcal{A}_{k}$ and $\mathcal{A}_{k+1}$, i.e. between the initial and the final outputs. 
If $\delta>\delta_0$, the model prediction is closer to the initial density profile than to the final one and hence fails to retrieve the simulated evolution. However, if $\delta_{\rm sim}$ is particularly small, the comparison between $\delta$ and $\delta_0$ loses its relevance.
We thus consider the model successful at predicting the evolution of the average collisionless density profile when (i) $\delta$ is below a certain threshold $\delta_{\rm max}$ and (ii) $\delta<\delta_0$ unless $\delta_{\rm sim}$ is too small to differenciate between $\delta$ and $\delta_0$. In terms of logical operators, the success condition we choose can be expressed as 
\be
\label{eq:criterion}
(\delta\leq \delta_{\rm max}) ~~\texttt{AND}~~ (\delta \leq \delta_0 ~~\texttt{OR}~~ \delta_{\rm sim}\leq \delta_{\rm min}),
\ee 
where $\delta_{\rm min}$ corresponds to the minimum discrepancy between the two outputs we deem physically meaningful. 
In the following, we take  $\delta_{\rm max}=10\%$ as success threshold and $\delta_{\rm min} = 3\%$. 
This value of $\delta_{\rm max}$ enables to consider the case shown in Fig.~\ref{fig:profile30} as a clear success. 
A different threshold would change our results quantitatively (a lower value of $\delta_{\rm max}$ leading to fewer reported successes and vice-versa) but would not change the qualitative trends reported in the next sections.

The \cite{Dekel2017} parametrization of the average density profile further enables to measure the success of the model in terms of the  errors in inner logarithmic slope 
\be
\label{eq:Dsin}
\Delta s=s_{\rm 0, model}-s_{\rm 0, f}
\ee
and effective concentration parameter 
\be
\label{eq:Dcmax}
\Delta c=\log(c_{\rm max, model}/c_{\rm max, f}). 
\ee
Since the success measure $\delta$ aims at capturing changes in the inner part of the average density profile, it correlates strongly with $\Delta s$ (with a Pearson correlation coefficient $r=0.71$ between $\delta$ and $\log|\Delta s|$), less with $\Delta c$
($r=0.43$ between $\delta$ and $\log|\Delta c| $), while $\Delta s$ and $\Delta c$ are themselves moderately correlated ($r=0.53$ between the logarithms of their absolute values). Figures showing these correlations can be found in Appendix~\ref{appendix:correlations} (Fig.~\ref{fig:correlations_success}).

Table~\ref{tab:params} summarizes the fiducial parameters used for fitting the average density profiles, implementing the model and assessing its success. 

\begin{table}
	\centering
	\caption{Fiducial parameters used when implementing the model and assessing its success. 
	}
	\label{tab:params}
	\begin{tabular}{lll} 
		\hline\noalign{\vskip 1mm} 
		Parameter & Description & Value \\
		\hline\noalign{\vskip 1mm} 
		$t_{\rm min}$ & earliest time considered & 3 Gyr \\
		$b$ & fixed profile parameter & 2 \\
		$g$ & fixed profile parameter & 3 \\
		$r_0/R_\vir$ & radius for evaluating $s_0$  and $\overline{\rho}_0$ & 0.01 \\
		$\Delta \log(r/R_\vir)_{\rm fit}$ & radius range for the profile fits & $[-2,0]$\\
		$\Delta \log(r/R_\vir)_{\rm model}$ & radius range for the model & $[-1.75,0]$\\
		$\Delta \log(r/R_\vir)_{\rm success}$ & radius range  to define $\delta$ & [-2,-1.5]\\
		$\log(\overline{\rho}/\overline{\rho}_0)_{\rm min}$ & floor to define $\delta$ & -1\\
		$r_{\rm merger}/R_\vir$ & radius for $f_{\rm merger}$ & $0.15$\\
		$f_{\rm merger,min}$ & threshold for $f_{\rm merger}$ & $10\%$\\
		$f_{\rm min}$ & threshold for $|f|_{\rm RMS}$ & 7\% \\
		$\delta_{\rm min}$ & minimum distinguishable $\delta_{\rm sim}$ & 3\% \\
		$\delta_{\rm max}$ & maximum $\delta$ for a success & 10\% \\
		$\Delta s_{\rm max}$ & maximum $|\Delta s|$ for a success & 0.10 \\
		\hline
	\end{tabular}
\end{table}	

\subsection{Model success results}
\label{subsection:results}

\begin{figure*}
	\centering
	\includegraphics[width=1\linewidth,trim={1.cm 39.8cm 9.5cm 18.3cm},clip]{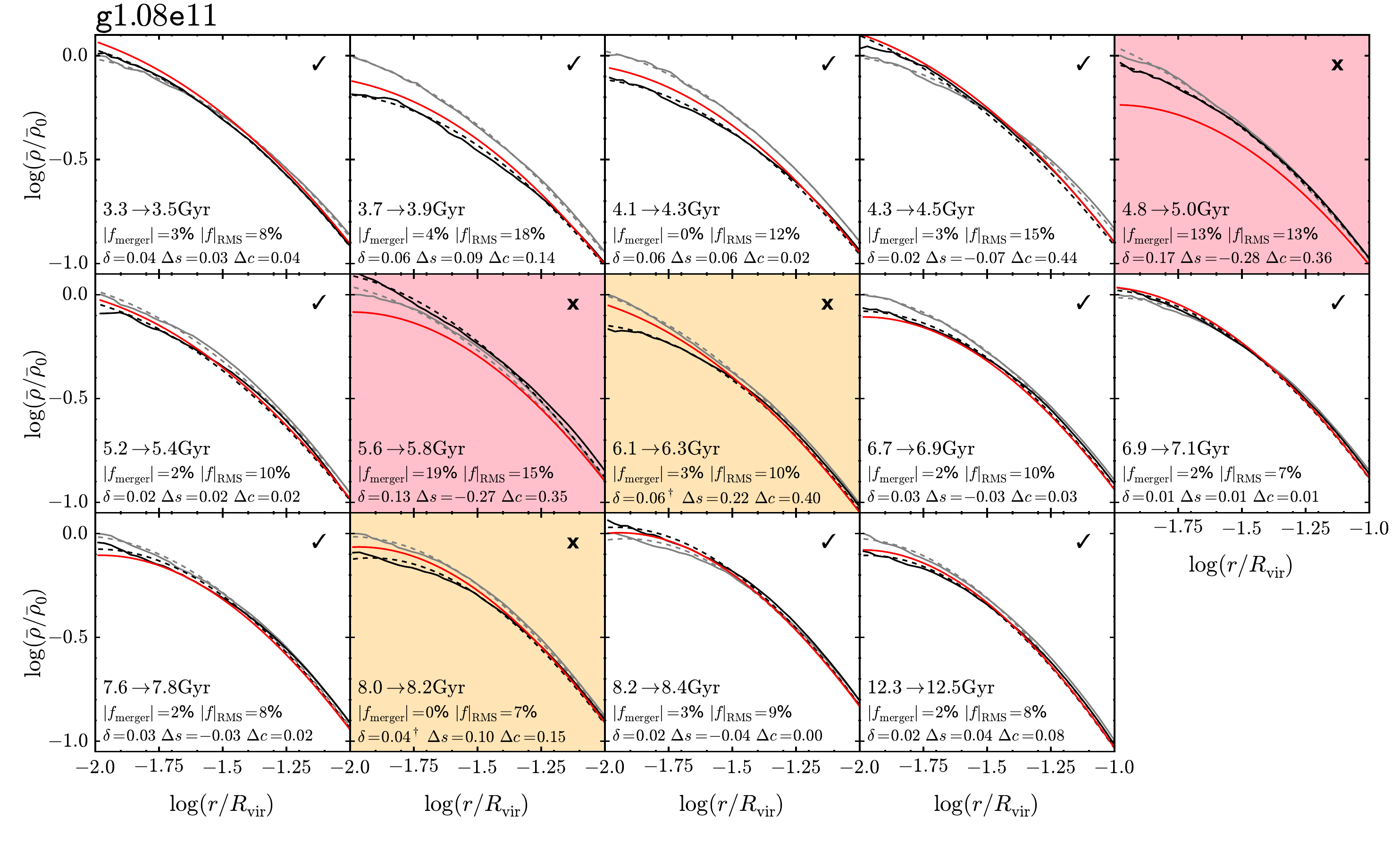}\\
	\includegraphics[width=1\linewidth,trim={1.cm 39.8cm 9.5cm 18.3cm},clip]{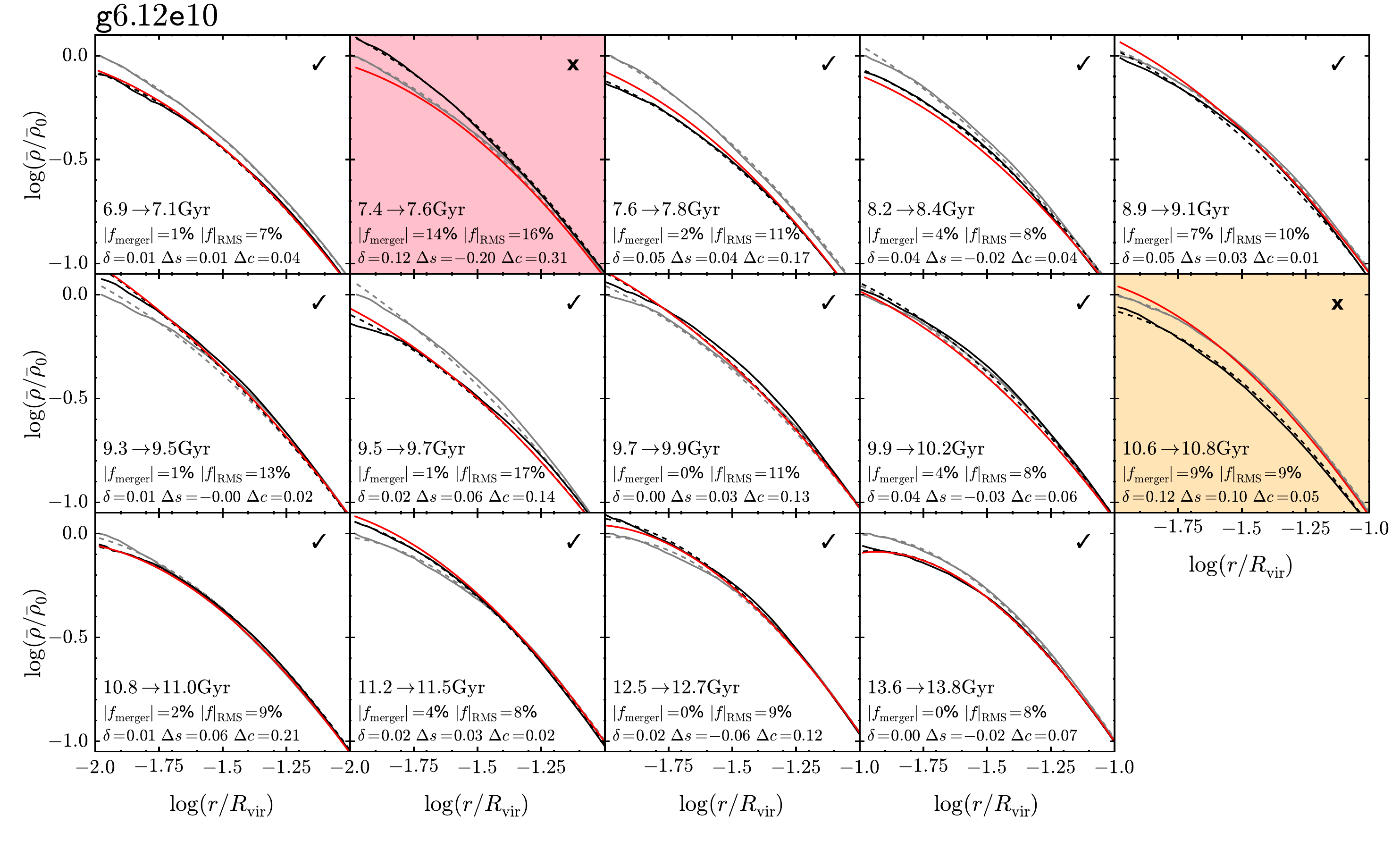}
	\vspace{-0.8cm}
	\caption{
		Evolution of the inner part of the average dark matter density profile between successive snapshots of NIHAO galaxies \texttt{g1.08e11} and \texttt{g6.12e10} after the first 3 and 6 Gyr when $|f|_{\rm RMS}>7\%$ compared to the instant mass change model prediction. The initial profile is in grey, the final profile in black and the prediction in red. Fits according to the \protect\cite{Dekel2017} parametrization (Eq.~(\ref{eq:rho32})) are indicated as dashed lines. The quantities $f_{\rm merger}$, $|f|_{\rm RMS}$, $\delta$, $\Delta s$, and $\Delta c$ are indicated at the bottom. As summarized in Table~\ref{tab:params}, we consider that $f_{\rm mergers}>10\%$ indicates a merger, $|f|_{\rm RMS}>7\%$ a case where we expect a significant change in the density profile, and $\delta>10\%$ an unequivocal failure. A case where $\delta<10\%$ but the model prediction is closer to the initial average density profile than to the final one while $\delta_{\rm sim} \leq 3\%$ following Eq.~(\ref{eq:criterion}) is also considered as a failure and marked with a dagger ($\dagger$). Successes and failures are respectively indicated by a check mark or a cross at the upper right corner. The background color further highlights mergers (red), successes (white), and failures (orange). 
		While the model fails in all merger cases, it is able to predict the evolution of the dark matter density profile with some scatter given the initial profile and the mass change at each radius. 
	}
	\label{fig:example_density}
\end{figure*}

Fig.~\ref{fig:example_density} shows the evolution of the inner part of the average dark matter average density profile between successive outputs of the fiducial NIHAO simulations \texttt{g1.08e11} and \texttt{g6.12e10} after the perturbed earliest gigayears.
Only cases where the RMS value of the mass ratio between $0.01 R_\vir$ and $R_\vir$, $|f|_{\rm RMS}$, is above a certain threshold $f_{\rm min}=7\%$ are shown, since these are the cases for which we expect a significant change in the collisionless density profile.
With the success criterion indicated above (Eq.~(\ref{eq:criterion})), the success rate of the model amongst the cases with $|f|_{\rm RMS}>7\%$ shown in Fig.~\ref{fig:example_density} is 79\%, with 22 successes out of 28. 
We note that the model systematically fails during the three merger cases. 
Excluding mergers, the success rate when $|f|_{\rm RMS}>7\%$ becomes 88\% (22/25).
Amongst the cases with $|f|_{\rm RMS}\leq 7\%$ that are not shown in Fig.~\ref{fig:example_density}, the success rate is 89\% (48/54), but these cases display little evolution. 
Including all successive outputs of both \texttt{g1.08e11} and \texttt{g6.12e11}, the overall success rate is 82\% (68/83) and 88\% (68/77) without mergers. 
The mass change $m$ is the main input to the model with the initial parametrization of the density profile and the ratio $M_{\rm tot}/M$ between the total mass and that of the dark matter component. Fig.~\ref{fig:example_fprofile} in the appendices shows the radial profiles of the ratio $f=m/M$ between the total mass change including all components and the enclosed dark matter mass together with its RMS average $|f|_{\rm RMS}$ for the cases shown in Fig.~\ref{fig:example_density}. We note that mergers coincide with large values of $|f|_{\rm RMS}$.

For the whole sample of simulated galaxies whose stellar mass lies between $5 \times 10^7$ and $5 \times 19^9~\rm M_\odot$ at $z=0$, the success rate given the chosen success criterion is  $p_{\rm all}=67\%$ overall and $p_{\rm no~mergers} = 74\%$ excluding mergers. In this latter case (excluding mergers), the success rate is $p_{>f_{\rm min}} = 65\%$ when $|f|_{\rm RMS}\geq f_{\rm min}$ and $p_{<f_{\rm min}}= 80\%$ when $|f|_{\rm RMS}<f_{\rm min}$. 
Table~\ref{tab:sims} in the appendices summarizes these trends galaxy per galaxy and for the whole sample. 
The standard deviation of the success rate is about 20\%, with significant variations from one galaxy to another. 
Successful evolutions according to Eq.~(\ref{eq:criterion}) systematically correspond to 
\be
\label{eq:Ds010}
|\Delta s| \leq 0.10,
\ee
which can thus provide a relatively equivalent success criterion (cf. also Fig.~\ref{fig:correlations_success}). In contrast, $\Delta c$ fluctuates more since it captures the overall dark matter density profile and not specifically its inner part. 
The success criterion defined by Eq.~(\ref{eq:Ds010}) leads to similar success rates as indicated above, with 68\% cases with $|\Delta s| \leq 0.10$ overall and 73\% excluding mergers.

Since the halo dynamical time after 3 Gyr spans over more than 3 time steps, the merger criterion may not exclude all the outputs affected by mergers. 
Considering that mergers are galaxies whose $f_{\rm merger}$ is above 10\% both during the evolution at stake (outputs $k \rightarrow k+1$) and during the previous one (outputs $k-1 \rightarrow k$) leads to 
$p_{\rm no ~mergers}=76\percent$.
Further extending this criterion to another previous evolution (outputs $k-2 \rightarrow k-1$)  leads to 
$p_{\rm no ~mergers}=77\percent$.
These alternative merger criteria remove some failures of the model that coincide with mergers but their incidence on the success rates is small and we thus prefer to retain our initial, simpler merger criterion. 
We note that the success rate for the fiducial galaxies \texttt{g1.08e11} and \texttt{g6.12e10} are higher than average since they were selected to have a relatively quiet merging histories. 
In the next section, we indeed show that the success rate of the model primarily depends on the merger indicator $f_{\rm merger}$.

\begin{figure*}
	\centering
	\includegraphics[width=0.45\linewidth,trim={0.75cm 0cm 4.cm 0cm},clip]{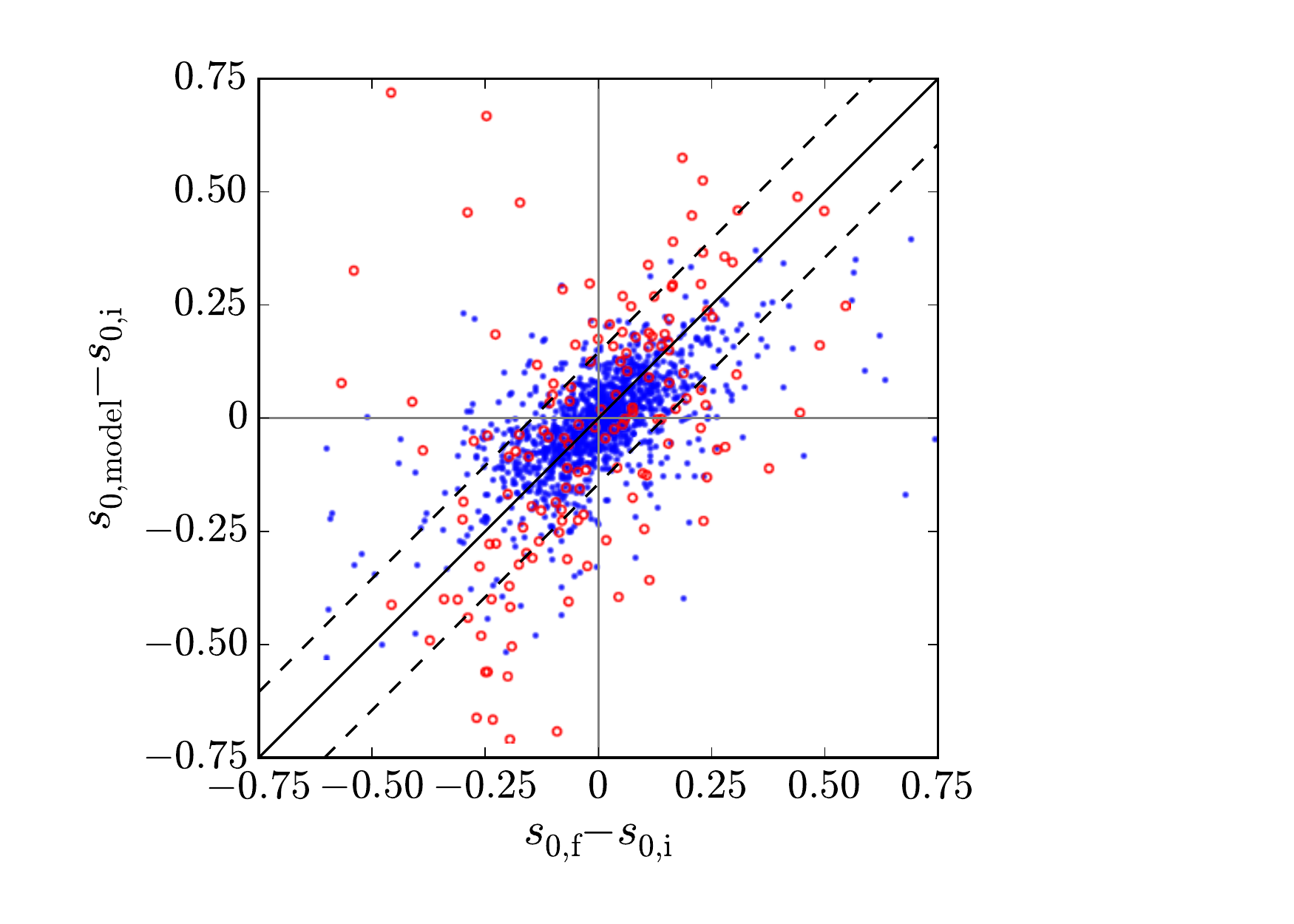}
	\hfill
	\includegraphics[width=0.45\linewidth,trim={0.75cm 0cm 4cm 0cm},clip]{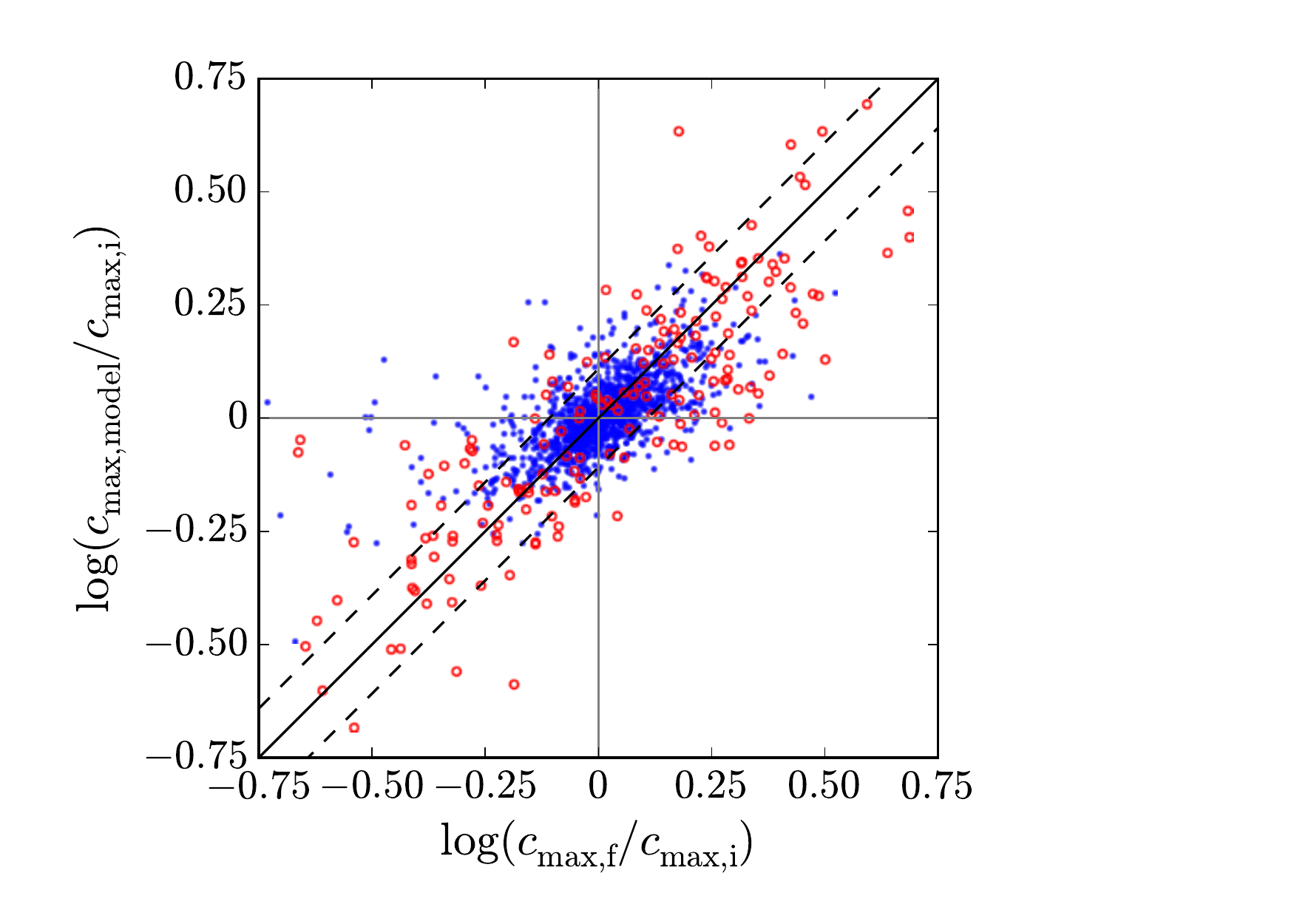}
	\vspace{-0.4cm}
	\caption{
		 Comparison between the variation in inner logarithmic slope $s_{\rm 0}$ \textit{(left)} and concentration  $c_{\rm max}$ \textit{(right)} between successive outputs of the NIHAO simulations whose stellar mass lies between $5\times 10^{7}$ and $5\times 10^{9}~ M_\odot$ at $z=0$, and that predicted by the model. The earliest phase of galaxy evolution before 3 Gyr is excluded. The inner logarithmic slope is evaluated at $0.01 R_\vir$ using Eq.~(\ref{eq:s}) and the concentration parameter using Eq.~(\ref{eq:cmax}) from the \protect\cite{Dekel2017} fit parameters. The Pearson correlations coefficiants within the boundaries of the x-axes (i.e., excluding outliers in $s_{\rm 0,f}-s_{\rm 0, i}$ and $\log(c_{\rm max, f}/c_{\rm max, i})$) are 0.55 for $s_{\rm 0}$ and 0.72 for $c_{\rm max}$ while the corresponding scatters are 0.15 and 0.11, respectively. These scatters are shown by the dashed lines around the first bisector. When considering only merger (non-merger) cases, the correlation coefficients are 0.48 (0.61) and 0.83 (0.63), the scatters 0.29 (0.12) and 0.17 (0.10).  
		 Open red points highlight mergers with $f_{\rm merger}>10\%$.  
		 As illustrated for one simulated galaxy in Fig.~\ref{fig:example_density} and further characterized here, our model qualitatively recovers the evolution of the dark matter density profile from the mass change between two successive snapshots, although with a non-negligible scatter.  
	}
	\label{fig:sin_cmax}
\end{figure*}

Fig.~\ref{fig:sin_cmax} compares the changes in $s_{\rm 0}$ and $c_{\rm max}$ between two successive simulation outputs with those predicted by the model. 
The relatively high Pearson correlation coefficients show that the model captures the overall evolution of the dark matter density profile, but the non-negligible scatter hinders the capacity of the model to yield precise, quantitative predictions for the profile parameters. Both points are already illustrated in Fig.~\ref{fig:example_density} for two of the simulated galaxies, since the predicted average density profiles is in most cases comparable to the final ones but not fully overlaping. In this figure, the correlations between $\delta$, $\Delta s$ and $\delta c$ (cf. Appendix~\ref{appendix:correlations}) are apparent, since $|\Delta s|>0.10$ always coincides with failures while all successes but one have $|\Delta c|<0.20$.  
For merger cases, the variations in $s_{\rm 0}$ and $c_{\rm max}$ are on average bigger than for non-merger cases, but they are still correlated to the actual variations. 
For non-merger cases, the relation between $s_{\rm 0,model}-s_{\rm 0, i}$ and $s_{\rm 0,f}-s_{\rm 0, i}$ is significantly tighter than for the whole sample (and for merger cases) as indicated both by the correlation coefficients and the scatter. 
Although the correlation coefficient of the relation between $\log(c_{\rm max, model}/c_{\rm max,i})$ and $\log(c_{\rm max, f}/c_{\rm max,i})$ is higher for merger cases, the scatter is significantly higher as well. 
We note that the quantities plotted against each other in Fig.~\ref{fig:sin_cmax} are not formally independent (they both involve $s_{\rm in, i}$ in the left panel and $\log(c_{\rm max, i})$ in the right panel). If we consider directly the relation between $s_{\rm 0, model}$ and $s_{\rm 0, f}$ (and between $\log(c_{\rm max, model})$ and $\log(c_{\rm max, f})$), the Pearson correlation coefficient yields 0.85 (0.95) but this is mostly due to the fact that the change between two successive outputs is relatively small compared to the initial slope (concentration). 
The null hypothesis of no change between two successive outputs would indeed yield a strong correlation between $s_{\rm 0, model}$ and $s_{\rm 0, f}$ (or between $\log(c_{\rm max, model})$ and $\log(c_{\rm max, f})$) since $s_{\rm 0, f}$ and $s_{\rm 0, i}$ are correlated with a correlation coefficient 0.73 (0.88 between $\log(c_{\rm max, f})$ and $\log(c_{\rm max, i})$). 
In the next section, we characterize successes and failures in terms of $\delta$, $\Delta s$ and $\Delta c$ and try to identify the main causes of failure of the model to characterize its validity domain and suggest possible future improvements. 

\section{Discussion}
\label{section:discussion}
\subsection{Mergers as the main cause of failure}
\label{section:mergers}

\begin{figure*}
	\centering
	\includegraphics[height=0.28\linewidth,trim={.8cm 0 4.5cm 0.6cm},clip]{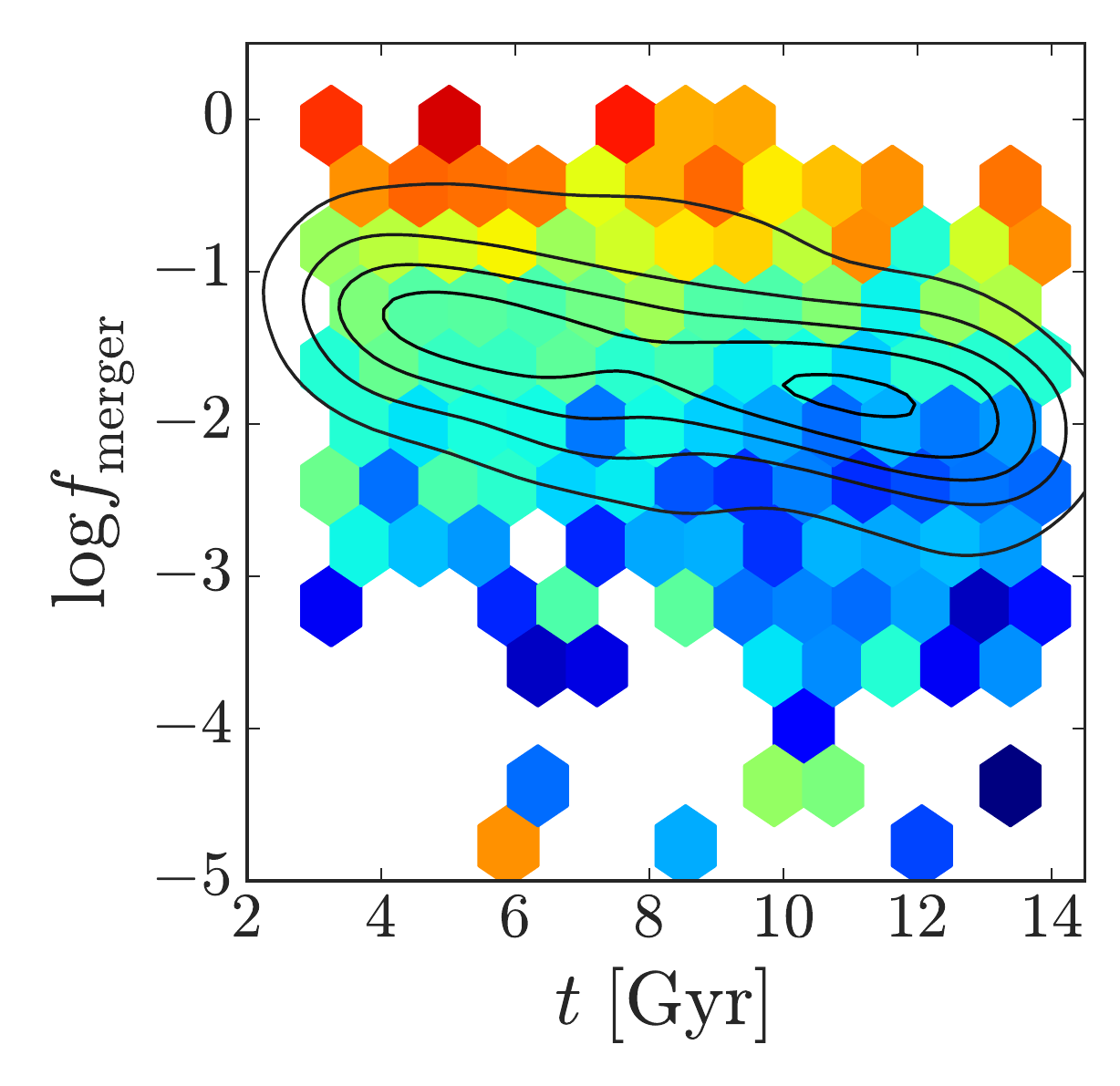}\hfill
	\includegraphics[height=0.28\linewidth,trim={.8cm 0 4.5cm 0.6cm},clip]{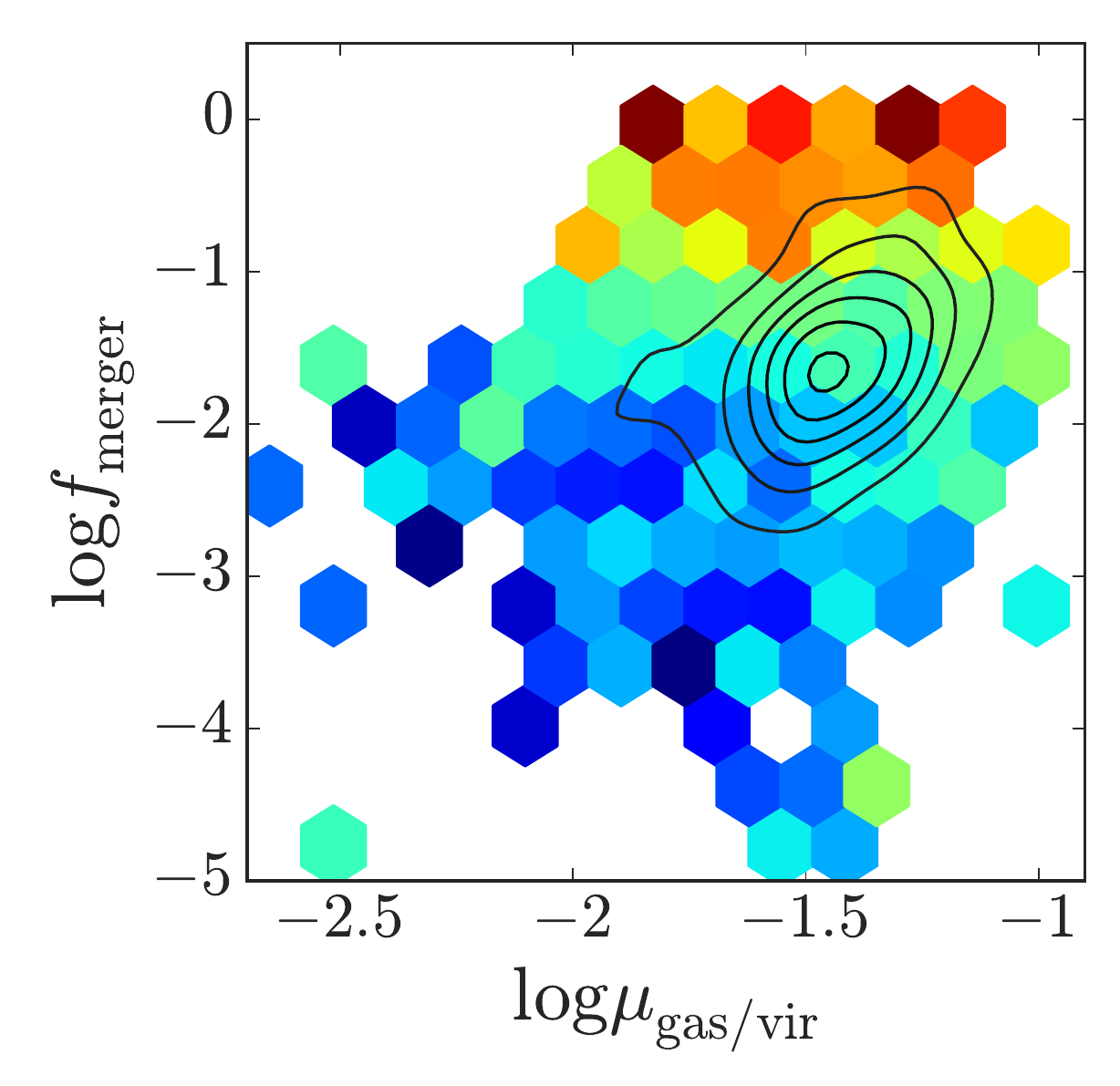}\hfill
	\includegraphics[height=0.28\linewidth,trim={0.8cm 0 0.2cm 0.6cm},clip]{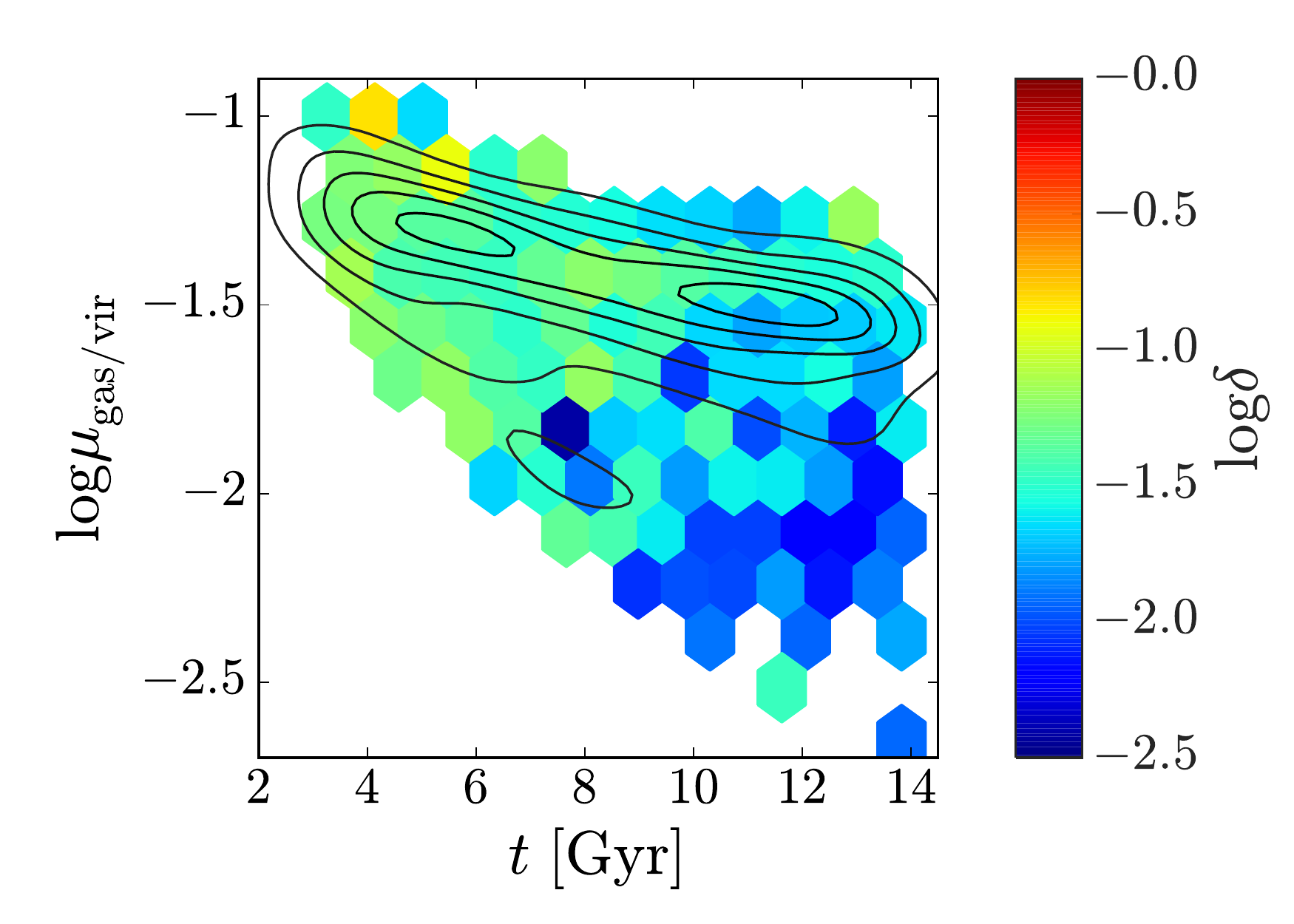}
	\includegraphics[height=0.28\linewidth,trim={.8cm 0 4.5cm 0.6cm},clip]{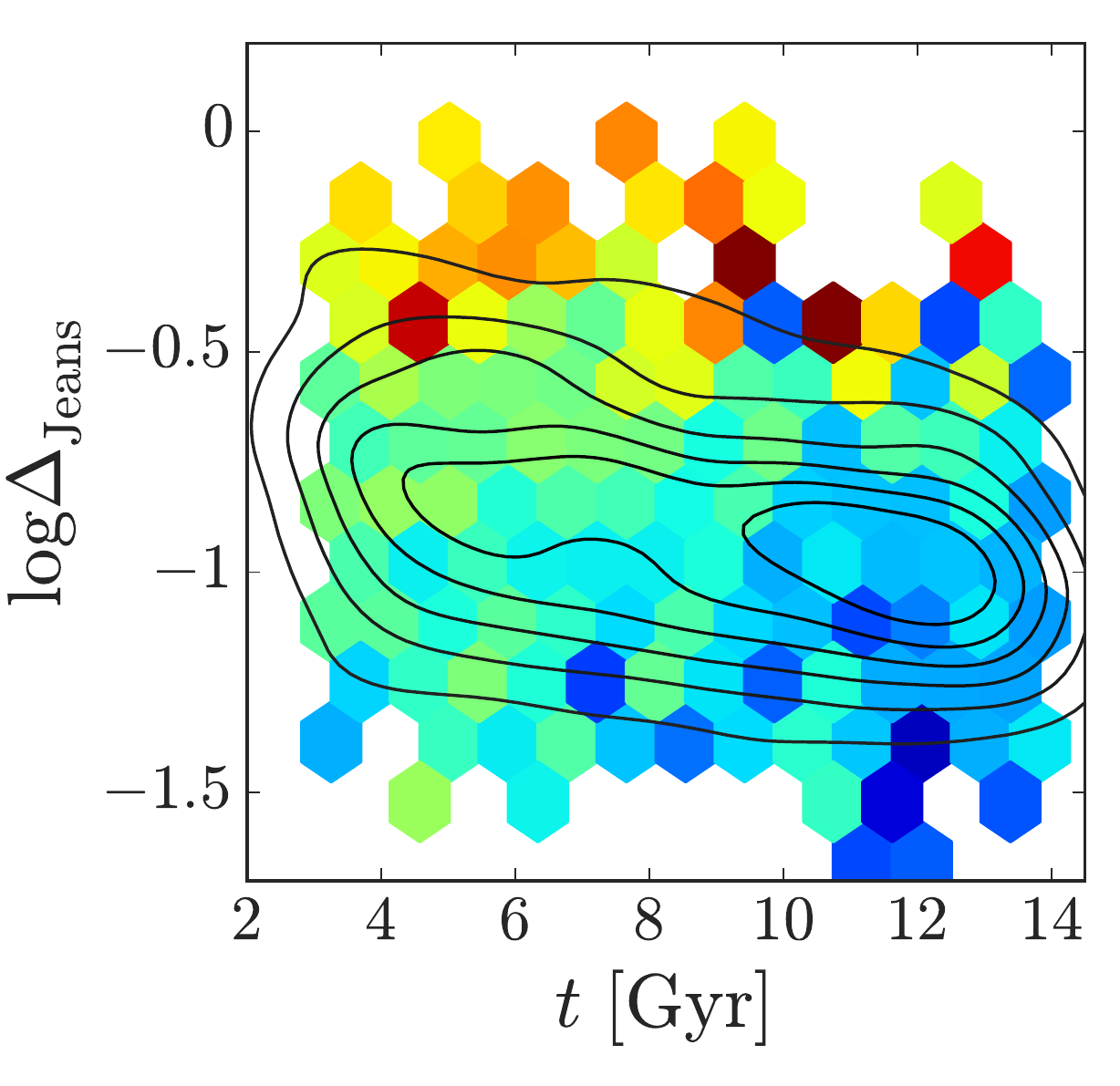}\hfill
	\includegraphics[height=0.28\linewidth,trim={.8cm 0 4.5cm 0.6cm},clip]{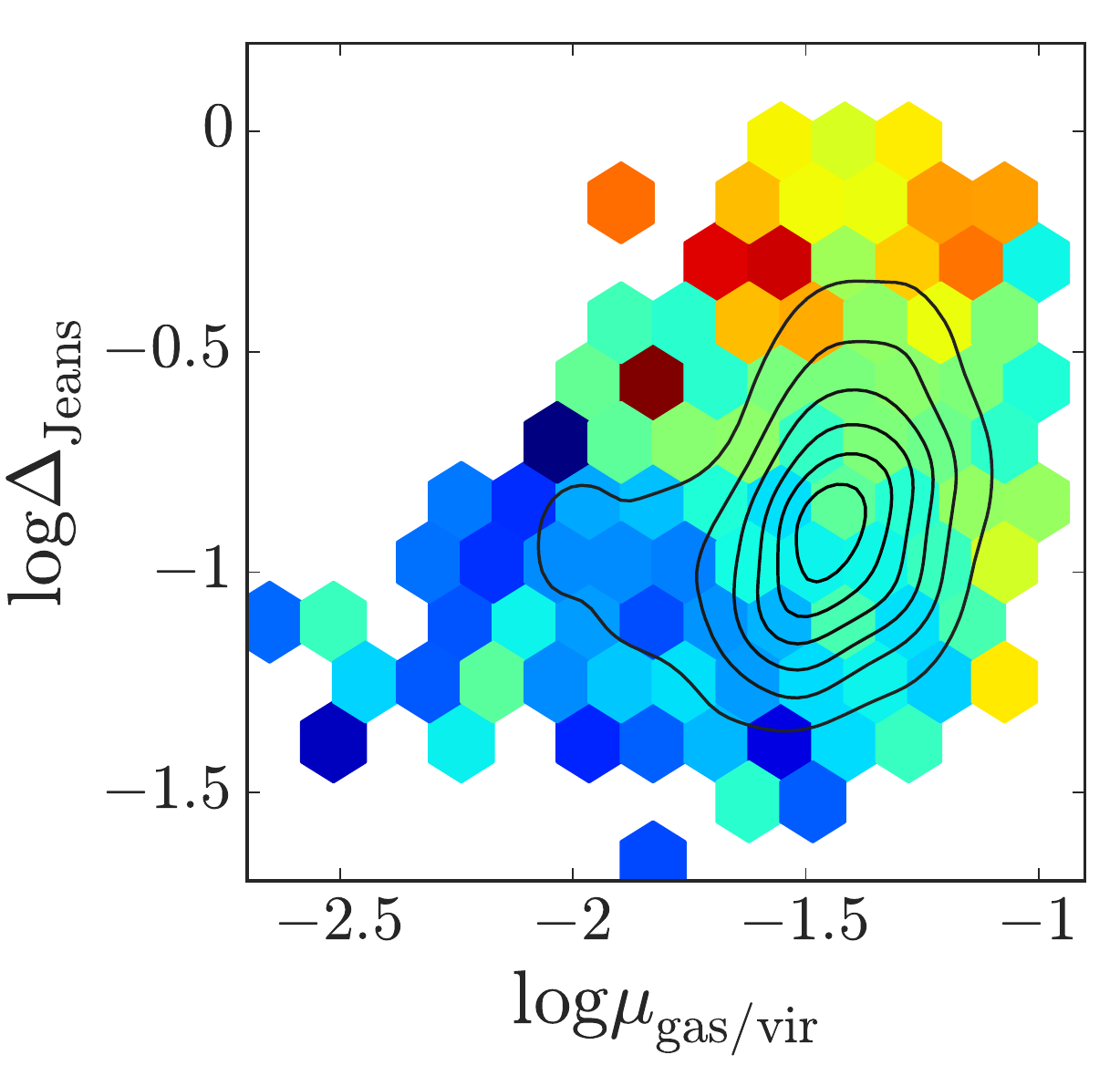}\hfill
	\includegraphics[height=0.28\linewidth,trim={0.8cm 0 0.2cm 0.6cm},clip]{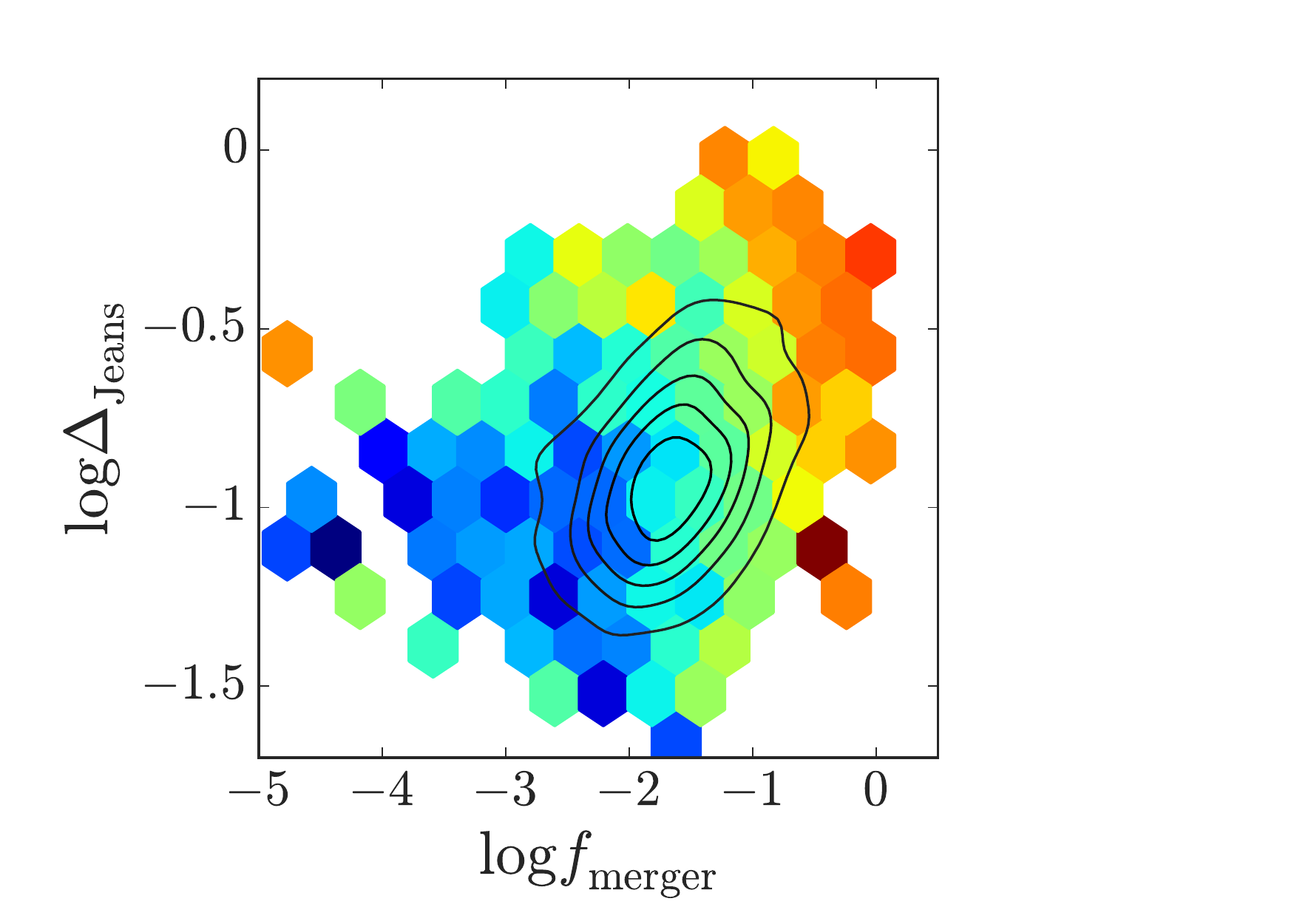}
	\vspace{-0.4cm }
	\caption{
		Distribution of the success measure $\log\delta$ as a function of time $t$, gas-to-virial mass ratio $\mu_{\rm gas/vir} = M_{\rm gas}/M_\vir$, merger indicator $f_{\rm merger}$ and deviation from Jeans equilibrium indicator $\Delta_{\rm Jeans}$. The color corresponds to the median value in each bin while contours show how the sample is distributed. Similar plots for the relative errors in inner slope and concentration, $\log|\Delta s|$ and $\log|\Delta c|$, are shown in Appendix~\ref{appendix:correlations}. Since $t$, $\mu_{\rm gas/vir}$, $f_{\rm merger}$, and $\Delta_{\rm Jeans}$ are the four main parameters on which the model success depends (cf. Fig.~\ref{fig:correlations_all}), this figure shows that mergers and the deviation from Jeans equilibrium are the main cause of failure, with secondary trends in gas content and time. 
	}
	\label{fig:correlations_hex}
\end{figure*}

To identify systematic causes for the failures of the model after the first 3 Gyr of evolution, we follow how the success measure $\delta$ (and the errors in inner slope and concentration, $\Delta s$ and $\Delta c$) depends on the following quantities relevent to disk and halo dynamics: time, masses of the dark matter, stellar and gas components, stellar size, gas fraction, as well as the merger indicator $f_{\rm merger}$ and a measure of the deviation from Jeans equilibrium, $\Delta_{\rm Jeans}$. This latter quantity corresponds to the RMS average of the relative difference between the two sides of the Jeans equation (Eq.~\ref{eq:appendix_Tjeans1}) for the dark matter component between 2\% and 60\% of the virial radius, namely $\log(r/R_\vir)$ between $-1.75$ and $-0.25$. This radius range covers most of the halo but avoids regions where the anisotropy parameter $\beta$ and the logarithmic slopes $\alpha$ and $\gamma$ entering the Jeans equation fluctuate more significantly (cf. Fig.\ref{fig:kinetic_beta} and the right panel of Fig.~\ref{fig:Tjeans}). 
Given the success condition and threshold, we further derive the probability $p$ for $\delta\leq \delta_{\rm max}$ in equally-spaced bins of these quantities to highlight the systematic trends: in each bin, $p$ is the ratio of the number of cases with $\delta\leq \delta_{\rm max}$ to the total number of cases in the bin\footnote{The probability $p$ differs from the success rates defined in Section~\ref{subsection:results} since it does not take into account whether the model prediction is closer to the final density profile than to the initial one, i.e., it does not take into account the second part of the success criterion explicited in Eq.~(\ref{eq:criterion}).}.
The figure with the different correlations is shown in Appendix~\ref{appendix:correlations} (Fig.~\ref{fig:correlations_all}). 
We find that $\delta$ correlates primarily with 
$f_{\rm merger}$ (with a Pearson correlation coefficiant $r=0.56$) 
and then with 
\textbf{$\Delta_{\rm Jeans}$ ($r=0.39$)}, 
time ($r=0.28$), and 
the gas-to-virial mass ratio $\mu_{\rm gas/vir} = M_{\rm gas}/M_{\vir}$ ($r=0.25$), which relates to the gas fraction. 
In contrast, it correlates weakly with 
the gas mass ($r=0.11$) and 
the stellar mass ($r=0.15$), and very weakly with 
the virial mass ($r=0.03$) and 
the stellar radius ($r=0.03$). 
Similarly, the errors on the inner slope and concentration correlate first with $f_{\rm merger}$ ($r=0.38$ for $\log|\Delta s|$ and $0.25$ for $\log|\Delta c|$), and then with $\Delta_{\rm Jeans}$ ($r=0.31$ and $r=0.21$), time ($r=0.21$ and $0.17$), $\mu_{\rm gas/vir}$ ($r=0.20$ and $0.14$) and $M_{\rm star}$ ($r=0.13$ and $0.16$). We note that the scatter in the different success measures increases significantly with $\mu_{\rm gas/vir}$. 
Given these correlations, the main causes for the failures of the model appear to be mergers, deviation from Jeans equilibrium, gas content, and time.

Fig.~\ref{fig:correlations_hex} projects the distribution of $\delta$ onto the  planes involving to these four main causes of failure  in order to disentangle their effect and highlight their relative importance.
Firstly, the figure clearly shows that amongst the four quantities at stake, the main drivers for failures (high values of $\delta$, in red) are mergers and deviations from Jeans equilibrium.
Secondly, we note that the four quantities are not uniformly sampled. 
In particular, the two upper left panels further indicate that mergers are more prevalent at high gas fraction and early time, which contributes to the dependences on $\mu_{\rm gas/vir}$ and~$t$. The median value of $\delta$ at $\log \mu_{\rm gas/vir}>-1.75$ is notably significantly more affected by mergers than below.
Finally, Fig.~\ref{fig:correlations_hex} shows secondary trends as a function of time and gas content: at fixed $f_{\rm merger}$ or $\Delta_{\rm Jeans}$, $\delta$ slightly decreases with time and increases with $\mu_{\rm gas/vir}$. 
Since $\mu_{\rm gas/vir}$ is inversely correlated with time (cf. Fig.~\ref{fig:correlations_all}, with $r=-52$), these two secondary trends may have the same physical origin. 
We further note in Fig.~\ref{fig:correlations_all} that while the scatter in $\delta$ is fairly uniform with time, it increases significantly with increasing gas content, hinting at a specific role for the gas.  
Plots comparable to Fig.~\ref{fig:correlations_hex} for $\log|\Delta s|$ and $\log|\Delta c|$ are shown in Appendix~\ref{appendix:correlations}, showing similar trends.

We conclude from this analysis that amongst the considered physical properties, (i) mergers and deviations from dynamical equilibrium are the main markers of failure of the model, (ii) that there are secondary trends in gas content and time, and (iii) that part of these secondary trends stem from the first two markers. As can be seen in Fig.~\ref{fig:correlations_all}, $f_{\rm merger}$ and $\Delta_{\rm Jeans}$ are themselves significantly correlated, with a Pearson correlation coefficient $r=0.47$, and it is thus difficult to fully disentangle their effects. Nevertheless, since mergers and fly-bys not only deviate the system from equilibrium but are also accompanied by significant non-sphericity that breaks another core assumptions of the model, they may constitute a better marker of failure -- which is notably reflected in the tighter correlation of $\delta$ with $f_{\rm merger}$ than with $\Delta_{\rm Jeans}$ ($r=0.56$ instead of $0.39$). 
Mergers are furthermore likely to invalidate energy conservation within shells enclosing a given collisionless mass during the relaxation. In the model, the sudden change of potential indeed imparts mean motions to the collisionless particles that are subsequently dissipated as the system relaxes to its new equilibrium. The change in the mass distribution during the relaxation may thus be envisioned to result from initial oscillations of the shells due to the imparted motions. If these oscillations are large, which would be the case during mergers, shell crossing would prevent energy conservation within shells enclosing the same collisionless mass. 
Since \cite{Pontzen2012} and \cite{El-Zant2016} invoke small successive mass changes to explain core formation, the model we propose here may thus also dynamically describe the situations they study and, in principle, provide a simple means to derive the evolution of the corresponding density profiles.
The secondary dependences of the success measure on time and gas fraction can be assumed to stem from the gas content, since the dissipationless nature of the gas makes it likely to violate the assumption of energy conservation during the relaxation phase. 
We also note in Fig.~\ref{fig:Kmulti} that an accumulation of gas around $0.1 R_\vir$ at the outskirts of the stellar disk induces a deviation in the ratio between the total mass and the dissipationless mass introduced in Section~\ref{sub:jeans}. Such deviations due to the gas distribution affect the parametrization of the kinetic energy and may thus also affect the ability of the model to predict the evolution of the dark matter density profile. 
Better accounting for the gas component, for example by parametrizing $M_{\rm tot}/M$ directly from the initial mass distribution instead of parametrizing it as in Eq.~(\ref{eq:Mratio}), may thus help improve the model in the future. 
We also further expect deviations from spherical symmetry to affect the success rate of the model independently of mergers.

\subsection{Contribution to the global change in inner density slope}
\label{section:sin}

Since our model describing the evolution of the density profile from inflow and outflow episodes fails during mergers, we assess the contribution of time steps devoid of mergers to the overall evolution of the inner slope $s_{\rm 0} = s(0.01R_\vir)$.
In this effect, we consider the inner slope variation between two successive time steps $\Delta s_{\rm 0}(t)= s_{\rm 0}(t)-s_{\rm 0}(t-\Delta t)$, where $\Delta t$ is the time step, and its absolute value $|\Delta s_{\rm 0}|$.

We find that merger-free time steps contribute to 71\% of $\Sigma \Delta s_{\rm 0}$ the total evolution of the inner slope for all galaxies of the sample (55\% for \texttt{g1.08e11}) and 81\% after 3 Gyr (83\% for \texttt{g1.08e11}).
Given these numbers and the high success rate of the model in the non-merger cases, our relatively simple model may thus be relevent to explain about 80\% of the inner slope variation after 3 Gyr of galaxies whose stellar mass at $z=0$ lies between $5\times 10^7$ and $5\times 10^9~M_\odot$ and which are making cores. 
If instead of $\Sigma \Delta s_{\rm 0}$ we consider $\Sigma |\Delta s_{\rm 0}|$ the sum of all inner slope variations in absolute value, we find that merger-free time steps contribute to 66\% for all galaxies of the sample (71\% for \texttt{g1.08e11}) and 70\% after 3 Gyr (94\% for \texttt{g1.08e11}).

However, in the current stage of the model, successful non-merger cases after 3 Gyr correspond only to 50\% of $\Sigma \Delta s_{\rm 0}$ (36\% for \texttt{g1.08e11}, where all the failures after 3 Gyr correspond to a decrease of $s_0$) and to 59\% of  $\Sigma |\Delta s_{\rm 0}|$ (78\% for \texttt{g1.08e11}). 
In the next section, we discuss the ability of the model to predict the evolution of the dark matter density profile over multiple time steps. 
We leave the description of the global time evolution of the inner logarithmic slope and concentration to future work, together with a sytematic study of the \cite{Dekel2017} parametrisation of the dark matter density profile in the NIHAO simulations.

\subsection{Multiple episodes}
\label{section:successive}

\begin{figure}
	\centering
	\includegraphics[width=1\linewidth,,trim={.3cm 1.5cm 0.3cm .3cm},clip]{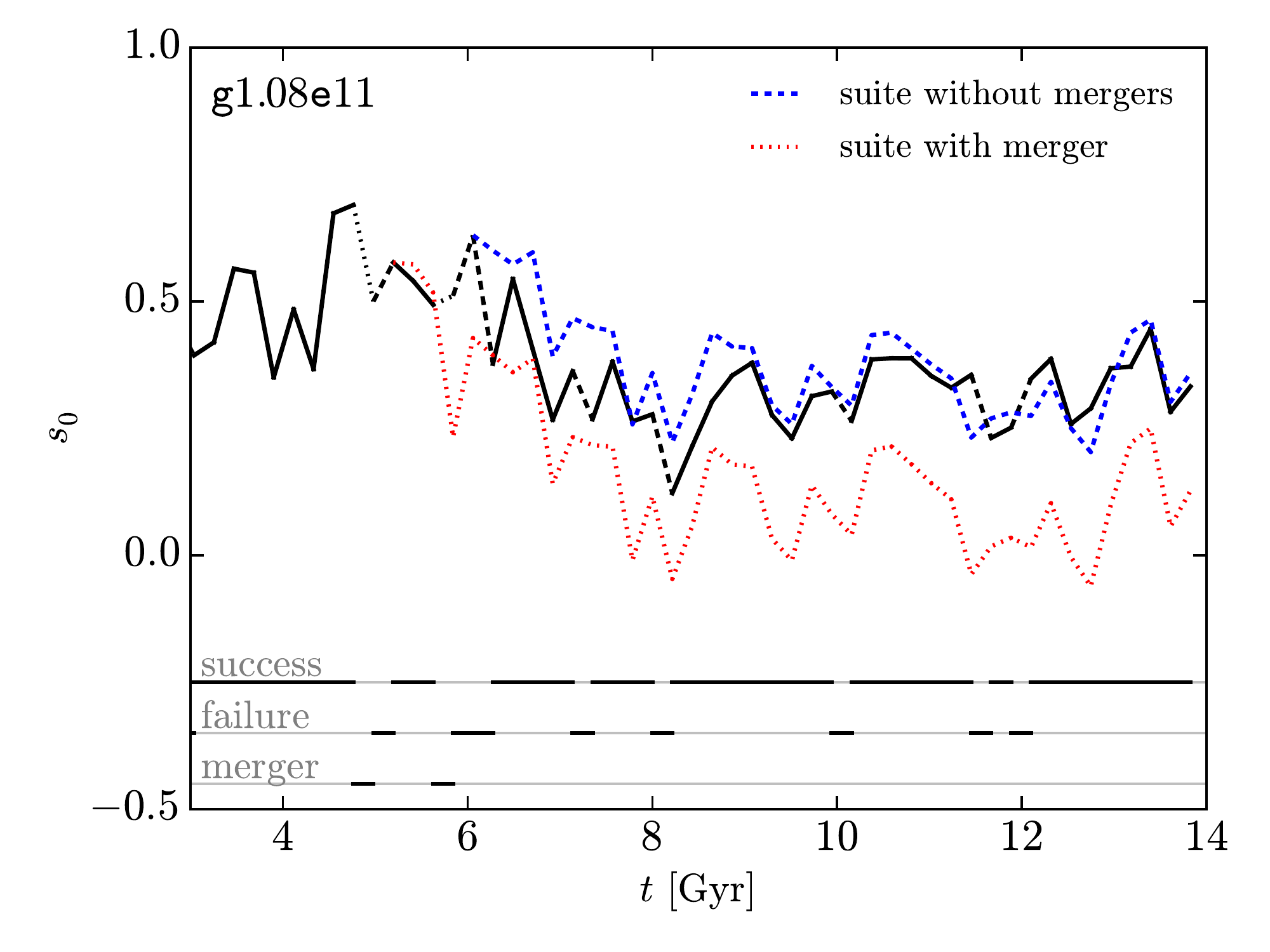}\\
	\vspace{0.2cm}
	\includegraphics[width=1\linewidth,,trim={.3cm 0.5cm 0.3cm .3cm},clip]{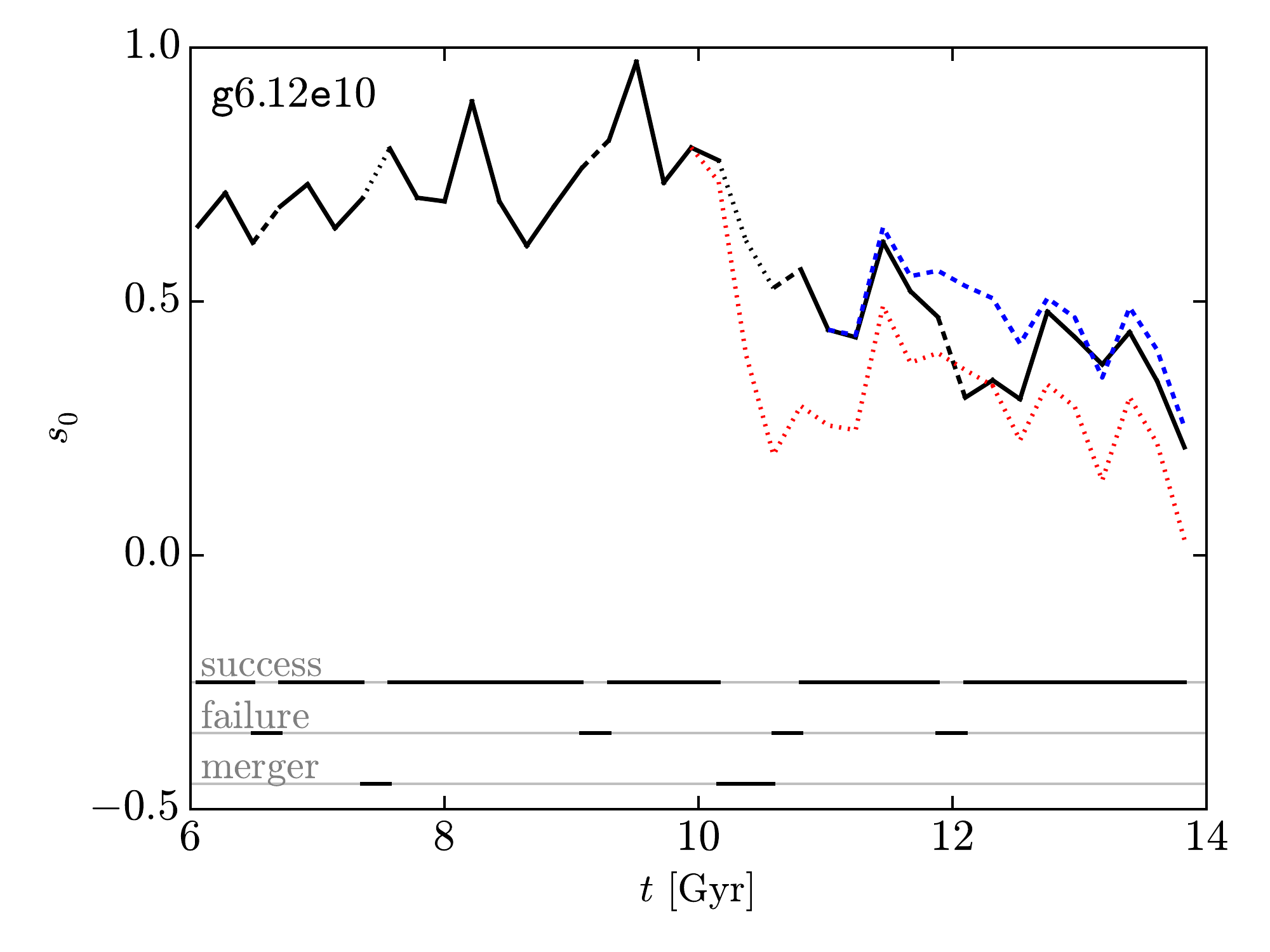}\hfill
	\vspace{-0.4cm}
	\caption{
		Evolution of the inner logarithmic slope of the dark matter density profile $s_0$ of NIHAO galaxies \texttt{g1.08e11} and \texttt{g6.12e10} (black lines) together with successive model predictions over suites of outputs without mergers (blue dashed lines) and interrupted by a merger (red dotted line). The style of the black line (plain, dahed, dotted) as well as the tracks at the bottom indicate model successes, failures and mergers (respectively). 
		In the absence of mergers, successive model predictions can account for the evolution of the inner slope of the dark matter density profile. 
	}
	\label{fig:successions}
\end{figure}

\begin{figure*}
	\centering
	\includegraphics[width=0.48\linewidth,trim={0.3cm .3cm 0.2cm .3cm},clip]{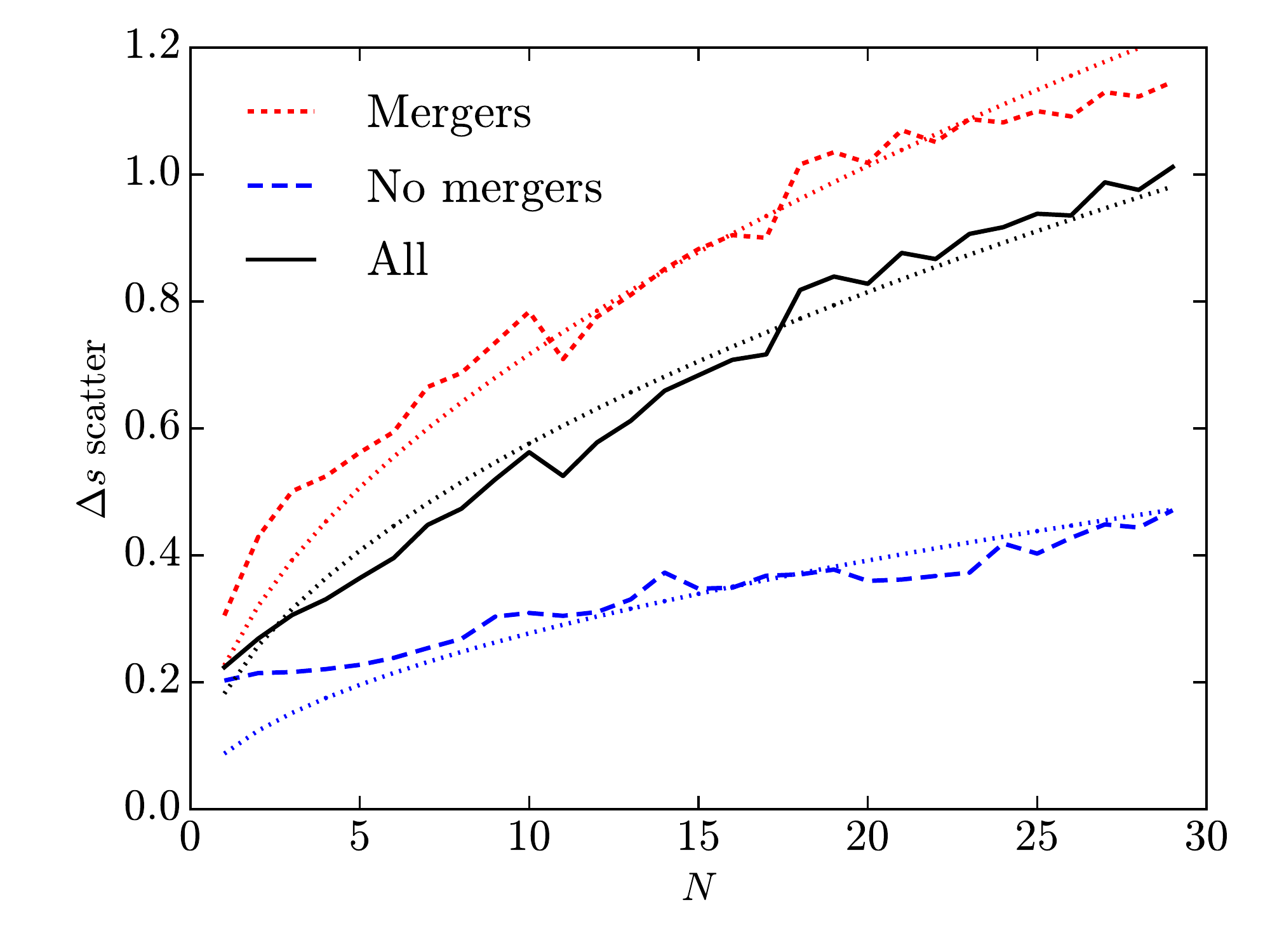}
	\includegraphics[width=0.48\linewidth,trim={0.3cm .3cm 0.2cm .3cm},clip]{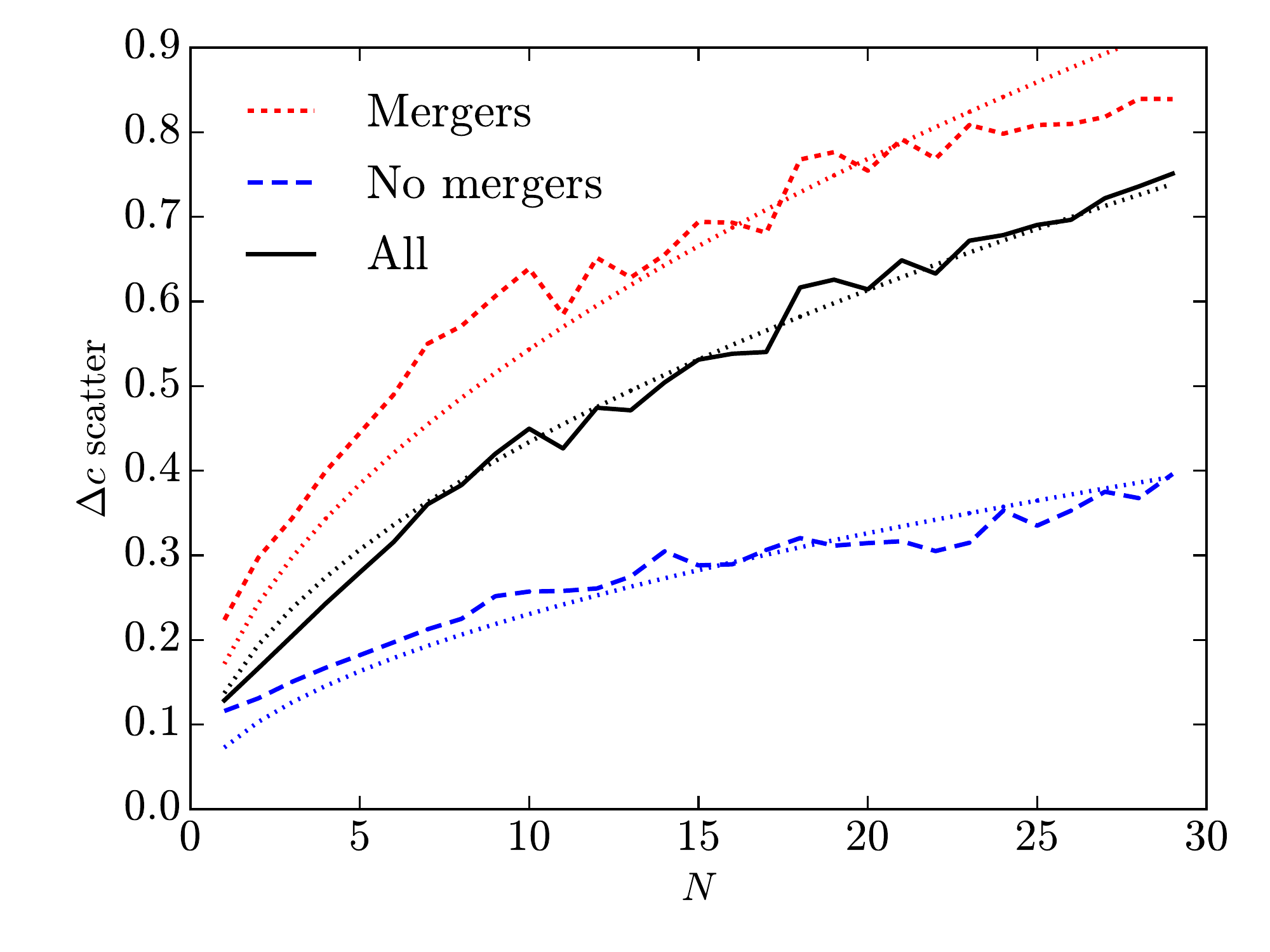}
	\vspace{-0.4cm}
	\caption{
		Evolution of the scatter in $\Delta s$ \textit{(left)} and $\Delta c$ \textit{(right)} when applying the model successively over $N$ consecutive outputs, for all the simulations of the sample after 3 Gyr. Dotted lines correspond to least-square fits proportional to $\sqrt{N}$. While the $\Delta s$ scatter doubles in 5 Gyr in the absence of mergers, it increases by a factor four in the presence of mergers. 
	}
	\label{fig:succession_scatter}
\end{figure*}

\begin{figure}
	\centering
	\includegraphics[width=1\linewidth,trim={0.3cm .3cm 0.2cm .3cm},clip]{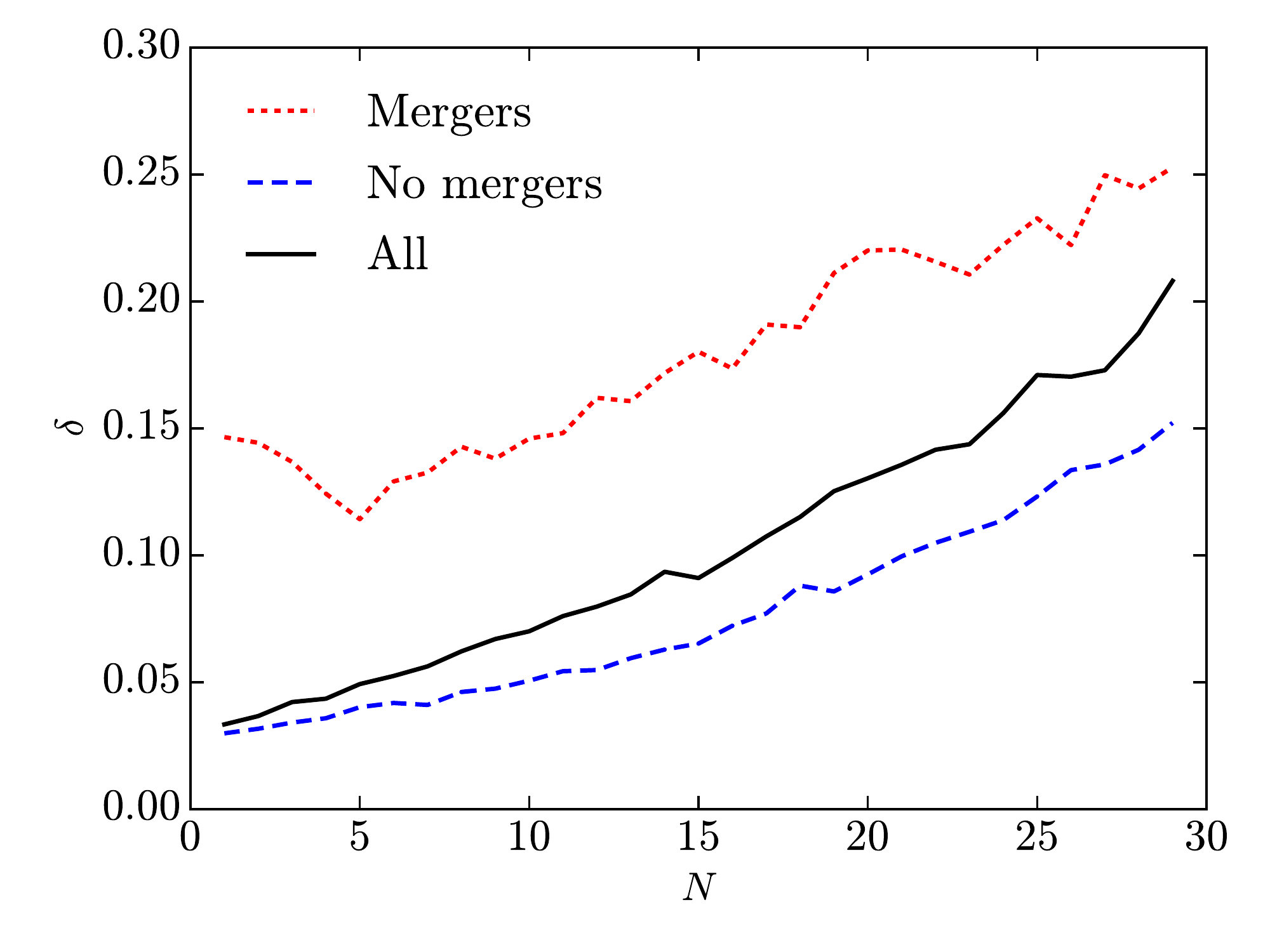}
	\vspace{-0.4cm}
	\caption{
		Evolution of the median success measure $\delta$ when applying the model successively over $N$ consecutive outputs, for all the simulations of the sample after 3 Gyr.
		The median $\delta$ remains below $\delta_{\rm max}$ until $N\sim 15$, i.e., for evolutions spanning about 3 Gyr. 
	}
	\label{fig:succession_delta}
\end{figure}

The model developed in Section~\ref{section:halo} was tested so far on successive outputs of the NIHAO simulations, separated by a timestep $\Delta t = 216~\rm Myr$. We showed in Section~\ref{subsection:results} that this model was successful at predicting the evolution of the dark matter density profile in about 74\% of non-merger cases given the success criterion in terms of $\delta$ stated in Eq.~(\ref{eq:criterion}) (or similarly that in terms of $\Delta s$ indicated in Eq.~(\ref{eq:Ds010}))
and in Section~\ref{section:mergers} that mergers were the main cause of failure. In the previous Section~\ref{section:sin}, we further showed that a large part of the evolution of the inner slope of the dark matter density profile (about 70\%) occured in the absence of mergers and hence in the validity domain of the model. However, to what extend is the model relevent to predict the evolution of the density profile over times larger than one timestep? 
Although we leave the full analysis to future work, we explore here the relevance of the model over multiple timesteps.

In this effect, we consider all suites of outputs within the sample, differentiating those that are interupted by mergers and those that are not. For each of these suites, we first apply the model to the initial dark matter distribution with the first mass change over $\Delta t$, then to the resulting prediction with the second mass change, and so on to the successive predictions. 
If $t_0$ is the initial time, we compare the dark matter density profiles parametrised by the successive parameters $p_n$ at $t=t_0+n\Delta t$ predicted by the model with the actual profiles.

Fig.~\ref{fig:successions} shows the time evolution of the inner logarithmic slope $s_0$ for simulated galaxies \texttt{g1.08e11} and \texttt{g6.12e10} together with two suites of successive predictions of the model for each galaxy, one uninterrupted by mergers and one including a merger. The correlations between the different curves and their slopes highlight the ability of the model to predict the evolution of the inner dark matter density profile between two successive outputs, as already shown in Section~\ref{subsection:success}, while the deviations between the curves relates to the scatter shown in Fig.~\ref{fig:sin_cmax}. 
The suites of successive predictions uninterupted by mergers follow closely the actual evolution of the inner logarithmic slopes over a large number of outputs. In contrast, the stochasticity introduced by the merger in the other suites leads the model predictions to deviate significantly from the actual values. Nevertheless, we note the consistency of the inner slope changes between the different suites. 

Fig.~\ref{fig:succession_scatter} shows how the scatters in $\Delta s$ and $\Delta c$, namely the scatter in Fig.~\ref{fig:sin_cmax} or the difference between the curves in Fig.~\ref{fig:successions}, vary with the number of successive outputs $N$ taken into account. Similarly, Fig.~\ref{fig:succession_delta} shows how the success measure $\delta$ varies with $N$. 
As already seen in Fig.~\ref{fig:successions}, suites of outputs interrupted by mergers first display much more scatter and a much larger success measure than those that are not interrupted by mergers. 
Scatters and success measures all increases with $N$ as a result of the accumulated uncertainties, however the evolution is significantly milder for the merger-free suites. Indeed, while the evolution of the scatter in the merger cases is compatible with an evolution in $\sqrt{N}$, it is flatter in the non-merger cases, in particular for $s_0$.  
Finally, the median $\delta$ remains below $\delta_{\rm max}=10\%$ until $N\sim 15$, so the probability associated to $\delta \leq \delta_{\rm max}$ remains above 50\% until this $N$, i.e., until about 3 Gyr. 
We thus conclude that while our simple theoretical model is unable to predict the global evolution of the dark matter density profile over more than a few Gyr because of the cumulative effect of mergers, of the scatter and of the progressive increase of $\delta$, it is able to predict with a non-negligible scatter the evolution of the dark matter density profile from episodes of inflows and outflows up to a few Gyr in the absence of mergers.

\section{Conclusion}
\label{section:conclusion}

In this article we presented a simple model describing the response of a dissipationless spherical system to an instantaneous mass change, intended to describe the formation of flat cores in dark matter haloes and UDGs by outflow episodes induced by stellar feedback. 
The model applies rigorously to a mass added or removed at the center of the  system but can be extended to mass changes at different radii under certain assumptions.
In addition to describing the effect of bulk mass changes, this model may further describe core formation from small successive mass variations, as envisioned by \cite{Pontzen2012} and \cite{El-Zant2016}.
The response of the system is divided into an instantaneous change of the gravitational potential to adapt to the mass variation while the velocities are frozen to their initial values, followed by relaxation to a new equilibrium. 
The key assumption of the model is that the local energy is conserved during the relaxation phase for shells enclosing a given dissipationless mass. 
This assumption is not formally justified in case of shell crossing, so its validity is tested by the success of the model in reproducing the simulation results. 
Other assumptions include spherical symmetry, isotropic velocities, and Jeans equilibrium for the initial and final systems. 
In order to follow the evolution of the dissipationless density profile with time, we parametrize it through a two-parameter function with a variable inner slope and an analytic gravitational potential profile following \cite{Dekel2017} and we derive analytic expressions for the resulting local kinetic energy. 
The contribution of the added or removed mass to the kinetic energy is determined assuming that most of it is located inside the spherical dark matter shells that are considered, which is notably valid when mass is added or removed directly at the center. 
To account for the different components of multi-component systems, such as in hydrodynamical simulations, we further use a simple parametrisation of the ratio between total and dissipationless enclosed masses.

The model is tested against NIHAO cosmological zoom-in simulations \citep{Wang2015} in the context of dark matter core formation. 
Firstly, we build upon \cite{Dekel2017} showing that their parametrisation of the dark matter density profile matches the simulated profiles significantly better than the usual NFW parametrisation. 
Secondly, we test the assumptions of spherical symmetry, velocity isotropy, Jeans equilibrium and the simple parametrization of the ratio between total and dissipationless masses. We show that they enable to recover the local kinetic energy with a relative error lower than 10\%. 
We note however that the assumption of energy conservation in shells encompassing a given collisionless mass during the relaxation phase can not be tested directly in the simulations since it is not possible to separate the two phases of evolution assumed in the model.
Finally, we compare the model predictions for the evolution of the dark matter density profile given the mass change during each 216 Myr timestep with the actual simulation outputs for the 31 galaxies whose stellar mass at $z=0$ is comprised between $5\times 10^7$ and $5\times 19^9~\rm M_\odot$. This is indeed the range where we expect dark matter core formation according to \cite{Tollet2016} and \cite{Dutton2016}. 
Focusing on times after the relatively perturbed first 3 Gyr of evolution and excluding mergers, we obtain a $\sim$74\% success rate, whether success is defined by a relative measure of the proximity between the inner density profile predicted by the model and that of the simulation output ($\delta\leq 10\%$) or by the accuracy of the predicted inner slope variation ($|\Delta s|\leq 10\%$). 

We show that mergers and deviations from Jeans equilibirum are the main cause of failure for the model, together with a small influence of the gas mass fraction -- higher mass fractions leading to more failures. Mergers are indeed expected to break the assumed spherical symmetry and perturb the equilibrium of the system, in particular through dynamical friction and gravitational torques. 
The presence of a large dissipative gas component makes the assumption of energy conservation during the relaxation phase less valid.
Nevertheless, the overall success of the model in reproducing the halo response in the simulations justifies a posteriori the assumption of energy conservation during the relaxation phase. 
Since about 70\% of the evolution of the inner dark matter density slope occurs without mergers, our model is relevent to describe a large part of this evolution given its success rate. 
We further test the model over times larger than 216 Myr, spanning multiple mass change episodes. We find that although the scatter increases with time, the success rate remains above 50\% over time intervals below 3 Gyr in the absence of mergers.

The model we propose provides a simple understanding of the formation of dark-matter cores and UDGs by supernova-driven outflow episodes. 
We note that the time step over which the model was tested (216 Myr) is above the dynamical time at the center of the halo, which may contribute to the success of the model since the system has enough time to relax between two successive outputs.
Instead of complex hydrodynamical simulations, idealized dissipationless simulations may further help testing this model and its assumptions in fully controlled experiments both for single mass changes and for multiple episodes. 
Possible improvements and extensions of the model were suggested throughout the text and are left to future studies. They include extending the model to cylindrical systems, adding a disk component to the spherical halo, taking into account the anisotropy of the velocity distribution, better accounting for the difference between total and dark matter mass and for the radial distribution of the mass change. 
Since simulations suggest that bursty star formation histories can lead both to dark matter core formation and to the expansion of the stellar distribution in UDGs, such galaxies may provide future observational tests for the model proposed here.


\section*{Acknowledgements}

We thank A. El-Zant for his careful reviewing, F. Combes, A. Nusser, F. van den Bosch, O. Ginzburg, T. Nussbaum, Y. Birnboim and M. Kretschmer for stimulating discussions and physical insights. 
This work was partly supported by the grants France-Israel PICS,
I-CORE Program of the
PBC/ISF 1829/12, BSF 2014-273, NSF AST-1405962,
GIF I-1341-303.7/2016, and DIP STE1869/2-1 GE625/17-1.
NIHAO simulations were carried out at the Gauss Centre for Super-computing e.V. (\url{www.gauss-centre.eu}) at the GCS Supercomputer SuperMUCat Leibniz Supercomputing Centre (\url{www.lrz.de}) and on the High Performance Computing resources at New York University Abu Dhabi. 




\bibliographystyle{mnras}
\bibliography{freundlich_cuspcore} 

\begin{thebibliography}{}
\makeatletter
\relax
\def\mn@urlcharsother{\let\do\@makeother \do\$\do\&\do\#\do\^\do\_\do\%\do\~}
\def\mn@doi{\begingroup\mn@urlcharsother \@ifnextchar [ {\mn@doi@}
  {\mn@doi@[]}}
\def\mn@doi@[#1]#2{\def\@tempa{#1}\ifx\@tempa\@empty \href
  {http://dx.doi.org/#2} {doi:#2}\else \href {http://dx.doi.org/#2} {#1}\fi
  \endgroup}
\def\mn@eprint#1#2{\mn@eprint@#1:#2::\@nil}
\def\mn@eprint@arXiv#1{\href {http://arxiv.org/abs/#1} {{\tt arXiv:#1}}}
\def\mn@eprint@dblp#1{\href {http://dblp.uni-trier.de/rec/bibtex/#1.xml}
  {dblp:#1}}
\def\mn@eprint@#1:#2:#3:#4\@nil{\def\@tempa {#1}\def\@tempb {#2}\def\@tempc
  {#3}\ifx \@tempc \@empty \let \@tempc \@tempb \let \@tempb \@tempa \fi \ifx
  \@tempb \@empty \def\@tempb {arXiv}\fi \@ifundefined
  {mn@eprint@\@tempb}{\@tempb:\@tempc}{\expandafter \expandafter \csname
  mn@eprint@\@tempb\endcsname \expandafter{\@tempc}}}

\bibitem[\protect\citeauthoryear{{Adams} et~al.,}{{Adams}
  et~al.}{2014}]{Adams2014}
{Adams} J.~J.,  et~al., 2014, \mn@doi [\apj] {10.1088/0004-637X/789/1/63}, 789,
  63

\bibitem[\protect\citeauthoryear{{Amorisco} \& {Loeb}}{{Amorisco} \&
  {Loeb}}{2016}]{Amorisco2016}
{Amorisco} N.~C.,  {Loeb} A.,  2016, \mn@doi [\mnras] {10.1093/mnrasl/slw055},
  459, L51

\bibitem[\protect\citeauthoryear{{Amorisco}, {Monachesi}, {Agnello}  \&
  {White}}{{Amorisco} et~al.}{2018}]{Amorisco2018}
{Amorisco} N.~C.,  {Monachesi} A.,  {Agnello} A.,   {White} S.~D.~M.,  2018,
  \mn@doi [\mnras] {10.1093/mnras/sty116}, 475, 4235

\bibitem[\protect\citeauthoryear{{Bar-Or}, {Fouvry}  \& {Tremaine}}{{Bar-Or}
  et~al.}{2018}]{Bar-Or2018}
{Bar-Or} B.,  {Fouvry} J.-B.,   {Tremaine} S.,  2018, preprint (\mn@eprint
  {arXiv} {1809.07673})

\bibitem[\protect\citeauthoryear{{Barnes} \& {Hernquist}}{{Barnes} \&
  {Hernquist}}{1991}]{Barnes1991}
{Barnes} J.~E.,  {Hernquist} L.~E.,  1991, \mn@doi [\apjl] {10.1086/185978},
  370, L65

\bibitem[\protect\citeauthoryear{{Beasley} \& {Trujillo}}{{Beasley} \&
  {Trujillo}}{2016}]{Beasley2016b}
{Beasley} M.~A.,  {Trujillo} I.,  2016, \mn@doi [\apj]
  {10.3847/0004-637X/830/1/23}, 830, 23

\bibitem[\protect\citeauthoryear{{Beasley}, {Romanowsky}, {Pota}, {Navarro},
  {Martinez Delgado}, {Neyer}  \& {Deich}}{{Beasley}
  et~al.}{2016}]{Beasley2016}
{Beasley} M.~A.,  {Romanowsky} A.~J.,  {Pota} V.,  {Navarro} I.~M.,  {Martinez
  Delgado} D.,  {Neyer} F.,   {Deich} A.~L.,  2016, \mn@doi [\apjl]
  {10.3847/2041-8205/819/2/L20}, 819, L20

\bibitem[\protect\citeauthoryear{{Behroozi}, {Wechsler}  \&
  {Conroy}}{{Behroozi} et~al.}{2013}]{Behroozi2013}
{Behroozi} P.~S.,  {Wechsler} R.~H.,   {Conroy} C.,  2013, \mn@doi [\apj]
  {10.1088/0004-637X/770/1/57}, 770, 57

\bibitem[\protect\citeauthoryear{{Binney} \& {Tremaine}}{{Binney} \&
  {Tremaine}}{2008}]{BinneyTremaine2008}
{Binney} J.,  {Tremaine} S.,  2008, {Galactic Dynamics: Second Edition}.
Princeton University Press

\bibitem[\protect\citeauthoryear{{Blumenthal}, {Faber}, {Flores}  \&
  {Primack}}{{Blumenthal} et~al.}{1986}]{Blumenthal1986}
{Blumenthal} G.~R.,  {Faber} S.~M.,  {Flores} R.,   {Primack} J.~R.,  1986,
  \mn@doi [\apj] {10.1086/163867}, 301, 27

\bibitem[\protect\citeauthoryear{{Bode}, {Ostriker}  \& {Turok}}{{Bode}
  et~al.}{2001}]{Bode2001}
{Bode} P.,  {Ostriker} J.~P.,   {Turok} N.,  2001, \mn@doi [\apj]
  {10.1086/321541}, 556, 93

\bibitem[\protect\citeauthoryear{{Bouch{\'e}} et~al.,}{{Bouch{\'e}}
  et~al.}{2010}]{Bouche2010}
{Bouch{\'e}} N.,  et~al., 2010, \mn@doi [\apj] {10.1088/0004-637X/718/2/1001},
  718, 1001

\bibitem[\protect\citeauthoryear{{Boylan-Kolchin}, {Bullock}  \&
  {Kaplinghat}}{{Boylan-Kolchin} et~al.}{2011}]{Boylan-Kolchin2011}
{Boylan-Kolchin} M.,  {Bullock} J.~S.,   {Kaplinghat} M.,  2011, \mn@doi
  [\mnras] {10.1111/j.1745-3933.2011.01074.x}, 415, L40

\bibitem[\protect\citeauthoryear{{Bryan} \& {Norman}}{{Bryan} \&
  {Norman}}{1998}]{Bryan1998}
{Bryan} G.~L.,  {Norman} M.~L.,  1998, \mn@doi [\apj] {10.1086/305262}, 495, 80

\bibitem[\protect\citeauthoryear{{Burkert}}{{Burkert}}{2000}]{Burkert2000}
{Burkert} A.,  2000, \mn@doi [\apjl] {10.1086/312674}, 534, L143

\bibitem[\protect\citeauthoryear{{Carleton}, {Errani}, {Cooper}, {Kaplinghat},
  {Pe{\~n}arrubia}  \& {Guo}}{{Carleton} et~al.}{2019}]{Carleton2019}
{Carleton} T.,  {Errani} R.,  {Cooper} M.,  {Kaplinghat} M.,  {Pe{\~n}arrubia}
  J.,   {Guo} Y.,  2019, \mn@doi [\mnras] {10.1093/mnras/stz383}, 485, 382

\bibitem[\protect\citeauthoryear{{Chan}, {Kere{\v s}}, {O{\~n}orbe}, {Hopkins},
  {Muratov}, {Faucher-Gigu{\`e}re}  \& {Quataert}}{{Chan}
  et~al.}{2015}]{Chan2015}
{Chan} T.~K.,  {Kere{\v s}} D.,  {O{\~n}orbe} J.,  {Hopkins} P.~F.,  {Muratov}
  A.~L.,  {Faucher-Gigu{\`e}re} C.-A.,   {Quataert} E.,  2015, ArXiv:
  1507.02282

\bibitem[\protect\citeauthoryear{{Chan}, {Kere{\v{s}}}, {Wetzel}, {Hopkins},
  {Faucher-Gigu{\`e}re}, {El-Badry}, {Garrison-Kimmel}  \&
  {Boylan-Kolchin}}{{Chan} et~al.}{2018}]{Chan2018}
{Chan} T.~K.,  {Kere{\v{s}}} D.,  {Wetzel} A.,  {Hopkins} P.~F.,
  {Faucher-Gigu{\`e}re} C.~A.,  {El-Badry} K.,  {Garrison-Kimmel} S.,
  {Boylan-Kolchin} M.,  2018, \mn@doi [\mnras] {10.1093/mnras/sty1153}, 478,
  906

\bibitem[\protect\citeauthoryear{{Chandrasekhar}}{{Chandrasekhar}}{1943}]{Chan%
drasekhar1943}
{Chandrasekhar} S.,  1943, \mn@doi [\apj] {10.1086/144517}, 97, 255

\bibitem[\protect\citeauthoryear{{Chevalier}}{{Chevalier}}{1977}]{Chevalier197%
7}
{Chevalier} R.~A.,  1977, \mn@doi [\araa]
  {10.1146/annurev.aa.15.090177.001135}, 15, 175

\bibitem[\protect\citeauthoryear{{Cole}, {Dehnen}  \& {Wilkinson}}{{Cole}
  et~al.}{2011}]{Cole2011}
{Cole} D.~R.,  {Dehnen} W.,   {Wilkinson} M.~I.,  2011, \mn@doi [\mnras]
  {10.1111/j.1365-2966.2011.19110.x}, 416, 1118

\bibitem[\protect\citeauthoryear{{Col{\'{\i}}n}, {Avila-Reese}  \&
  {Valenzuela}}{{Col{\'{\i}}n} et~al.}{2000}]{Colin2000}
{Col{\'{\i}}n} P.,  {Avila-Reese} V.,   {Valenzuela} O.,  2000, \mn@doi [\apj]
  {10.1086/317057}, 542, 622

\bibitem[\protect\citeauthoryear{{Dekel} \& {Mandelker}}{{Dekel} \&
  {Mandelker}}{2014}]{Dekel2014}
{Dekel} A.,  {Mandelker} N.,  2014, \mn@doi [\mnras] {10.1093/mnras/stu1427},
  444, 2071

\bibitem[\protect\citeauthoryear{{Dekel} \& {Silk}}{{Dekel} \&
  {Silk}}{1986}]{Dekel1986}
{Dekel} A.,  {Silk} J.,  1986, \mn@doi [\apj] {10.1086/164050}, 303, 39

\bibitem[\protect\citeauthoryear{{Dekel}, {Ishai}, {Dutton}  \&
  {Maccio}}{{Dekel} et~al.}{2017}]{Dekel2017}
{Dekel} A.,  {Ishai} G.,  {Dutton} A.~A.,   {Maccio} A.~V.,  2017, \mn@doi
  [\mnras] {10.1093/mnras/stx486}, 468, 1005

\bibitem[\protect\citeauthoryear{{Destri}, {de Vega}  \& {Sanchez}}{{Destri}
  et~al.}{2013}]{Destri2013}
{Destri} C.,  {de Vega} H.~J.,   {Sanchez} N.~G.,  2013, \mn@doi [\na]
  {10.1016/j.newast.2012.12.003}, 22, 39

\bibitem[\protect\citeauthoryear{{Di Cintio}, {Brook}, {Macci{\`o}}, {Stinson},
  {Knebe}, {Dutton}  \& {Wadsley}}{{Di Cintio} et~al.}{2014a}]{DiCintio2014}
{Di Cintio} A.,  {Brook} C.~B.,  {Macci{\`o}} A.~V.,  {Stinson} G.~S.,  {Knebe}
  A.,  {Dutton} A.~A.,   {Wadsley} J.,  2014a, \mn@doi [\mnras]
  {10.1093/mnras/stt1891}, 437, 415

\bibitem[\protect\citeauthoryear{{Di Cintio}, {Brook}, {Dutton}, {Macci{\`o}},
  {Stinson}  \& {Knebe}}{{Di Cintio} et~al.}{2014b}]{DiCintio2014b}
{Di Cintio} A.,  {Brook} C.~B.,  {Dutton} A.~A.,  {Macci{\`o}} A.~V.,
  {Stinson} G.~S.,   {Knebe} A.,  2014b, \mn@doi [\mnras]
  {10.1093/mnras/stu729}, 441, 2986

\bibitem[\protect\citeauthoryear{{Di Cintio}, {Brook}, {Dutton}, {Macci{\`o}},
  {Obreja}  \& {Dekel}}{{Di Cintio} et~al.}{2017}]{DiCintio2017}
{Di Cintio} A.,  {Brook} C.~B.,  {Dutton} A.~A.,  {Macci{\`o}} A.~V.,  {Obreja}
  A.,   {Dekel} A.,  2017, \mn@doi [\mnras] {10.1093/mnrasl/slw210}, 466, L1

\bibitem[\protect\citeauthoryear{{Dutton} \& {Macci{\`o}}}{{Dutton} \&
  {Macci{\`o}}}{2014}]{Dutton2014}
{Dutton} A.~A.,  {Macci{\`o}} A.~V.,  2014, \mn@doi [\mnras]
  {10.1093/mnras/stu742}, 441, 3359

\bibitem[\protect\citeauthoryear{{Dutton}, {Macci{\`o}}, {Frings}, {Wang},
  {Stinson}, {Penzo}  \& {Kang}}{{Dutton} et~al.}{2016a}]{Dutton2016a}
{Dutton} A.~A.,  {Macci{\`o}} A.~V.,  {Frings} J.,  {Wang} L.,  {Stinson}
  G.~S.,  {Penzo} C.,   {Kang} X.,  2016a, \mn@doi [\mnras]
  {10.1093/mnrasl/slv193}, 457, L74

\bibitem[\protect\citeauthoryear{{Dutton} et~al.,}{{Dutton}
  et~al.}{2016b}]{Dutton2016}
{Dutton} A.~A.,  et~al., 2016b, \mn@doi [\mnras] {10.1093/mnras/stw1537}, 461,
  2658

\bibitem[\protect\citeauthoryear{{Einasto}}{{Einasto}}{1965}]{Einasto1965}
{Einasto} J.,  1965, Trudy Astrofizicheskogo Instituta Alma-Ata, 5, 87

\bibitem[\protect\citeauthoryear{{El-Badry}, {Wetzel}, {Geha}, {Hopkins},
  {Kere{\v s}}, {Chan}  \& {Faucher-Gigu{\`e}re}}{{El-Badry}
  et~al.}{2016}]{El-Badry2016}
{El-Badry} K.,  {Wetzel} A.,  {Geha} M.,  {Hopkins} P.~F.,  {Kere{\v s}} D.,
  {Chan} T.~K.,   {Faucher-Gigu{\`e}re} C.-A.,  2016, \mn@doi [\apj]
  {10.3847/0004-637X/820/2/131}, 820, 131

\bibitem[\protect\citeauthoryear{{El-Zant}, {Shlosman}  \& {Hoffman}}{{El-Zant}
  et~al.}{2001}]{El-Zant2001}
{El-Zant} A.,  {Shlosman} I.,   {Hoffman} Y.,  2001, \mn@doi [\apj]
  {10.1086/322516}, 560, 636

\bibitem[\protect\citeauthoryear{{El-Zant}, {Hoffman}, {Primack}, {Combes}  \&
  {Shlosman}}{{El-Zant} et~al.}{2004}]{El-Zant2004}
{El-Zant} A.~A.,  {Hoffman} Y.,  {Primack} J.,  {Combes} F.,   {Shlosman} I.,
  2004, \mn@doi [\apjl] {10.1086/421938}, 607, L75

\bibitem[\protect\citeauthoryear{{El-Zant}, {Freundlich}  \&
  {Combes}}{{El-Zant} et~al.}{2016}]{El-Zant2016}
{El-Zant} A.~A.,  {Freundlich} J.,   {Combes} F.,  2016, \mn@doi [\mnras]
  {10.1093/mnras/stw1398}, 461, 1745

\bibitem[\protect\citeauthoryear{{El-Zant}, {Freundlich}, {Combes}  \&
  {Halle}}{{El-Zant} et~al.}{2019}]{El-Zant2019}
{El-Zant} A.~A.,  {Freundlich} J.,  {Combes} F.,   {Halle} A.,  2019, in prep.

\bibitem[\protect\citeauthoryear{{Famaey} \& {McGaugh}}{{Famaey} \&
  {McGaugh}}{2012}]{Famaey2012}
{Famaey} B.,  {McGaugh} S.~S.,  2012, \mn@doi [Living Reviews in Relativity]
  {10.12942/lrr-2012-10}, 15, 10

\bibitem[\protect\citeauthoryear{{Fensch} et~al.,}{{Fensch}
  et~al.}{2017}]{Fensch2017}
{Fensch} J.,  et~al., 2017, \mn@doi [\mnras] {10.1093/mnras/stw2920}, 465, 1934

\bibitem[\protect\citeauthoryear{{Flores} \& {Primack}}{{Flores} \&
  {Primack}}{1994}]{Flores1994}
{Flores} R.~A.,  {Primack} J.~R.,  1994, \mn@doi [\apjl] {10.1086/187350}, 427,
  L1

\bibitem[\protect\citeauthoryear{{Gentile}, {Famaey}  \& {de Blok}}{{Gentile}
  et~al.}{2011}]{Gentile2011}
{Gentile} G.,  {Famaey} B.,   {de Blok} W.~J.~G.,  2011, \mn@doi [\aap]
  {10.1051/0004-6361/201015283}, 527, A76

\bibitem[\protect\citeauthoryear{{Gill}, {Knebe}  \& {Gibson}}{{Gill}
  et~al.}{2004}]{Gill2004}
{Gill} S.~P.~D.,  {Knebe} A.,   {Gibson} B.~K.,  2004, \mn@doi [\mnras]
  {10.1111/j.1365-2966.2004.07786.x}, 351, 399

\bibitem[\protect\citeauthoryear{{Girardi} et~al.,}{{Girardi}
  et~al.}{2010}]{Girardi2010}
{Girardi} L.,  et~al., 2010, \mn@doi [\apj] {10.1088/0004-637X/724/2/1030},
  724, 1030

\bibitem[\protect\citeauthoryear{{Goerdt}, {Moore}, {Read}  \&
  {Stadel}}{{Goerdt} et~al.}{2010}]{Goerdt2010}
{Goerdt} T.,  {Moore} B.,  {Read} J.~I.,   {Stadel} J.,  2010, \mn@doi [\apj]
  {10.1088/0004-637X/725/2/1707}, 725, 1707

\bibitem[\protect\citeauthoryear{{Goodman}}{{Goodman}}{2000}]{Goodman2000}
{Goodman} J.,  2000, \mn@doi [\na] {10.1016/S1384-1076(00)00015-4}, 5, 103

\bibitem[\protect\citeauthoryear{{Governato} et~al.,}{{Governato}
  et~al.}{2010}]{Governato2010}
{Governato} F.,  et~al., 2010, \mn@doi [\nat] {10.1038/nature08640}, 463, 203

\bibitem[\protect\citeauthoryear{{Governato} et~al.,}{{Governato}
  et~al.}{2012}]{Governato2012}
{Governato} F.,  et~al., 2012, \mn@doi [\mnras]
  {10.1111/j.1365-2966.2012.20696.x}, 422, 1231

\bibitem[\protect\citeauthoryear{{Greco} et~al.,}{{Greco}
  et~al.}{2018}]{Greco2018}
{Greco} J.~P.,  et~al., 2018, \mn@doi [\apj] {10.3847/1538-4357/aab842}, 857,
  104

\bibitem[\protect\citeauthoryear{{Hu}, {Barkana}  \& {Gruzinov}}{{Hu}
  et~al.}{2000}]{Hu2000}
{Hu} W.,  {Barkana} R.,   {Gruzinov} A.,  2000, \mn@doi [Physical Review
  Letters] {10.1103/PhysRevLett.85.1158}, 85, 1158

\bibitem[\protect\citeauthoryear{{Janowiecki} et~al.,}{{Janowiecki}
  et~al.}{2015}]{Janowiecki2015}
{Janowiecki} S.,  et~al., 2015, \mn@doi [\apj] {10.1088/0004-637X/801/2/96},
  801, 96

\bibitem[\protect\citeauthoryear{{Janssens}, {Abraham}, {Brodie}, {Forbes},
  {Romanowsky}  \& {van Dokkum}}{{Janssens} et~al.}{2017}]{Janssens2017}
{Janssens} S.,  {Abraham} R.,  {Brodie} J.,  {Forbes} D.,  {Romanowsky} A.~J.,
   {van Dokkum} P.,  2017, \mn@doi [\apjl] {10.3847/2041-8213/aa667d}, 839, L17

\bibitem[\protect\citeauthoryear{{Jiang}, {Dekel}, {Freundlich}, {Romanowsky},
  {Dutton}, {Macci{\`o}}  \& {Di Cintio}}{{Jiang} et~al.}{2019}]{Jiang2018}
{Jiang} F.,  {Dekel} A.,  {Freundlich} J.,  {Romanowsky} A.~J.,  {Dutton}
  A.~A.,  {Macci{\`o}} A.~V.,   {Di Cintio} A.,  2019, \mn@doi [\mnras]
  {10.1093/mnras/stz1499}, 487, 5272

\bibitem[\protect\citeauthoryear{{Keller}, {Wadsley}, {Benincasa}  \&
  {Couchman}}{{Keller} et~al.}{2014}]{Keller2014}
{Keller} B.~W.,  {Wadsley} J.,  {Benincasa} S.~M.,   {Couchman} H.~M.~P.,
  2014, \mn@doi [\mnras] {10.1093/mnras/stu1058}, 442, 3013

\bibitem[\protect\citeauthoryear{{Knollmann} \& {Knebe}}{{Knollmann} \&
  {Knebe}}{2009}]{Knollmann2009}
{Knollmann} S.~R.,  {Knebe} A.,  2009, \mn@doi [\apjs]
  {10.1088/0067-0049/182/2/608}, 182, 608

\bibitem[\protect\citeauthoryear{{Koda}, {Yagi}, {Yamanoi}  \&
  {Komiyama}}{{Koda} et~al.}{2015}]{Koda2015}
{Koda} J.,  {Yagi} M.,  {Yamanoi} H.,   {Komiyama} Y.,  2015, \mn@doi [\apjl]
  {10.1088/2041-8205/807/1/L2}, 807, L2

\bibitem[\protect\citeauthoryear{{Kuzio de Naray} \& {Spekkens}}{{Kuzio de
  Naray} \& {Spekkens}}{2011}]{Kuzio2011}
{Kuzio de Naray} R.,  {Spekkens} K.,  2011, \mn@doi [\apjl]
  {10.1088/2041-8205/741/2/L29}, 741, L29

\bibitem[\protect\citeauthoryear{{Leisman} et~al.,}{{Leisman}
  et~al.}{2017}]{Leisman2017}
{Leisman} L.,  et~al., 2017, \mn@doi [\apj] {10.3847/1538-4357/aa7575}, 842,
  133

\bibitem[\protect\citeauthoryear{{Lilly}, {Carollo}, {Pipino}, {Renzini}  \&
  {Peng}}{{Lilly} et~al.}{2013}]{Lilly2013}
{Lilly} S.~J.,  {Carollo} C.~M.,  {Pipino} A.,  {Renzini} A.,   {Peng} Y.,
  2013, \mn@doi [\apj] {10.1088/0004-637X/772/2/119}, 772, 119

\bibitem[\protect\citeauthoryear{{Lim}, {Peng}, {C{\^o}t{\'e}}, {Sales}, {den
  Brok}, {Blakeslee}  \& {Guhathakurta}}{{Lim} et~al.}{2018}]{Lim2018}
{Lim} S.,  {Peng} E.~W.,  {C{\^o}t{\'e}} P.,  {Sales} L.~V.,  {den Brok} M.,
  {Blakeslee} J.~P.,   {Guhathakurta} P.,  2018, \mn@doi [\apj]
  {10.3847/1538-4357/aacb81}, 862, 82

\bibitem[\protect\citeauthoryear{{Macci{\`o}}, {Paduroiu}, {Anderhalden},
  {Schneider}  \& {Moore}}{{Macci{\`o}} et~al.}{2012a}]{Maccio2012a}
{Macci{\`o}} A.~V.,  {Paduroiu} S.,  {Anderhalden} D.,  {Schneider} A.,
  {Moore} B.,  2012a, \mn@doi [\mnras] {10.1111/j.1365-2966.2012.21284.x}, 424,
  1105

\bibitem[\protect\citeauthoryear{{Macci{\`o}}, {Stinson}, {Brook}, {Wadsley},
  {Couchman}, {Shen}, {Gibson}  \& {Quinn}}{{Macci{\`o}}
  et~al.}{2012b}]{Maccio2012b}
{Macci{\`o}} A.~V.,  {Stinson} G.,  {Brook} C.~B.,  {Wadsley} J.,  {Couchman}
  H.~M.~P.,  {Shen} S.,  {Gibson} B.~K.,   {Quinn} T.,  2012b, \mn@doi [\apjl]
  {10.1088/2041-8205/744/1/L9}, 744, L9

\bibitem[\protect\citeauthoryear{{Madau}, {Shen}  \& {Governato}}{{Madau}
  et~al.}{2014}]{Madau2014}
{Madau} P.,  {Shen} S.,   {Governato} F.,  2014, \mn@doi [\apjl]
  {10.1088/2041-8205/789/1/L17}, 789, L17

\bibitem[\protect\citeauthoryear{{Marigo}, {Girardi}, {Bressan}, {Groenewegen},
  {Silva}  \& {Granato}}{{Marigo} et~al.}{2008}]{Marigo2008}
{Marigo} P.,  {Girardi} L.,  {Bressan} A.,  {Groenewegen} M.~A.~T.,  {Silva}
  L.,   {Granato} G.~L.,  2008, \mn@doi [\aap] {10.1051/0004-6361:20078467},
  482, 883

\bibitem[\protect\citeauthoryear{{Marsh} \& {Niemeyer}}{{Marsh} \&
  {Niemeyer}}{2018}]{Marsh2018}
{Marsh} D. J.~E.,  {Niemeyer} J.~C.,  2018, arXiv e-prints, p. arXiv:1810.08543

\bibitem[\protect\citeauthoryear{{Marsh} \& {Silk}}{{Marsh} \&
  {Silk}}{2014}]{Marsh2014}
{Marsh} D.~J.~E.,  {Silk} J.,  2014, \mn@doi [\mnras] {10.1093/mnras/stt2079},
  437, 2652

\bibitem[\protect\citeauthoryear{{Mart{\'{\i}}n-Navarro}
  et~al.,}{{Mart{\'{\i}}n-Navarro} et~al.}{2019}]{Martin-Navarro2019}
{Mart{\'{\i}}n-Navarro} I.,  et~al., 2019, \mn@doi [\mnras]
  {10.1093/mnras/stz252}, 484, 3425

\bibitem[\protect\citeauthoryear{{Mart{\'{\i}}nez-Delgado}
  et~al.,}{{Mart{\'{\i}}nez-Delgado} et~al.}{2016}]{Martinez-Delgado2016}
{Mart{\'{\i}}nez-Delgado} D.,  et~al., 2016, \mn@doi [\aj]
  {10.3847/0004-6256/151/4/96}, 151, 96

\bibitem[\protect\citeauthoryear{{Martizzi}, {Teyssier}  \& {Moore}}{{Martizzi}
  et~al.}{2013}]{Martizzi2013}
{Martizzi} D.,  {Teyssier} R.,   {Moore} B.,  2013, \mn@doi [\mnras]
  {10.1093/mnras/stt297}, 432, 1947

\bibitem[\protect\citeauthoryear{{Mashchenko}, {Couchman}  \&
  {Wadsley}}{{Mashchenko} et~al.}{2006}]{Mashchenko2006}
{Mashchenko} S.,  {Couchman} H.~M.~P.,   {Wadsley} J.,  2006, \mn@doi [\nat]
  {10.1038/nature04944}, 442, 539

\bibitem[\protect\citeauthoryear{{Mashchenko}, {Wadsley}  \&
  {Couchman}}{{Mashchenko} et~al.}{2008}]{Mashchenko2008}
{Mashchenko} S.,  {Wadsley} J.,   {Couchman} H.~M.~P.,  2008, \mn@doi [Science]
  {10.1126/science.1148666}, 319, 174

\bibitem[\protect\citeauthoryear{{McGaugh} \& {de Blok}}{{McGaugh} \& {de
  Blok}}{1998}]{McGaugh1998a}
{McGaugh} S.~S.,  {de Blok} W.~J.~G.,  1998, \mn@doi [\apj] {10.1086/305612},
  499, 41

\bibitem[\protect\citeauthoryear{{McKee} \& {Ostriker}}{{McKee} \&
  {Ostriker}}{1977}]{McKee1977}
{McKee} C.~F.,  {Ostriker} J.~P.,  1977, \mn@doi [\apj] {10.1086/155667}, 218,
  148

\bibitem[\protect\citeauthoryear{{Merritt}, {van Dokkum}, {Danieli}, {Abraham},
  {Zhang}, {Karachentsev}  \& {Makarova}}{{Merritt} et~al.}{2016}]{Merritt2016}
{Merritt} A.,  {van Dokkum} P.,  {Danieli} S.,  {Abraham} R.,  {Zhang} J.,
  {Karachentsev} I.~D.,   {Makarova} L.~N.,  2016, \mn@doi [\apj]
  {10.3847/1538-4357/833/2/168}, 833, 168

\bibitem[\protect\citeauthoryear{{Mihos} \& {Hernquist}}{{Mihos} \&
  {Hernquist}}{1996}]{Mihos1996}
{Mihos} J.~C.,  {Hernquist} L.,  1996, \mn@doi [\apj] {10.1086/177353}, 464,
  641

\bibitem[\protect\citeauthoryear{{Mihos} et~al.,}{{Mihos}
  et~al.}{2015}]{Mihos2015}
{Mihos} J.~C.,  et~al., 2015, \mn@doi [\apjl] {10.1088/2041-8205/809/2/L21},
  809, L21

\bibitem[\protect\citeauthoryear{{Mihos}, {Harding}, {Feldmeier}, {Rudick},
  {Janowiecki}, {Morrison}, {Slater}  \& {Watkins}}{{Mihos}
  et~al.}{2017}]{Mihos2017}
{Mihos} J.~C.,  {Harding} P.,  {Feldmeier} J.~J.,  {Rudick} C.,  {Janowiecki}
  S.,  {Morrison} H.,  {Slater} C.,   {Watkins} A.,  2017, \mn@doi [\apj]
  {10.3847/1538-4357/834/1/16}, 834, 16

\bibitem[\protect\citeauthoryear{{Milgrom}}{{Milgrom}}{1983}]{Milgrom1983a}
{Milgrom} M.,  1983, \mn@doi [\apj] {10.1086/161130}, 270, 365

\bibitem[\protect\citeauthoryear{{Moore}}{{Moore}}{1994}]{Moore1994}
{Moore} B.,  1994, \mn@doi [\nat] {10.1038/370629a0}, 370, 629

\bibitem[\protect\citeauthoryear{{Mowla}, {van Dokkum}, {Merritt}, {Abraham},
  {Yagi}  \& {Koda}}{{Mowla} et~al.}{2017}]{Mowla2017}
{Mowla} L.,  {van Dokkum} P.,  {Merritt} A.,  {Abraham} R.,  {Yagi} M.,
  {Koda} J.,  2017, \mn@doi [\apj] {10.3847/1538-4357/aa961b}, 851, 27

\bibitem[\protect\citeauthoryear{{Mu{\~n}oz} et~al.,}{{Mu{\~n}oz}
  et~al.}{2015}]{Munoz2015}
{Mu{\~n}oz} R.~P.,  et~al., 2015, \mn@doi [\apjl]
  {10.1088/2041-8205/813/1/L15}, 813, L15

\bibitem[\protect\citeauthoryear{{Navarro}, {Frenk}  \& {White}}{{Navarro}
  et~al.}{1996}]{NFW1996}
{Navarro} J.~F.,  {Frenk} C.~S.,   {White} S.~D.~M.,  1996, \mn@doi [\apj]
  {10.1086/177173}, 462, 563

\bibitem[\protect\citeauthoryear{{Navarro}, {Frenk}  \& {White}}{{Navarro}
  et~al.}{1997}]{NFW1997}
{Navarro} J.~F.,  {Frenk} C.~S.,   {White} S.~D.~M.,  1997, \mn@doi [\apj]
  {10.1086/304888}, 490, 493

\bibitem[\protect\citeauthoryear{{Newman}, {Treu}, {Ellis}, {Sand}, {Nipoti},
  {Richard}  \& {Jullo}}{{Newman} et~al.}{2013a}]{Newman2013a}
{Newman} A.~B.,  {Treu} T.,  {Ellis} R.~S.,  {Sand} D.~J.,  {Nipoti} C.,
  {Richard} J.,   {Jullo} E.,  2013a, \mn@doi [\apj]
  {10.1088/0004-637X/765/1/24}, 765, 24

\bibitem[\protect\citeauthoryear{{Newman}, {Treu}, {Ellis}  \& {Sand}}{{Newman}
  et~al.}{2013b}]{Newman2013b}
{Newman} A.~B.,  {Treu} T.,  {Ellis} R.~S.,   {Sand} D.~J.,  2013b, \mn@doi
  [\apj] {10.1088/0004-637X/765/1/25}, 765, 25

\bibitem[\protect\citeauthoryear{{Nipoti} \& {Binney}}{{Nipoti} \&
  {Binney}}{2015}]{Nipoti2015}
{Nipoti} C.,  {Binney} J.,  2015, \mn@doi [\mnras] {10.1093/mnras/stu2217},
  446, 1820

\bibitem[\protect\citeauthoryear{{Oh}, {de Blok}, {Brinks}, {Walter}  \&
  {Kennicutt}}{{Oh} et~al.}{2011}]{Oh2011}
{Oh} S.-H.,  {de Blok} W.~J.~G.,  {Brinks} E.,  {Walter} F.,   {Kennicutt} Jr.
  R.~C.,  2011, \mn@doi [\aj] {10.1088/0004-6256/141/6/193}, 141, 193

\bibitem[\protect\citeauthoryear{{Oh} et~al.,}{{Oh} et~al.}{2015}]{Oh2015}
{Oh} S.-H.,  et~al., 2015, \mn@doi [\aj] {10.1088/0004-6256/149/6/180}, 149,
  180

\bibitem[\protect\citeauthoryear{{Pandya} et~al.,}{{Pandya}
  et~al.}{2018}]{Pandya2018}
{Pandya} V.,  et~al., 2018, \mn@doi [\apj] {10.3847/1538-4357/aab498}, 858, 29

\bibitem[\protect\citeauthoryear{{Peirani} et~al.,}{{Peirani}
  et~al.}{2017}]{Peirani2017}
{Peirani} S.,  et~al., 2017, \mn@doi [\mnras] {10.1093/mnras/stx2099}, 472,
  2153

\bibitem[\protect\citeauthoryear{{Peng} \& {Lim}}{{Peng} \&
  {Lim}}{2016}]{Peng2016}
{Peng} E.~W.,  {Lim} S.,  2016, \mn@doi [\apjl] {10.3847/2041-8205/822/2/L31},
  822, L31

\bibitem[\protect\citeauthoryear{{Penoyre} \& {Haiman}}{{Penoyre} \&
  {Haiman}}{2018}]{Penoyre2018}
{Penoyre} Z.,  {Haiman} Z.,  2018, \mn@doi [\mnras] {10.1093/mnras/stx2469},
  473, 498

\bibitem[\protect\citeauthoryear{{Peter}, {Rocha}, {Bullock}  \&
  {Kaplinghat}}{{Peter} et~al.}{2013}]{Peter2013}
{Peter} A.~H.~G.,  {Rocha} M.,  {Bullock} J.~S.,   {Kaplinghat} M.,  2013,
  \mn@doi [\mnras] {10.1093/mnras/sts535}, 430, 105

\bibitem[\protect\citeauthoryear{{Planck Collaboration} et~al.,}{{Planck
  Collaboration} et~al.}{2014}]{Planck2014}
{Planck Collaboration} et~al., 2014, \mn@doi [\aap]
  {10.1051/0004-6361/201321591}, 571, A16

\bibitem[\protect\citeauthoryear{{Pontzen} \& {Governato}}{{Pontzen} \&
  {Governato}}{2012}]{Pontzen2012}
{Pontzen} A.,  {Governato} F.,  2012, \mn@doi [\mnras]
  {10.1111/j.1365-2966.2012.20571.x}, 421, 3464

\bibitem[\protect\citeauthoryear{{Pontzen} \& {Governato}}{{Pontzen} \&
  {Governato}}{2014}]{Pontzen2014}
{Pontzen} A.,  {Governato} F.,  2014, \mn@doi [\nat] {10.1038/nature12953},
  506, 171

\bibitem[\protect\citeauthoryear{{Pontzen}, {Ro{\v s}kar}, {Stinson}  \&
  {Woods}}{{Pontzen} et~al.}{2013}]{Pontzen2013}
{Pontzen} A.,  {Ro{\v s}kar} R.,  {Stinson} G.,   {Woods} R.,  2013, {pynbody:
  N-Body/SPH analysis for python}, Astrophysics Source Code Library (\mn@eprint
  {ascl} {1305.002})

\bibitem[\protect\citeauthoryear{{Read} \& {Gilmore}}{{Read} \&
  {Gilmore}}{2005}]{Read2005}
{Read} J.~I.,  {Gilmore} G.,  2005, \mn@doi [\mnras]
  {10.1111/j.1365-2966.2004.08424.x}, 356, 107

\bibitem[\protect\citeauthoryear{{Rom{\'a}n} \& {Trujillo}}{{Rom{\'a}n} \&
  {Trujillo}}{2017}]{Roman2017}
{Rom{\'a}n} J.,  {Trujillo} I.,  2017, \mn@doi [\mnras] {10.1093/mnras/stx438},
  468, 703

\bibitem[\protect\citeauthoryear{{Romano-D{\'{\i}}az}, {Shlosman}, {Hoffman}
  \& {Heller}}{{Romano-D{\'{\i}}az} et~al.}{2008}]{RomanoDiaz2008}
{Romano-D{\'{\i}}az} E.,  {Shlosman} I.,  {Hoffman} Y.,   {Heller} C.,  2008,
  \mn@doi [\apjl] {10.1086/592687}, 685, L105

\bibitem[\protect\citeauthoryear{{Rong}, {Guo}, {Gao}, {Liao}, {Xie}, {Puzia},
  {Sun}  \& {Pan}}{{Rong} et~al.}{2017}]{Rong2017}
{Rong} Y.,  {Guo} Q.,  {Gao} L.,  {Liao} S.,  {Xie} L.,  {Puzia} T.~H.,  {Sun}
  S.,   {Pan} J.,  2017, \mn@doi [\mnras] {10.1093/mnras/stx1440}, 470, 4231

\bibitem[\protect\citeauthoryear{Savitzky \& Golay}{Savitzky \&
  Golay}{1964}]{SavitzkyGolay1964}
Savitzky A.,  Golay M. J.~E.,  1964, \mn@doi [Analytical Chemistry]
  {10.1021/ac60214a047}, 36, 1627

\bibitem[\protect\citeauthoryear{{Shen}, {Wadsley}  \& {Stinson}}{{Shen}
  et~al.}{2010}]{Shen2010}
{Shen} S.,  {Wadsley} J.,   {Stinson} G.,  2010, \mn@doi [\mnras]
  {10.1111/j.1365-2966.2010.17047.x}, 407, 1581

\bibitem[\protect\citeauthoryear{{Shi} et~al.,}{{Shi} et~al.}{2017}]{Shi2017}
{Shi} D.~D.,  et~al., 2017, \mn@doi [\apj] {10.3847/1538-4357/aa8327}, 846, 26

\bibitem[\protect\citeauthoryear{{Spergel} \& {Steinhardt}}{{Spergel} \&
  {Steinhardt}}{2000}]{Spergel2000}
{Spergel} D.~N.,  {Steinhardt} P.~J.,  2000, \mn@doi [Physical Review Letters]
  {10.1103/PhysRevLett.84.3760}, 84, 3760

\bibitem[\protect\citeauthoryear{{Stinson}, {Seth}, {Katz}, {Wadsley},
  {Governato}  \& {Quinn}}{{Stinson} et~al.}{2006}]{Stinson2006}
{Stinson} G.,  {Seth} A.,  {Katz} N.,  {Wadsley} J.,  {Governato} F.,   {Quinn}
  T.,  2006, \mn@doi [\mnras] {10.1111/j.1365-2966.2006.11097.x}, 373, 1074

\bibitem[\protect\citeauthoryear{{Stinson}, {Brook}, {Macci{\`o}}, {Wadsley},
  {Quinn}  \& {Couchman}}{{Stinson} et~al.}{2013}]{Stinson2013}
{Stinson} G.~S.,  {Brook} C.,  {Macci{\`o}} A.~V.,  {Wadsley} J.,  {Quinn}
  T.~R.,   {Couchman} H.~M.~P.,  2013, \mn@doi [\mnras] {10.1093/mnras/sts028},
  428, 129

\bibitem[\protect\citeauthoryear{{Stinson} et~al.,}{{Stinson}
  et~al.}{2015}]{Stinson2015}
{Stinson} G.~S.,  et~al., 2015, \mn@doi [\mnras] {10.1093/mnras/stv1985}, 454,
  1105

\bibitem[\protect\citeauthoryear{{Teyssier}, {Pontzen}, {Dubois}  \&
  {Read}}{{Teyssier} et~al.}{2013}]{Teyssier2013}
{Teyssier} R.,  {Pontzen} A.,  {Dubois} Y.,   {Read} J.~I.,  2013, \mn@doi
  [\mnras] {10.1093/mnras/sts563}, 429, 3068

\bibitem[\protect\citeauthoryear{{Tollet} et~al.,}{{Tollet}
  et~al.}{2016}]{Tollet2016}
{Tollet} E.,  et~al., 2016, \mn@doi [\mnras] {10.1093/mnras/stv2856}, 456, 3542

\bibitem[\protect\citeauthoryear{{Tonini}, {Lapi}  \& {Salucci}}{{Tonini}
  et~al.}{2006}]{Tonini2006}
{Tonini} C.,  {Lapi} A.,   {Salucci} P.,  2006, \mn@doi [\apj]
  {10.1086/506431}, 649, 591

\bibitem[\protect\citeauthoryear{{Trujillo}, {Roman}, {Filho}  \& {S{\'a}nchez
  Almeida}}{{Trujillo} et~al.}{2017}]{Trujillo2017}
{Trujillo} I.,  {Roman} J.,  {Filho} M.,   {S{\'a}nchez Almeida} J.,  2017,
  \mn@doi [\apj] {10.3847/1538-4357/aa5cbb}, 836, 191

\bibitem[\protect\citeauthoryear{{Wadsley}, {Stadel}  \& {Quinn}}{{Wadsley}
  et~al.}{2004}]{Wadsley2004}
{Wadsley} J.~W.,  {Stadel} J.,   {Quinn} T.,  2004, \mn@doi [\na]
  {10.1016/j.newast.2003.08.004}, 9, 137

\bibitem[\protect\citeauthoryear{{Wadsley}, {Veeravalli}  \&
  {Couchman}}{{Wadsley} et~al.}{2008}]{Wadsley2008}
{Wadsley} J.~W.,  {Veeravalli} G.,   {Couchman} H.~M.~P.,  2008, \mn@doi
  [\mnras] {10.1111/j.1365-2966.2008.13260.x}, 387, 427

\bibitem[\protect\citeauthoryear{{Wadsley}, {Keller}  \& {Quinn}}{{Wadsley}
  et~al.}{2017}]{Wadsley2017}
{Wadsley} J.~W.,  {Keller} B.~W.,   {Quinn} T.~R.,  2017, \mn@doi [\mnras]
  {10.1093/mnras/stx1643}, 471, 2357

\bibitem[\protect\citeauthoryear{{Wang}, {Dutton}, {Stinson}, {Macci{\`o}},
  {Penzo}, {Kang}, {Keller}  \& {Wadsley}}{{Wang} et~al.}{2015}]{Wang2015}
{Wang} L.,  {Dutton} A.~A.,  {Stinson} G.~S.,  {Macci{\`o}} A.~V.,  {Penzo} C.,
   {Kang} X.,  {Keller} B.~W.,   {Wadsley} J.,  2015, \mn@doi [\mnras]
  {10.1093/mnras/stv1937}, 454, 83

\bibitem[\protect\citeauthoryear{{Yagi}, {Koda}, {Komiyama}  \&
  {Yamanoi}}{{Yagi} et~al.}{2016}]{Yagi2016}
{Yagi} M.,  {Koda} J.,  {Komiyama} Y.,   {Yamanoi} H.,  2016, \mn@doi [\apjs]
  {10.3847/0067-0049/225/1/11}, 225, 11

\bibitem[\protect\citeauthoryear{{Yozin} \& {Bekki}}{{Yozin} \&
  {Bekki}}{2015}]{Yozin2015}
{Yozin} C.,  {Bekki} K.,  2015, \mn@doi [\mnras] {10.1093/mnras/stv1073}, 452,
  937

\bibitem[\protect\citeauthoryear{{Zhao}}{{Zhao}}{1996}]{Zhao1996}
{Zhao} H.,  1996, \mn@doi [\mnras] {10.1093/mnras/278.2.488}, 278, 488

\bibitem[\protect\citeauthoryear{{Zolotov} et~al.,}{{Zolotov}
  et~al.}{2012}]{Zolotov2012}
{Zolotov} A.,  et~al., 2012, \mn@doi [\apj] {10.1088/0004-637X/761/1/71}, 761,
  71

\bibitem[\protect\citeauthoryear{{de Blok}}{{de Blok}}{2010}]{deBlok2010}
{de Blok} W.~J.~G.,  2010, \mn@doi [Advances in Astronomy]
  {10.1155/2010/789293}, 2010, 789293

\bibitem[\protect\citeauthoryear{{de Blok}, {Walter}, {Brinks}, {Trachternach},
  {Oh}  \& {Kennicutt}}{{de Blok} et~al.}{2008}]{deBlok2008}
{de Blok} W.~J.~G.,  {Walter} F.,  {Brinks} E.,  {Trachternach} C.,  {Oh}
  S.-H.,   {Kennicutt} Jr. R.~C.,  2008, \mn@doi [\aj]
  {10.1088/0004-6256/136/6/2648}, 136, 2648

\bibitem[\protect\citeauthoryear{{van Dokkum}, {Abraham}, {Merritt}, {Zhang},
  {Geha}  \& {Conroy}}{{van Dokkum} et~al.}{2015a}]{VanDokkum2015}
{van Dokkum} P.~G.,  {Abraham} R.,  {Merritt} A.,  {Zhang} J.,  {Geha} M.,
  {Conroy} C.,  2015a, \mn@doi [\apjl] {10.1088/2041-8205/798/2/L45}, 798, L45

\bibitem[\protect\citeauthoryear{{van Dokkum} et~al.,}{{van Dokkum}
  et~al.}{2015b}]{vanDokkum2015b}
{van Dokkum} P.~G.,  et~al., 2015b, \mn@doi [\apjl]
  {10.1088/2041-8205/804/1/L26}, 804, L26

\bibitem[\protect\citeauthoryear{{van Dokkum} et~al.,}{{van Dokkum}
  et~al.}{2016}]{vanDokkum2016}
{van Dokkum} P.,  et~al., 2016, \mn@doi [\apjl] {10.3847/2041-8205/828/1/L6},
  828, L6

\bibitem[\protect\citeauthoryear{{van Dokkum} et~al.,}{{van Dokkum}
  et~al.}{2017}]{vanDokkum2017}
{van Dokkum} P.,  et~al., 2017, \mn@doi [\apjl] {10.3847/2041-8213/aa7ca2},
  844, L11

\bibitem[\protect\citeauthoryear{{van den Bosch} \& {Swaters}}{{van den Bosch}
  \& {Swaters}}{2001}]{vandenBosch2001}
{van den Bosch} F.~C.,  {Swaters} R.~A.,  2001, \mn@doi [\mnras]
  {10.1046/j.1365-8711.2001.04456.x}, 325, 1017

\bibitem[\protect\citeauthoryear{{van der Burg}, {Muzzin}  \& {Hoekstra}}{{van
  der Burg} et~al.}{2016}]{vanderBurg2016}
{van der Burg} R.~F.~J.,  {Muzzin} A.,   {Hoekstra} H.,  2016, \mn@doi [\aap]
  {10.1051/0004-6361/201628222}, 590, A20

\makeatother
\end{thebibliography}

\bsp	
\label{lastpage}


\newpage

\appendix

\section{Density and potential for a Dekel et al. (2017) mean density profile}
\label{appendix:dekel}

The functional form for the average density proposed by \cite{Dekel2017} and presented in Eq.~(\ref{eq:rhob}) yields the following analytical expressions for the mass, velocity and force per unit mass profiles:
\begin{equation}
\label{eq:appendix_M}
M(r) = \frac{4\pi r^3}{3}\overline{\rho}(r) = \mu M_{\rm vir} x^3 \overline{\rho}(r)/\overline{\rho_c},
\end{equation}
\begin{equation}
\label{eq:appendix_V2}
V^2(r) = \frac{GM(r)}{r} = c \mu V_{\rm vir}^2 x^2 \overline{\rho}(r)/\overline{\rho_c}
\end{equation}
and 
\begin{equation}
\label{eq:appendix_F}
F(r) = -\frac{GM(r)}{r^2} = c^2 \mu F_{\rm vir} x \overline{\rho}(r)/\overline{\rho_c}
\end{equation}
where $x=r/r_c$, $c=R_\vir/r_c$, $\mu = c^{a-3} (1+c^{1/b})^{b(g-a)}$, $M_{\rm vir}$ is the virial mass, \mbox{$V_{\rm vir}^2 = G M_{\rm vir}/R_{\rm vir}$} and \mbox{$F_{\rm vir} = - G M_{\rm vir}/R_{\rm vir}^2$}. We recall that $\overline{\rho_c}=c^3 \mu \overline{\rho_\vir}$ with $\overline{\rho_\vir}$ the mean mass density within $R_\vir$.

The density profile is obtained by derivating the expression of the enclosed mass:
\begin{equation}
\label{eq:appendix_rho}
\rho(r) = \frac{1}{4\pi r^2} \frac{dM}{dr} = \frac{3-a}{3}\left( 1+\frac{3-g}{3-a} x^{1/b} \right) \frac{1}{1+x^{1/b}} \overline{\rho}(r).
\end{equation}
This expression is in most cases very different from that of the average density profile (Eq.~(\ref{eq:rhob})), but does yield a similar expression in the case $g=3$, with an outer asymptotic slope equal to $3+1/b$.

The potential energy at radius $r$ is the work done by the gravitational force on a unit mass from $r$ to infinity. To derive the expression of the potential energy per unit mass from the force per unit mass $F$, we assume that it vanishes at infinity and that the halo density profile is truncated at the virial radius, in which case Eq.~(\ref{eq:Ur}) yields
\begin{equation}
\label{eq:appendix_U}
\displaystyle U_{\rm DM}(r)  = -V_\vir^2~\Bigg(1 + c\mu  \int_{x}^{c} \frac{1}{z^{a-1}(1+z^{1/b})^{b(g-a)}}dz\Bigg), 
\end{equation} 
from which Eq.~(\ref{eq:U32}) is derived.

The logarithmic slope of the density profile expressed in Eq.~(\ref{eq:appendix_rho}) is 
\be
\label{eq:appendix_s}
s(r)=-\frac{d\ln \rho}{d \ln r}=\frac{a+(g+b^{-1})x^{1/b}}{1+x^{1/b}}, 
\ee
leading to Eq.~(\ref{eq:s}) when $g=3$ and $b=2$, while the velocity given by Eq.~(\ref{eq:appendix_V2}) peaks at 
\be
x_{\rm max}=\left(\frac{2-a}{g-2}\right)^b.
\ee
This defines a concentration parameter
\be
c_{\rm max}=\frac{R_{\rm vir}}{r_c x_{\rm max}} =c \left(\frac{g-2}{2-a}\right)^b, 
\ee
which coincides with $c$ when $a+g=4$ and leads to Eq.~(\ref{eq:cmax}) when $g=3$ and $b=2$. The logarithmic slope $s$ at the resolution limit ($0.01 R_\vir$ for the NIHAO simulations) and $c_{\rm max}$ can be used as effective inner slope and concentration when describing the density profile.

\section{Parametrizing the local kinetic energy with the Jeans equation}
\label{appendix:kinetic}

\subsection{Spherical symmetry and isotropy}
\label{appendix:symmetry}

\begin{figure}
	\centering
	\includegraphics[width=1\linewidth,trim={0.1cm 0.3cm 0.1cm 0.1cm},clip]{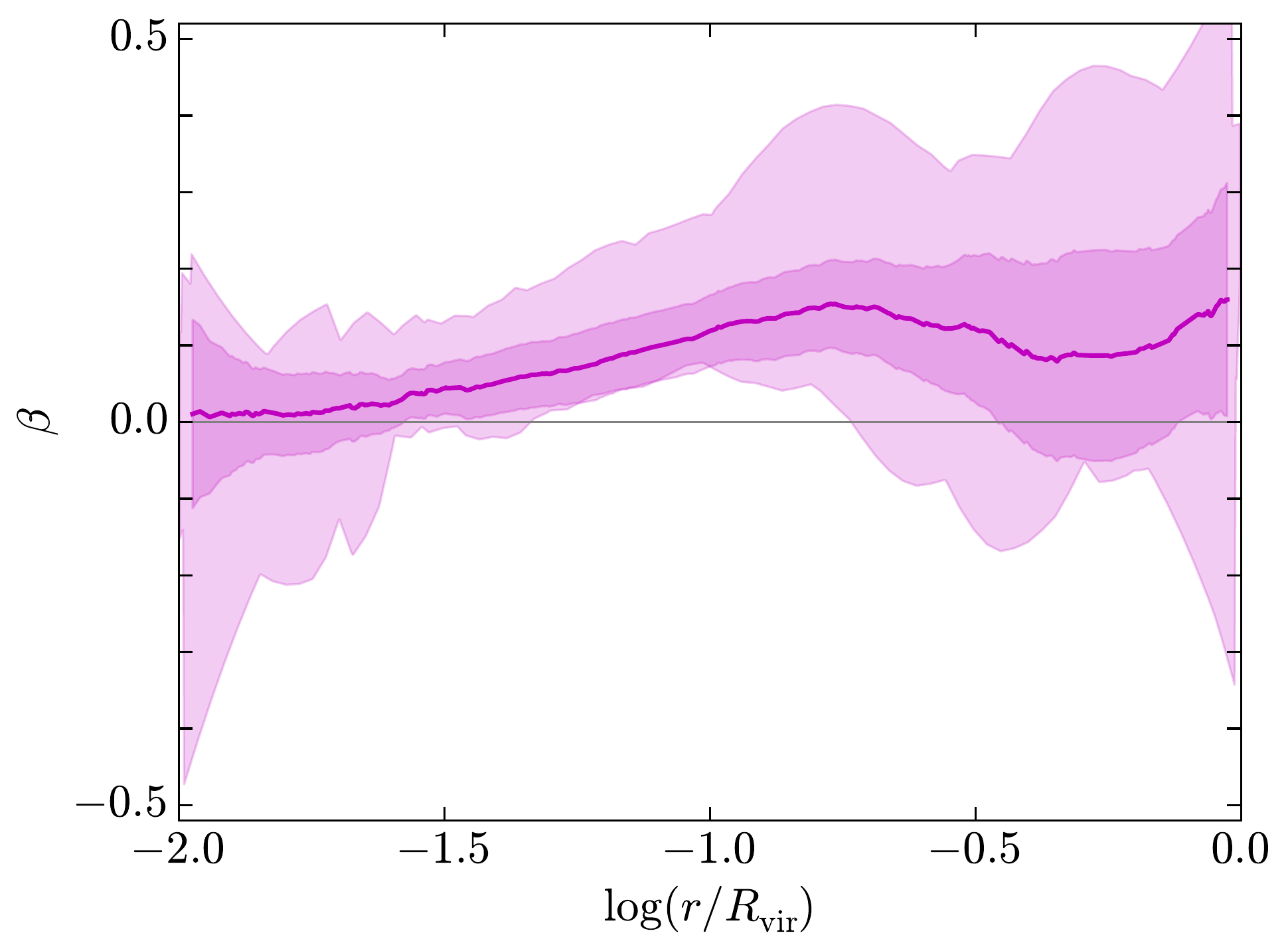}
	\caption{
		Anisotropy parameter $\beta$ in the different outputs of the NIHAO galaxy \texttt{g1.08e11} after 3 Gyr, smoothed with a Savitzky-Golay filter with a polynomial order $n=3$ and a window size $w=21$. 
		The line corresponds to the median curve while the shaded areas highlight the range and the standard deviation from the median.
		The RMS average of $\beta$ is 0.12 so $\beta \ll 1.5$ in most cases and in particular towards the center.
	}
	\label{fig:kinetic_beta}
\end{figure}

\begin{figure*}
	\centering
	\includegraphics[width=0.49\linewidth,trim={0.4cm 2.8cm 0.6cm 1.2cm},clip]{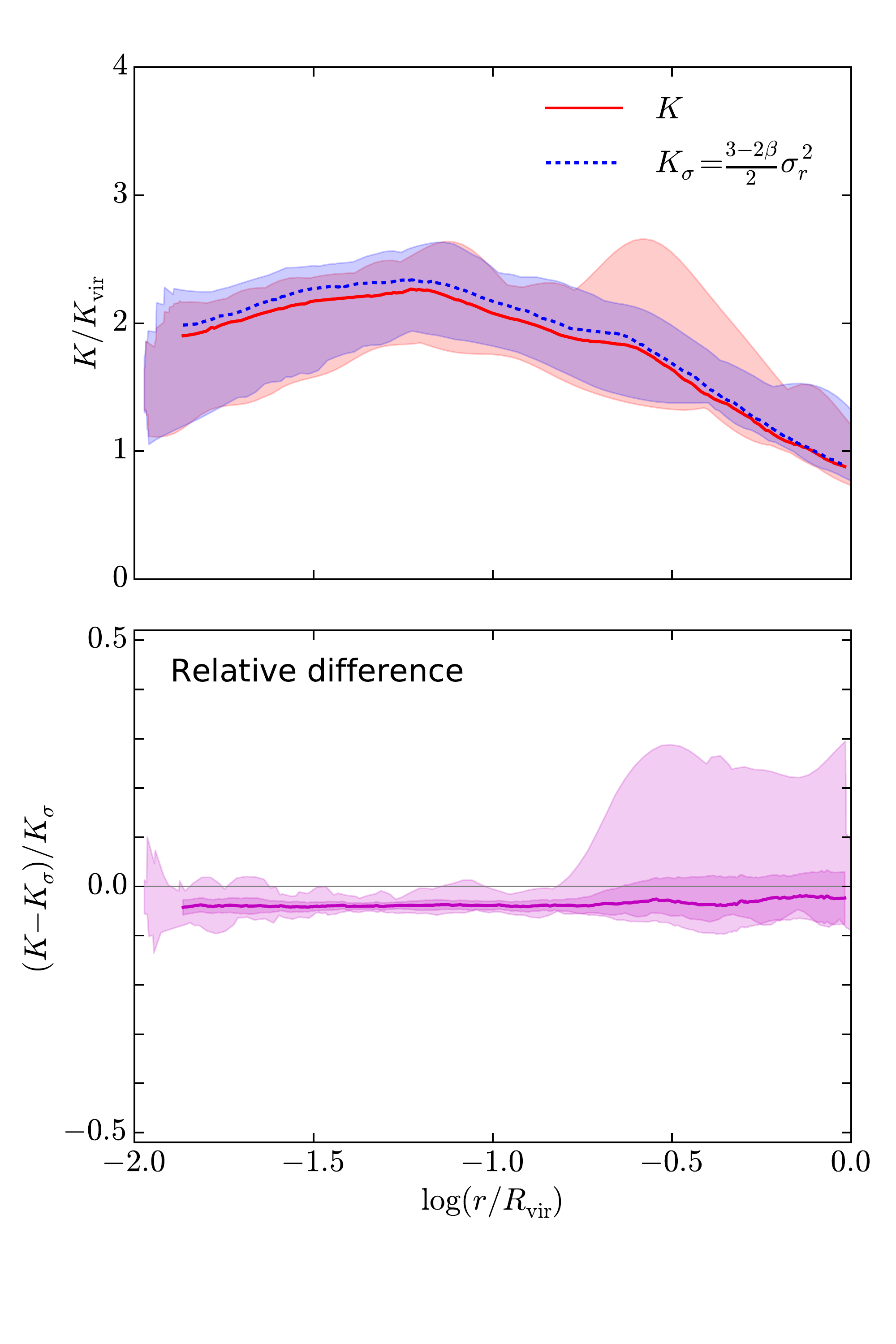}
	\hfill
	\includegraphics[width=0.49\linewidth,trim={0.4cm 2.8cm 0.6cm 1.2cm},clip]{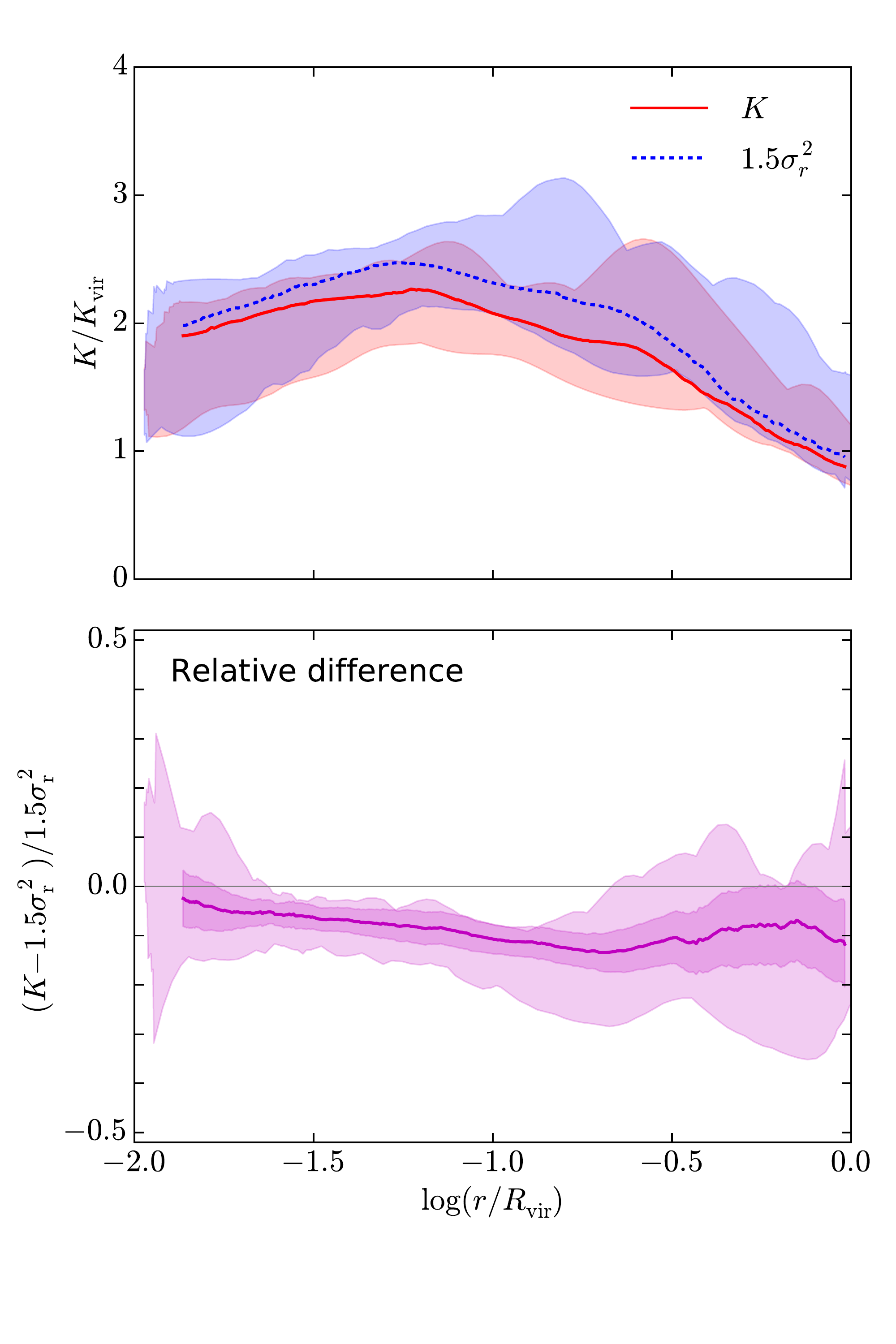}
	\caption{
		The local kinetic energy per unit mass $K$ compared to $K_\sigma=(1.5-\beta) \sigma_{\rm r}^2$ (\textit{Left}) and $1.5 \sigma_r$ (\textit{Right}) in the different outputs of NIHAO galaxy \texttt{g1.08e11} after 3 Gyr, $1.5 \sigma_r$ corresponding to $K_\sigma$ when $\beta=0$. 
		The mean RMS relative difference with $K_\sigma$ is 4.2\% while it is 10.6\% with $1.5 \sigma_r$. 
		Both the radial velocity dispersion $\sigma_r$ and the anisotropy parameter $\beta$ have been smoothed using a Savitzky-Golay filter with a polynomial order $n=3$ and a window size $w=21$. 
		The kinetic energy is in units of $K_{\rm vir} = G M_{\rm vir}/2R_{\rm vir}$. 
	}
	\label{fig:Tsigma}
\end{figure*}

Complete spherical symmetry yields Eq.~(\ref{eq:Ksim}) expressing the local kinetic energy per unit mass as a function of the anisotropy parameter $\beta$ and the radial velocity dispersion $\sigma_r$ while we subsequently assume $\beta=0$ to write $K(r) = 1.5 \sigma_r^2$. 
As shown in Fig.~\ref{fig:kinetic_beta}, $\beta$ is indeed close to zero at most radii, although with a small offset that tend to increase with radius.
In the left panel of Fig.~\ref{fig:Tsigma}, we test Eq.~(\ref{eq:Ksim}) for the outputs of one NIHAO galaxy, computing independently the local kinetic energy per unit mass $K=dK(<r)/dM$ on one side and $K_\sigma=(1.5-\beta)\sigma_{\rm r}^2$ on the other. The relative root mean square (RMS) difference between the two terms is below $5\%$, hence validating the assumption of complete spherical symmetry: we can neglect the global rotation of the halo when parametrizing the kinetic energy. 
In the right panel of Fig.~\ref{fig:Tsigma}, we show that further assuming $\beta=0$ in Eq.~(\ref{eq:Ksim}) still leads to an accurate estimate of the kinetic energy, with a 10\% relative difference.

\subsection{Jeans equilibrium}
\label{appendix:jeans}

\begin{figure*}
	\centering
	\includegraphics[width=0.49\linewidth,trim={0.4cm 2.8cm 0.6cm 1.2cm},clip]{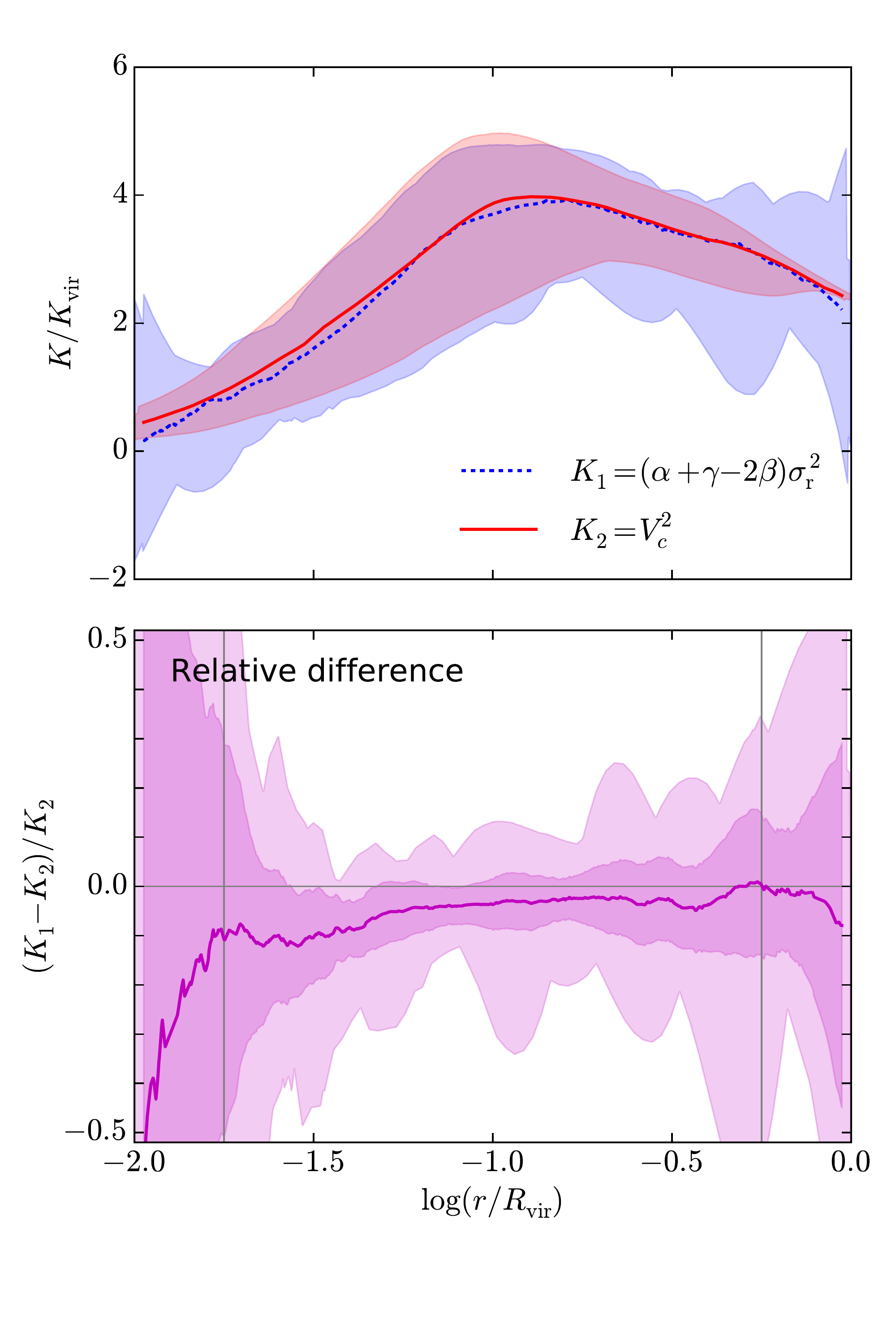}\hfill
	\includegraphics[width=0.49\linewidth,trim={0.4cm 2.8cm 0.6cm 1.2cm},clip]{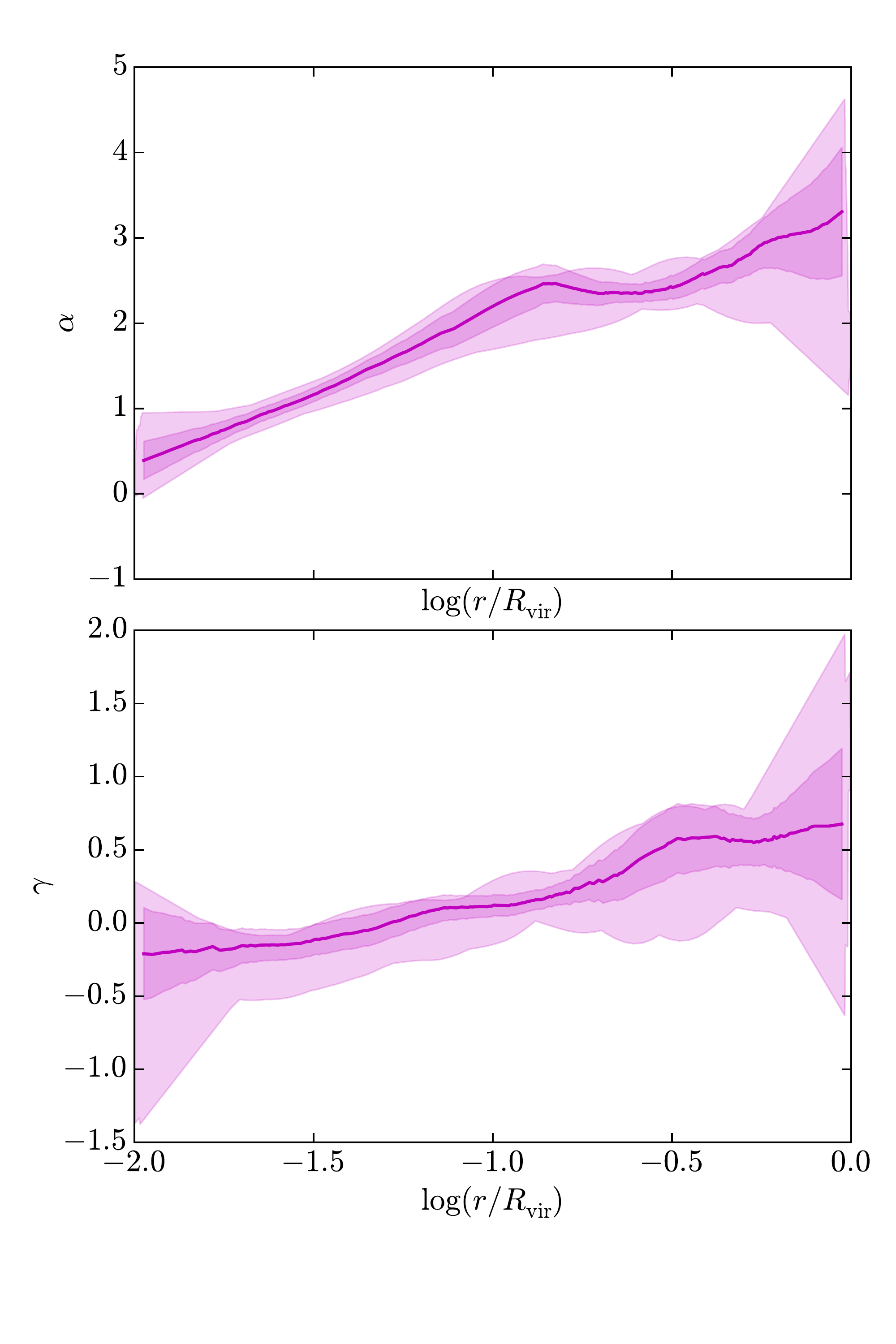}
	\caption{(\textit{Left}) Comparison between the two terms of the Jeans equation (Eq.~(\ref{eq:appendix_Tjeans1})) describing the steady-state equilibrium of a spherical collisionless system in the different outputs of NIHAO galaxy \texttt{g1.08e11} after 3 Gyr. 
	(\textit{Right}) The logarithmic slopes $\alpha$ and $\beta$ of the density and radial velocity dispersion profiles entering the left-hand term of the Jeans equation, as defined from Eq.~(\ref{eq:alpha_gamma}).
	The radial velocity dispersion $\sigma_r$, the anisotropy parameter $\beta$ and the logarithmic slopes $\alpha$ and $\gamma$ were smoothed using a Savitzky-Golay filter with polynomial order $n=3$ and window size $w=21$.
	The kinetic energy in the upper left panel is in units of $K_{\rm vir} = G M_{\rm vir}/2R_{\rm vir}$. The mean RMS relative difference between the two terms of the Jeans equation is 11\%  for $\log(r/R_{\rm vir})$ between -1.75 and -0.25 (highlighted by the vertical lines), which corresponds to the radius range where the fluctuations in $\alpha$, $\beta$ and $\gamma$ are less pronounced. Part of these fluctuations arise at the edges from the Savitzky-Golay filter.
	}
	\label{fig:Tjeans}
\end{figure*}

The steady-state equilibrium of a spherical collisionless system can be described by the 
spherical Jeans equation stemming from the Boltzmann equation \citep[][Eq.~4.215]{BinneyTremaine2008},
\be \label{eq:appendix_Jeans}
\frac{d(\rho \sigma_r^2)}{dr} +\frac{2\beta}{r}\rho\sigma_r^2 = -\rho \frac{d\phi}{dr},
\ee
where $\phi$ is the gravitational potential. Since $\sigma_r^2$ is related to the pressure support, this relation is analogous to hydrostatic equilibrium in hydrodynamics. 
When the anisotropy parameter $\beta$ is a constant, the Jeans equation corresponds to a differential equation in $\rho \sigma_r^2$ whose solution is 
\be
\label{eq:appendix_vrsqr}
\sigma_r^2 (r) = \frac{1}{r^{2\beta}\rho(r)}\int_{r}^{\infty}\rho(\rp)\frac{\rmd\phi}{\rmd\rp}\rp{}^{2\beta}\rmd\rp \nonumber \ee
assuming that it satisfies the boundary condition $\lim_{r\rightarrow +\infty} \sigma_r^2 = 0$. As the gravitational force is given by $ \rmd \phi / \rmd r = GM(r)/r^2$, this yields
\be
\sigma_r^2 (r) =
\frac{G}{r^{2\beta}\rho(r)}\int_{r}^{\infty}\rho(\rp)M(\rp)\rp{}^{2\beta-2}\rmd\rp, 
\ee 
leading to Eq.~(\ref{eq:sigmar2-0}) when $\beta =0$.
Introducing the logarithmic slopes 
\be
\label{eq:alpha_gamma}
\alpha(r)  \equiv -\frac{d \ln\rho}{d \ln r}~~\textrm{and}~~
\gamma(r)  \equiv -\frac{d \ln\sigma_r^2}{d \ln r}, 
\ee
the spherical Jeans equation can be rewritten
\be
\label{eq:appendix_Tjeans1}
\left[\alpha+\gamma -2\beta \right] \sigma_r^2 = V_c^2
\ee
where $V_c^2 \equiv r d\phi/dr = GM/r$.
In Fig.~\ref{fig:Tjeans}, we compare the two terms of Eq.~(\ref{eq:appendix_Tjeans1}) independently of any assumptions on the anisotropy parameter $\beta$. The figure shows that the Jeans equation describing the steady-state equilibrium of a spherical collisionless system is valid within a mean RMS difference of 13\% for $\log(r/R_{\rm vir})$ between $-1.75$ ($0.02 R_{\rm vir}$) and $-0.25$ ($0.56 R_\vir$). 
As $\beta \ll 1.5$ (Fig.~\ref{fig:kinetic_beta}), the sum $\alpha+\gamma-2\beta$ is dominated by the first two terms and hence barely affected by $\beta$.
The relative difference between the two terms of Eq.~(\ref{eq:appendix_Tjeans1}) in a given radius range, and in particular in the range $0.02-0.60~ R_\vir$, can be used to characterize the deviation from Jeans equilibrium, as mentioned in Section~\ref{section:mergers} and Appendix~\ref{appendix:correlations}.

\subsection{Analytical expressions of the local kinetic energy} 
\label{appendix:expressions}

Eq.~(\ref{eq:sigmar2-0}) enables to obtain analytical expressions for the local kinetic energy per unit mass for a spherical, isotropic halo in Jeans equilibrium which is not globally rotating given the mass and density profiles.

\subsubsection{Single-component halo}
\label{appendix:Ksingle}

For a single-component halo whose density profile follows the \cite{Dekel2017} parametrization with $g=3$ and $b=2$ as in Eq.~(\ref{eq:rho32}), the enclosed mass $M(r)$ is given by Eq.~(\ref{eq:appendix_M}), which yields
\begin{align} 
K_{\rm DM}(r)
&= \frac{3-a}{2} c\mu \frac{GM_{\vir}}{R_{\vir}} \frac{\overline{\rho_c}}{\rho(r)}
\int_x^\infty \frac{y^{1-2a}}{(1+y^{1/2})^{13-4a}} \rmd y
\\
&\label{eq:appendix_sigma_dekel_int}
 = c\mu \frac{GM_{\vir}}{R_{\vir}} \frac{(3-a)\overline{\rho_c}}{\rho(r)}
 \int_\chi^1 \zeta^{3-4a} (1-\zeta)^8 \rmd \zeta
\\
K_{\rm DM}(r) 
& 
\label{eq:appendix_sigma_dekel}
= c\mu \frac{GM_{\vir}}{R_{\vir}} \frac{(3-a)\overline{\rho_c}}{\rho(r)}
\Big[ \mathcal{B}(4-4a,9,\zeta)\Big]_\chi^1
\end{align} 
where $\mathcal{B}(a,b,x) = \int_0^x t^{a-1}(1-t)^{b-1} dt$ is the incomplete beta function and the outer brackets denote the difference of the enclosed function between 1 and $\chi$, i.e., $\left[ f(\zeta)\right]_\chi^1 \equiv f(1)-f(\chi)$.
As mentioned in section~\ref{section:kinetic}, we extend the definition of the incomplete beta function appearing inside the outer brackets to negative parameters since the integral of Eq.~(\ref{eq:appendix_sigma_dekel_int}) is well-defined as long as $\chi>0$. 
Following Eqs. 19 and A.9-11 of \cite{Zhao1996}, Eq.~(\ref{eq:sigma_dekel}) (Eq.~(\ref{eq:appendix_sigma_dekel})) can also be expressed \citep[see also][Eq. A.10]{Dekel2017} as the following sum: 
\be \label{eq:appendix_sigma_dekel2} 
K_{\rm DM}(r)
= 3 c\mu \frac{GM_{vir}}{R_{vir}}  x^a (1+x^{1/2})^{2(3.5-a)}
\sum_{i=0}^{8} \frac{(-1)^i 8!}{i!(8-i)!}  \frac{1-\chi^{4(1-a)+i}}{4(1-a)+i}.
\ee
Alternatively, noticing that $\mathcal{B}(a,b,x) = a^{-1} x^a (1-x)^{b-1}+a^{-1}(b-1) \times
\mathcal{B}(a+1,b-1,x)$,
we can deduce that 
\be
\mathcal{B}(a,9,x) = \sum_{i=0}^8 \frac{8!}{i!} \frac{\Gamma(a)}{\Gamma(a+9-i)} x^{a+8-i}(1-x)^i,
\ee
where $\Gamma$ denotes the usual gamma function. 
This enables to write the local kinetic energy as
\begin{multline}
\label{eq:appendix_sigma_dekel3}
K_{\rm DM}(r)
= 3 c\mu \frac{GM_{vir}}{R_{vir}}  x^a (1+x^{1/2})^{2(3.5-a)}
\Bigg[  
\frac{8! \Gamma(4(1-a))}{\Gamma(4(1-a)+9)}
\\
-\sum_{i=0}^{8} \frac{8!}{i!} \frac{\Gamma(4(1-a))}{\Gamma(4(1-a)+9-i)} \chi^{4(1-a)+8-i}(1-\chi)^i
\Bigg].
\end{multline}
The different expressions for $K_{\rm DM}$ are equivalent.

\subsubsection{Power-law multi-component halo}
\label{appendix:Kmulti}

For a multi-component halo in which the ratio between the enclosed total mass $M_{\rm tot}$ and the enclosed collisionless mass $M$ is parametrized as a power-law according to Eq.~(\ref{eq:Mratio}), substituting $M_{\rm tot}$ to $M$ in Eq.~(\ref{eq:sigmar2-0}) yields the local kinetic energy per unit mass
\begin{align} 
K_{\rm multi}(r)
&= \mu X_M c^{n+1} \frac{3-a}{2}  \frac{GM_{\vir}}{R_{\vir}} \frac{\overline{\rho_c}}{\rho(r)}
\int_x^\infty \frac{y^{1-2a-n}}{(1+y^{1/2})^{13-4a}} \rmd y
\\
& = \mu X_M c^{n+1}  \frac{GM_{\vir}}{R_{\vir}} \frac{(3-a)\overline{\rho_c}}{\rho(r)}
\int_\chi^1 \zeta^{3-4a-2n} (1-\zeta)^{8+2n} \rmd \zeta
\\
K_{\rm multi}(r) 
& 
\label{eq:appendix_Kmulti}
= \mu X_M c^{n+1} 
\frac{GM_{\vir}}{R_{\vir}} \frac{(3-a)\overline{\rho_c}}{\rho(r)}
\Big[ \mathcal{B}(4-4a-2n,9+2n,\zeta)\Big]_\chi^1.
\end{align}
As shown in Fig.~\ref{fig:Kmulti} this expression reproduces very well the kinetic energy profiles of the NIHAO simulations.  
For such a multi-component halo, the gravitational potential from Eq.~(\ref{eq:appendix_U}) further becomes
\begin{align}
	U_{\rm multi}(r) &= -X_M V_\vir^2 \left( 1+\mu c^{n+1} \int_x^c \frac{1}{y^{a+n-1} (1+y^{1/2})^{2(3-a)}} dy  \right) \\
	&= -X_M V_\vir^2 \left( 1+2\mu c^{n+1} \int_\chi^{\chi_c} \zeta^{3-2n-2a}(1-\zeta)^{1+2n}d\zeta \right)\\
	U_{\rm multi}(r) &= -X_M V_\vir^2 \left( 1+2\mu c^{n+1} \left[ \mathcal{B}(4-2n-2a,2+2n,\zeta) \right]_\chi^{\chi_c}\right), 
\end{align}
which reverts to Eq.~(\ref{eq:U32}) when $n=0$ and $X_M=1$.

\subsubsection{Additional central point mass}
\label{appendix:Km}

The contribution $\sigma_{m}$ of an additional mass $m$ to the velocity dispersion is expressed in Eq.~(\ref{eq:sigma_m}); its contribution to the local kinetic energy is $K_m=1.5 \sigma_m^2$ for an isotropic velocity field. 
For an additional point mass at the center, this yields
\begin{align}
K_{m} (r) 
& \displaystyle = \frac{3}{2} \frac{G m}{\rho(r)}\int_{r}^{\infty}\rho(\rp) r^{-2}\rmd\rp\\
& \displaystyle = \frac{3-a}{2} \frac{Gmc}{R_{\rm vir}} \frac{\overline{\rho_c}}{{\rho(r)}} 
\int_{x}^{\infty} \frac{1}{y^{2+a}(1+y^{1/2})^{2(3.5-a)}} dy\\
& \displaystyle = \frac{Gmc}{R_{\rm vir}} \frac{(3-a)\overline{\rho_c}}{{\rho(r)}} 
\int_{\chi}^{1} \zeta^{-3-2a} (1-\zeta)^8 d\zeta\\
\label{eq:appendix_Km}
\displaystyle K_{m} (r) 
& \displaystyle = \frac{Gmc}{R_{\rm vir}} \frac{(3-a)\overline{\rho_c}}{{\rho(r)}} 
\Big[ \mathcal{B}(-2-2a,9,\zeta)\Big]_\chi^1. 
\end{align}
This expression is approximately valid for an additional mass distribution dominated by the regions inside radius $r$, as notably assumed in Eq.~(\ref{eq:m(r)}).

\subsubsection{Additional uniform sphere}
\label{appendix:m_uniform}

For an additional (removed) mass $m$ uniformly spread between the halo center and a given radius $r_m$, the additional mass enclosed within a radius $r$ is $m(r/r_m)^3$ for $r<r_m$ and $m$ for $r\geq r_m$. 
If $r\geq r_m$, its contribution to the kinetic energy corresponds to that of a point mass $m$ as in Eq.~(\ref{eq:appendix_Km}).
However, if $r\leq r_m$, Eq.~(\ref{eq:sigma_m}) yields
\be
\label{eq:Km_uniform_leq}
K_m (r) = \frac{3}{2}\frac{G}{\rho(r)} \left[ \int_r^{r_m} \rho(\rp) m\left( \frac{\rp}{r_m}\right)^3 {\rp}^{-2}d\rp +\int_{r_m}^{\infty} \rho(\rp) m {\rp}^{-2} d\rp\right], 
\ee
where the second term corresponds to the contribution of a point mass $m$ evaluated at $r_m$. Hence if $r\leq r_m$, 
\begin{align}
K_{m}(r)
& = \frac{Gmc}{R_\vir} \frac{(3\!-\!a)\overline{\rho_c}}{\rho(r)} 
\left(
\frac{1}{2 x_m^3} \!\int_x^{x_m} \!\!\!\!\frac{y dy}{y^a(1+y^{1/2})^{2(3.5-a)}}
+\Big[\mathcal{B}(-\!2\!-\!2a,9,\zeta)\Big]_{\chi_m}^1
\right)\\
& =\frac{Gmc}{R_\vir} \frac{(3-a)\overline{\rho_c}}{\rho(r)} 
\left(
\frac{1}{x_m^3}  \int_{\chi}^{\chi_m} \zeta^{3-2a} (1-\zeta)^2 d\zeta
+\Big[ \mathcal{B}(-2-2a,9,\zeta)\Big]_{\chi_m}^1
\right)\\
K_m(r)
& = \frac{Gmc}{R_\vir} \frac{(3-a)\overline{\rho_c}}{\rho(r)}  
\left(
\frac{1}{x_m^3} \Big[ \mathcal{B}(4-2a,3,\zeta)\Big]_\chi^{\chi_m}
+ \Big[ \mathcal{B}(-2-2a,9,\zeta)\Big]_{\chi_m}^1
\right), 
\end{align}
where $x_m=r_m/r_c$ and $\chi_m=\chi(x_m)$.

\subsubsection{Additional singular isothermal sphere}
\label{appendix:m_SIS}

The mass distribution of a singular isothermal sphere of total mass $m$ and radius $r_m$ is $m(r/r_m)$ for $r\leq r_m$ and $m$ for $r\geq r_m$. 
If $r\geq r_m$, its contribution to the kinetic energy corresponds to that of a point mass $m$ as in Eq.~(\ref{eq:appendix_Km}). 
If $r\leq r_m$, Eq.~(\ref{eq:sigma_m}) yields 
\begin{align}
K_m(r) & = \frac{Gmc}{R_{\rm vir}} \frac{(3-a)\overline{\rho_c}}{{\rho(r)}} 
\Bigg(
\frac{1}{2 x_m} \int_x^{x_m} \frac{y^{1-a}dy}{(1+y^{1/2})^{7-2a}}
+\Big[ \mathcal{B}(-2-2a,9,\zeta)\Big]_{\chi_m}^1
\Bigg)\\
& = \!\frac{Gmc}{R_{\rm vir}} \frac{(3-a)\overline{\rho_c}}{{\rho(r)}} 
\Bigg(
\frac{1}{x_m} \int_\chi^{\chi_m} \zeta^{-1-2a}(1-\zeta)^6 d\zeta 
+\Big[ \mathcal{B}(-2-2a,9,\zeta)\Big]_{\chi_m}^1
\Bigg)\\
K_m(r) &= \frac{Gmc}{R_{\rm vir}} \frac{(3-a)\overline{\rho_c}}{{\rho(r)}} 
\Bigg(
\frac{1}{x_m} \Big[ \mathcal{B}(-2a,7,\zeta)\Big]_{\chi}^{\chi_m}
+\Big[ \mathcal{B}(-2-2a,9,\zeta)\Big]_{\chi_m}^1
\Bigg). 
\end{align}

\subsubsection{Additional Dekel at al. (2017) sphere}
\label{appendix:m_Dekel}

If the additional mass is rather spread according to a \cite{Dekel2017} density profile $\rho_m(r)$ as in Eq.~(\ref{eq:rho32}) with a total mass $m$ within $R_\vir$, the additional enclosed mass within $r$ is $ \mu m x^3 \overline{\rho(r)}/\overline{\rho_c}$ and the calculation of $K_m$
follow that of $K_{\rm DM}$, replacing $M_{\vir}$ by $m$ so that 
\be
\displaystyle K_{m} (r) 
=  \frac{G\mu mc}{R_{\vir}} \frac{(3\!-\!a)\overline{\rho_c}}{{\rho(r)}} 
\Big[ \mathcal{B}(4-4a,9,\zeta)\Big]_\chi^1.
\ee

\section{Additional figures and table}
\label{appendix:correlations}

\begin{figure*}
	\centering
	\includegraphics[width=1\linewidth,trim={1.cm 39.8cm 9.5cm 18.3cm},clip]{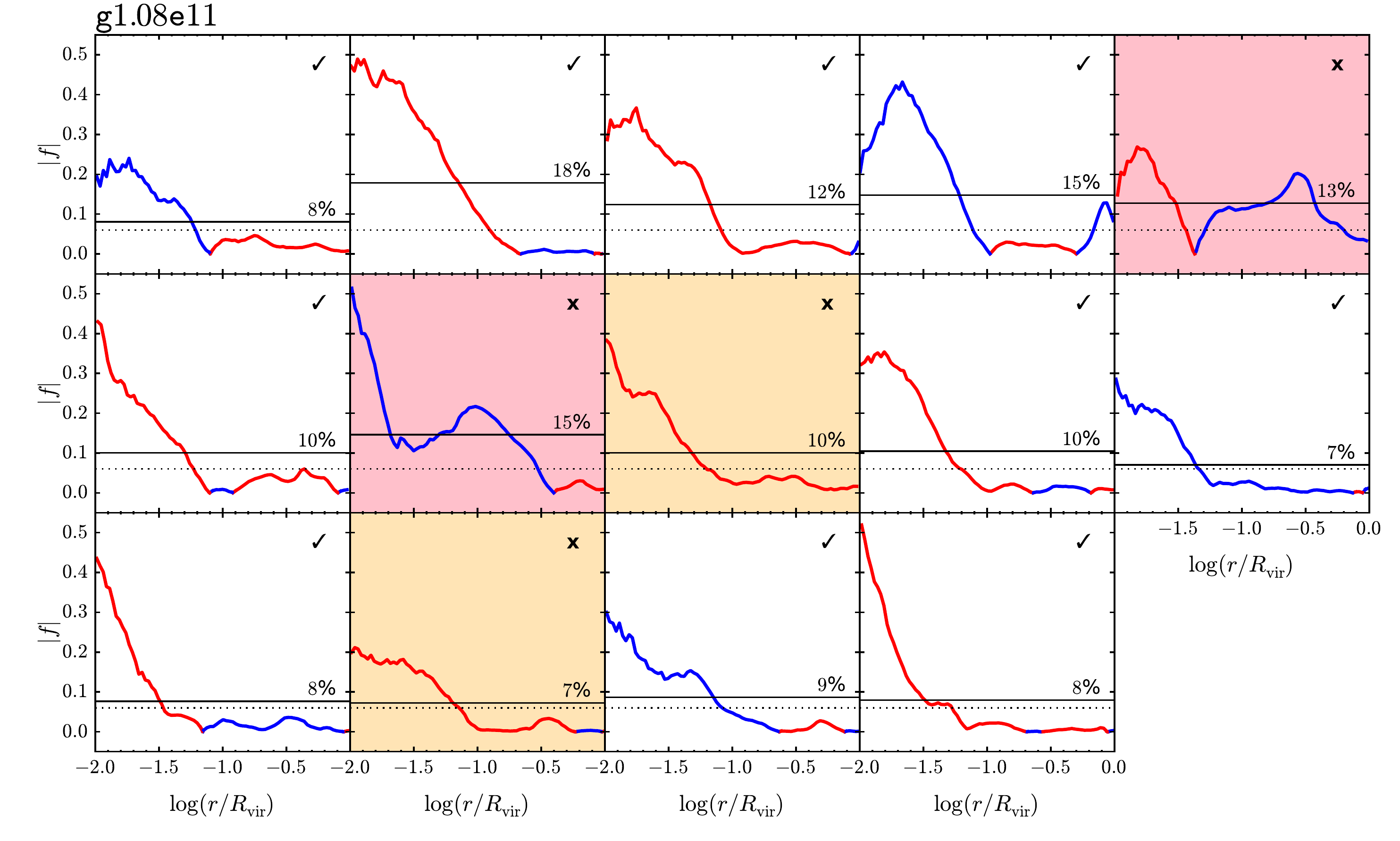}\\
	\includegraphics[width=1\linewidth,trim={1.cm 39.8cm 9.5cm 18.3cm},clip]{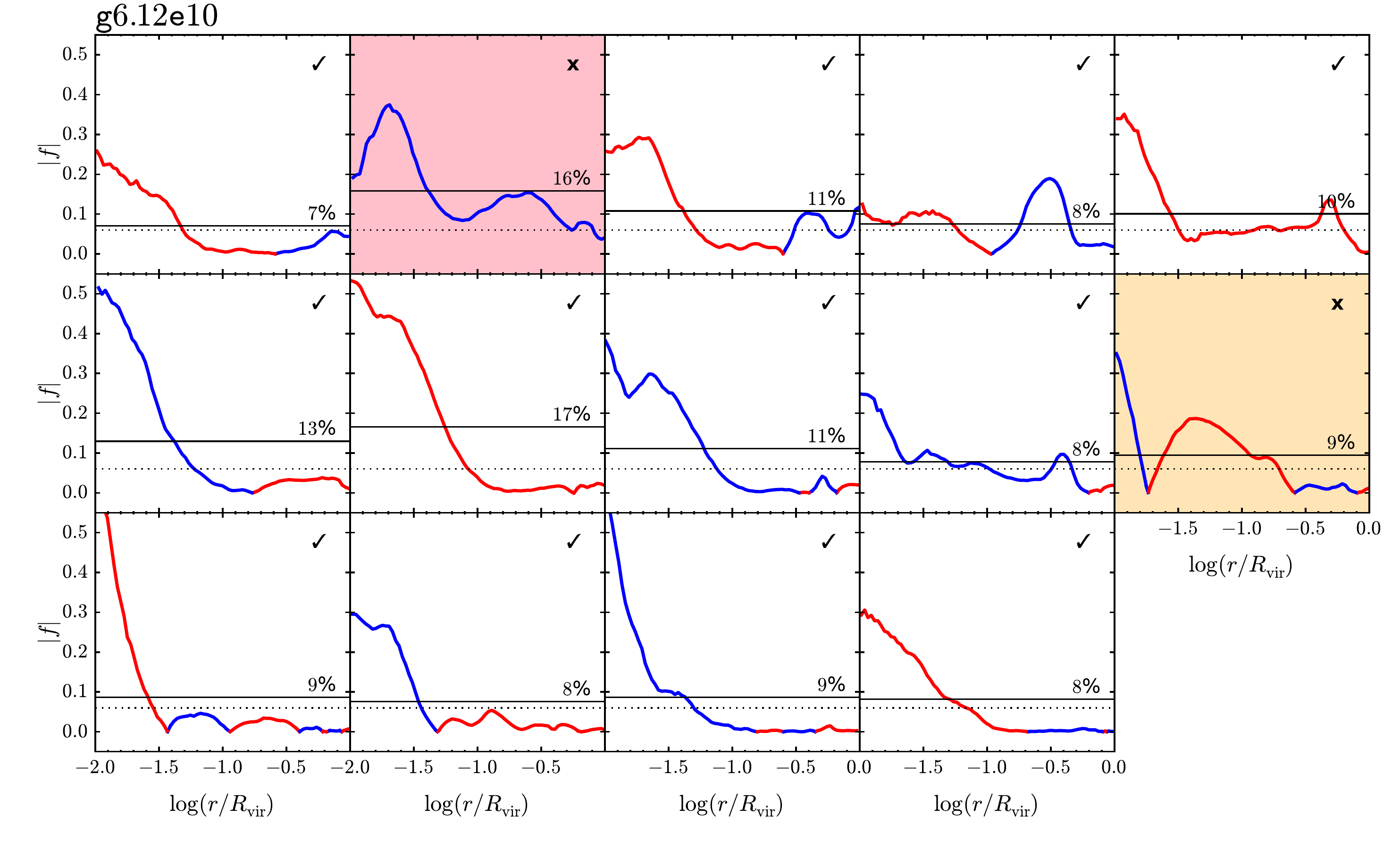}
	\vspace{-0.8cm}
	\caption{
		Absolute value of the mass ratio $f=m/M$ entering the model as a function of radius for the successive ouputs plotted in Fig.~\ref{fig:example_density}. 
		Red portions of the curve correspond to outflows ($m<0$), blue portions to inflows ($m>0$). Successes and failures are respectively indicated by a check mark or a cross at the upper right corner while the background color further indicates mergers (red) and highlights whether the model is successful (white) or not (red and orange) as in Fig.~\ref{fig:example_density}. The plain horizontal line indicates $|f|_{\rm RMS}$, whose value is quoted next to it, while the dotted line indicates the $|f|_{\rm RMS}$ threshold equal to $7\%$.  
		These $f$ profiles are used as inputs for the model, giving the mass changes at each radius. Mergers, which are characterized by $f_{\rm merger}\equiv|f|_{r=0.15 R_\vir}>10\%$, are systematically associated to failures of the model for these two simulated galaxies.
	}
	\label{fig:example_fprofile}
\end{figure*}

\begin{table*}
	\centering
	\caption{Success rate of the instant mass change model for the 33 NIHAO simulated galaxies whose stellar mass lies between $5\times 10^{7}$ and $5\times 10^{9}~ M_\odot$ at $z=0$, excluding the earliest phase of galaxy evolution before 3 Gyr: $p_{\rm all}$ for all such outputs; $p_{\rm no mergers}$ excluding mergers as indicated in section~\ref{subsection:implementation}; $p_{>f_{\rm min}}$ and $p_{<f_{\rm min}}$ further focusing either on cases where we expect a significant profile change ($|f|_{\rm RMS}>f_{\rm min}$) or on cases where we expect negligible changes ($|f|_{\rm RMS}<f_{\rm min}$). The simulation ID follows the nomenclature used by \protect\cite{Wang2015} but the indicated virial and stellar masses $M_\vir$ and $M_\star$ as well as the virial radius $R_\vir$ at $z=0$ slightly differ from their values since we determined the virial radius $R_\vir$ following \protect\cite{Bryan1998} instead of using $R_{200}$ and calculated $M_\star$ within $0.15\times R_\vir$ instead of $0.2 \times R_\vir$. The stellar half-light radius $r_e$ and the V-band central surface brightness $\mu_{0,V}$ at $z=0$ are also indicated. Daggers ($\dagger$) highlight UDGs. Galaxies are ordered by halo mass; fiducial galaxies \texttt{g6.12e10} and \texttt{g1.08e11} are highlighted in bold.
	}
	\label{tab:sims}
	\begin{tabular}{@{}llllllllll} 
	\hline
	\noalign{\vskip 1mm} ID & $M_\vir$ & $R_\vir$ & $M_\star$ & $r_e$ & $\mu_{0,V}$  & $p_{\rm all}$ & $p_{\rm no~mergers}$ & $p_{>f_{\rm min}}$ & $p_{<f_{\rm min}}$\\
	&$\rm [M_\odot]$&$\rm [kpc]$&$\rm [M_\odot]$&$\rm [kpc]$& $\rm [mag/as^{2}]$&&&&\\
	\hline
	\noalign{\vskip 1mm}

\texttt{g2.94e10}$^\dagger$& $3.80 \times 10^{10}$& $88.8$& $5.86 \times 10^{7}$& $1.96$ & $24.26$& 22/27 = 81\percent& 22/27 = 81\percent& 06/09 = 67\percent& 16/18 = 89\percent \\

\texttt{g5.05e10}$^\dagger$& $4.74 \times 10^{10}$& $95.6$& $9.47 \times 10^{7}$& $2.09$ & $24.28$& 27/38 = 71\percent& 26/33 = 79\percent& 09/12 = 75\percent& 17/21 = 81\percent \\

\textbf{\texttt{g6.12e10}}$^\dagger$& $\mathbf{5.50 \times 10^{10}}$& $\mathbf{100.5}$& $\mathbf{9.13 \times 10^{7}}$& $\mathbf{2.67}$ & $\mathbf{24.03}$& \textbf{27/33 = 82\percent}& \textbf{27/30 = 90\percent}& \textbf{10/11 = 91\percent}& \textbf{17/19 = 89\percent} \\
 
\texttt{g4.27e10}$^\dagger$& $5.54 \times 10^{10}$& $100.7$& $6.15 \times 10^{7}$& $2.83$ & $24.21$& 15/20 = 75\percent& 15/19 = 79\percent& 09/11 = 82\percent& 06/08 = 75\percent \\

\texttt{g4.99e10}$^\dagger$& $5.71 \times 10^{10}$& $101.8$& $1.22 \times 10^{8}$& $3.01$ & $24.16$& 31/43 = 72\percent& 31/42 = 74\percent& 13/23 = 57\percent& 18/19 = 95\percent \\

\texttt{g4.86e10}$^\dagger$& $5.79 \times 10^{10}$& $102.2$& $1.22 \times 10^{8}$& $2.36$ & $24.37$& 45/50 = 90\percent& 45/50 = 90\percent& 00/02 = 0\percent& 45/48 = 94\percent \\

\texttt{g4.94e10}& $5.93 \times 10^{10}$& $103.0$& $1.11 \times 10^{8}$& $2.42$ & $23.89$& 25/38 = 66\percent& 25/33 = 76\percent& 04/10 = 40\percent& 21/23 = 91\percent \\

\texttt{g3.44e10}$^\dagger$& $6.65 \times 10^{10}$& $107.0$& $6.32 \times 10^{7}$& $2.54$ & $24.43$& 21/33 = 64\percent& 19/26 = 73\percent& 13/15 = 87\percent& 06/11 = 55\percent \\

\texttt{g4.48e10}& $6.77 \times 10^{10}$& $107.7$& $1.37 \times 10^{8}$& $3.62$ & $23.92$& 16/31 = 52\percent& 16/24 = 67\percent& 06/11 = 55\percent& 10/13 = 77\percent \\

\texttt{g6.91e10}& $7.82 \times 10^{10}$& $113.0$& $2.50 \times 10^{8}$& $2.54$ & $23.49$& 39/50 = 78\percent& 39/49 = 80\percent& 07/12 = 58\percent& 32/37 = 86\percent \\

\texttt{g6.96e10}& $1.08 \times 10^{11}$& $125.9$& $3.64 \times 10^{8}$& $3.49$ & $23.34$& 34/47 = 72\percent& 34/42 = 81\percent& 14/17 = 82\percent& 20/25 = 80\percent \\

\texttt{g6.77e10}& $1.09 \times 10^{11}$& $126.1$& $4.83 \times 10^{8}$& $4.37$ & $23.22$& 24/42 = 57\percent& 21/35 = 60\percent& 19/27 = 70\percent& 02/08 = 25\percent \\

\texttt{g8.89e10}& $1.10 \times 10^{11}$& $126.6$& $4.02 \times 10^{8}$& $3.12$ & $23.10$& 42/50 = 84\percent& 42/48 = 88\percent& 07/10 = 70\percent& 35/38 = 92\percent \\

\texttt{g9.59e10}$^\dagger$& $1.13 \times 10^{11}$& $127.9$& $2.75 \times 10^{8}$& $4.91$ & $24.48$& 15/50 = 30\percent& 13/39 = 33\percent& 13/33 = 39\percent& 00/06 = 0\percent \\

\texttt{g3.23e11}$^\dagger$& $1.13 \times 10^{11}$& $127.7$& $3.60 \times 10^{8}$& $4.85$ & $24.05$& 25/49 = 51\percent& 24/43 = 56\percent& 22/36 = 61\percent& 02/07 = 29\percent \\

\texttt{g1.05e11}& $1.32 \times 10^{11}$& $134.6$& $5.66 \times 10^{8}$& $4.99$ & $23.39$& 25/44 = 57\percent& 24/38 = 63\percent& 15/26 = 58\percent& 09/12 = 75\percent \\

\texttt{g6.37e10}$^\dagger$& $1.35 \times 10^{11}$& $135.7$& $2.11 \times 10^{8}$& $4.19$ & $24.10$& 17/41 = 41\percent& 16/30 = 53\percent& 10/16 = 62\percent& 06/14 = 43\percent \\

\textbf{\texttt{g1.08e11}}$^\dagger$& $\mathbf{1.36 \times 10^{11}}$& $\mathbf{135.9}$& $\mathbf{8.47 \times 10^{8}}$& $\mathbf{4.41}$ & $\mathbf{24.10}$& \textbf{41/50 = 82\percent}& \textbf{41/47 = 87\percent}& \textbf{10/12 = 83\percent}& \textbf{31/35 = 89\percent} \\

\texttt{g2.19e11}& $1.52 \times 10^{11}$& $140.9$& $9.27 \times 10^{8}$& $4.52$ & $23.01$& 27/47 = 57\percent& 27/41 = 66\percent& 20/31 = 65\percent& 07/10 = 70\percent \\

\texttt{g1.37e11}& $1.69 \times 10^{11}$& $146.2$& $2.02 \times 10^{9}$& $3.46$ & $22.51$& 44/50 = 88\percent& 44/50 = 88\percent& 03/05 = 60\percent& 41/45 = 91\percent \\

\texttt{g1.57e11}& $1.78 \times 10^{11}$& $148.6$& $1.15 \times 10^{9}$& $5.42$ & $23.62$& 29/50 = 58\percent& 24/36 = 67\percent& 18/23 = 78\percent& 06/13 = 46\percent \\

\texttt{g1.52e11}$^\dagger$& $1.81 \times 10^{11}$& $149.5$& $7.90 \times 10^{8}$& $5.81$ & $24.29$& 32/50 = 64\percent& 32/50 = 64\percent& 02/07 = 29\percent& 30/43 = 70\percent \\

\texttt{g1.59e11}$^\dagger$& $2.00 \times 10^{11}$& $154.6$& $6.69 \times 10^{8}$& $6.21$ & $24.97$& 39/50 = 78\percent& 39/48 = 81\percent& 16/19 = 84\percent& 23/29 = 79\percent \\

\texttt{g2.04e11}& $2.43 \times 10^{11}$& $164.9$& $4.70 \times 10^{9}$& $3.07$ & $20.18$& 43/50 = 86\percent& 43/46 = 93\percent& 18/21 = 86\percent& 25/25 = 100\percent \\

\texttt{g1.64e11}& $2.45 \times 10^{11}$& $165.4$& $9.12 \times 10^{8}$& $5.72$ & $22.41$& 10/33 = 30\percent& 08/21 = 38\percent& 07/16 = 44\percent& 01/05 = 20\percent \\

\texttt{g2.41e11}& $2.85 \times 10^{11}$& $173.8$& $4.10 \times 10^{9}$& $3.97$ & $21.48$& 36/49 = 73\percent& 34/43 = 79\percent& 12/15 = 80\percent& 22/28 = 79\percent \\

\texttt{g2.54e11}& $3.00 \times 10^{11}$& $177.0$& $3.50 \times 10^{9}$& $3.60$ & $19.84$& 28/46 = 61\percent& 27/38 = 71\percent& 17/25 = 68\percent& 10/13 = 77\percent \\

\texttt{g4.90e11}& $3.85 \times 10^{11}$& $192.3$& $3.43 \times 10^{9}$& $5.77$ & $22.95$& 37/50 = 74\percent& 34/43 = 79\percent& 07/08 = 88\percent& 27/35 = 77\percent \\

\texttt{g5.46e11}& $3.87 \times 10^{11}$& $192.5$& $3.77 \times 10^{9}$& $5.29$ & $22.47$& 37/48 = 77\percent& 36/43 = 84\percent& 07/09 = 78\percent& 29/34 = 85\percent \\

\texttt{g3.21e11}& $3.95 \times 10^{11}$& $193.9$& $3.67 \times 10^{9}$& $4.88$ & $21.29$& 25/48 = 52\percent& 25/39 = 64\percent& 18/29 = 62\percent& 07/10 = 70\percent \\

\texttt{g3.59e11}& $4.44 \times 10^{11}$& $201.6$& $4.36 \times 10^{9}$& $5.10$ & $21.76$& 26/46 = 57\percent& 24/33 = 73\percent& 14/19 = 74\percent& 10/14 = 71\percent \\

\texttt{g3.49e11}& $5.04 \times 10^{11}$& $210.4$& $3.96 \times 10^{9}$& $5.23$ & $22.02$& 39/50 = 78\percent& 39/49 = 80\percent& 07/13 = 54\percent& 32/36 = 89\percent \\

\texttt{g3.55e11}& $5.04 \times 10^{11}$& $210.3$& $3.85 \times 10^{9}$& $6.48$ & $22.44$& 19/43 = 44\percent& 16/33 = 48\percent& 11/24 = 46\percent& 05/09 = 56\percent \\

\hline
\noalign{\vskip 1mm} 
\textbf{All}& & & & & & \textbf{962/1446 = 67\percent}& \textbf{932/1268 = 74\percent}& \textbf{364/557 = 65\percent}& \textbf{568/711 = 80\percent} \\
\hline
\end{tabular}
\end{table*}

\begin{figure*}
	\centering
	\includegraphics[width=1\linewidth,trim={4cm 5.cm 5.cm 5.5cm},clip]{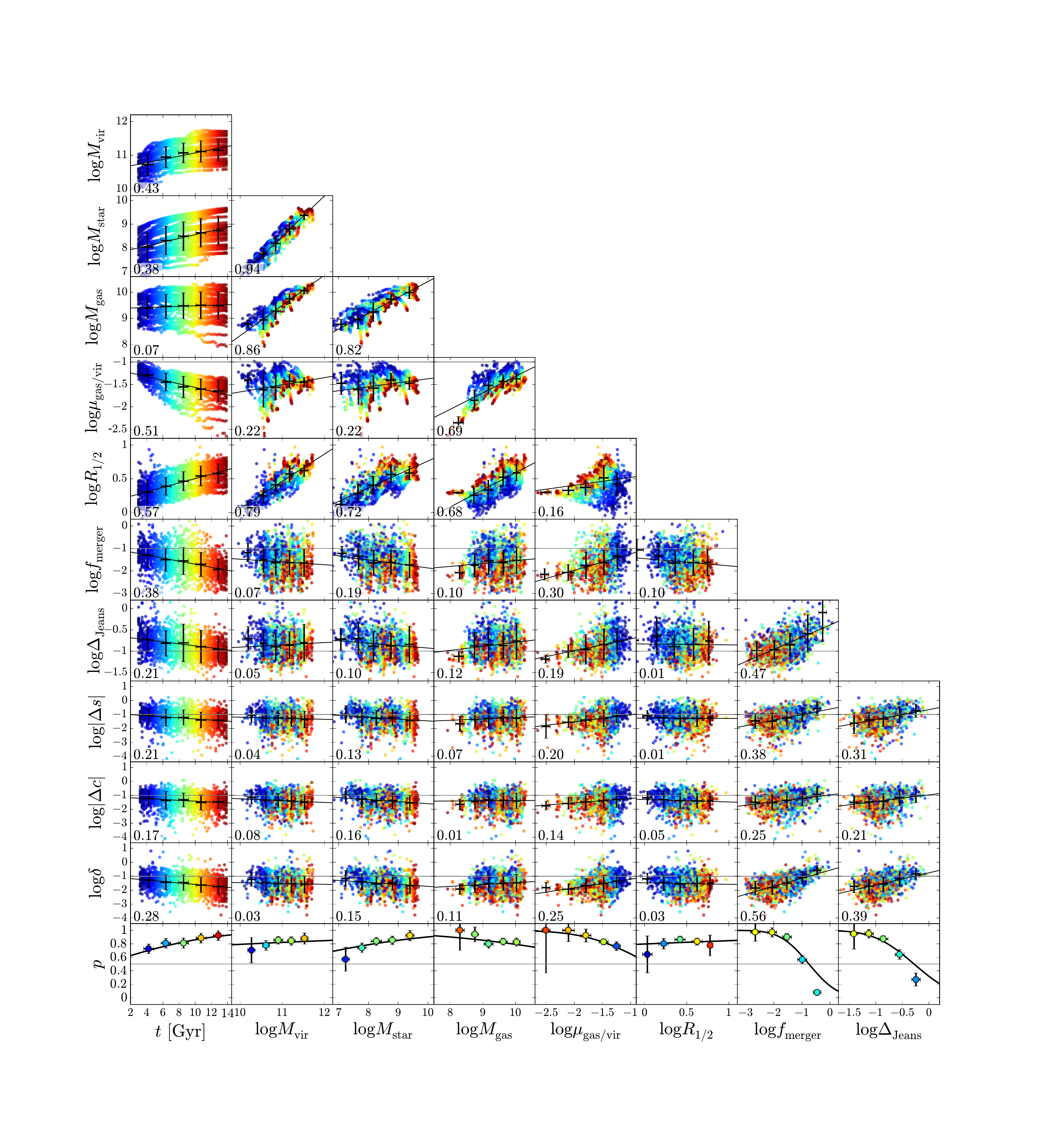}
	\vspace{-0.4cm }
	\caption{
		Correlations between different quantities relevent to the simulated NIHAO galaxies whose stellar mass lies between $5\times 10^{7}$ and $5\times 10^{9}~ M_\odot$ at $z=0$ and their dark matter haloes, together with the dependence of $\Delta s$, $\Delta c$, and $\delta$ measuring the agreement between model prediction and simulation output as well as $f_{\rm merger}$ the merger indicator, $\Delta_{\rm Jeans}$ measuring the deviation from Jeans equilibrium, and $p$ the probability associated to $\delta\leq\delta_{\rm max}$ on these quantities. The first 3 Gyr of evolution are excluded. In each panel, Poisson error bars correspond to equally-spaced bins along the x-axis and the solid line to a linear least-square fit of the form $\texttt{y}=\texttt{slope} \times \texttt{x} + \texttt{intercept}$ to the data. The absolute value of the Pearson correlation coefficient is indicated at the bottom left of each panel. Color corresponds to time as in the first column. In the bottom panels, the probability $p$ is defined in each bin as the ratio of the number of cases with $\delta\leq\delta_{\rm max}$ on the total number of cases. The lines in this panel correspond to the probability associated with the linear fit on $\log \delta$ assuming a Gaussian distribution at each x-axis value: $\texttt{y}=0.5 \times [1+\texttt{erf}((\log\delta_{\rm max}-\texttt{slope} \times \texttt{x} - \texttt{intercept})/\sqrt{2\texttt{sigma}^2})]$, where $\texttt{sigma}$ is the standard deviation of $\log\delta$ from its linear fit. 
	}
	\label{fig:correlations_all}
\end{figure*}

\begin{figure*}
	\centering
	\includegraphics[height=0.28\linewidth,trim={1.2cm 0 4.5cm 0.6cm},clip]{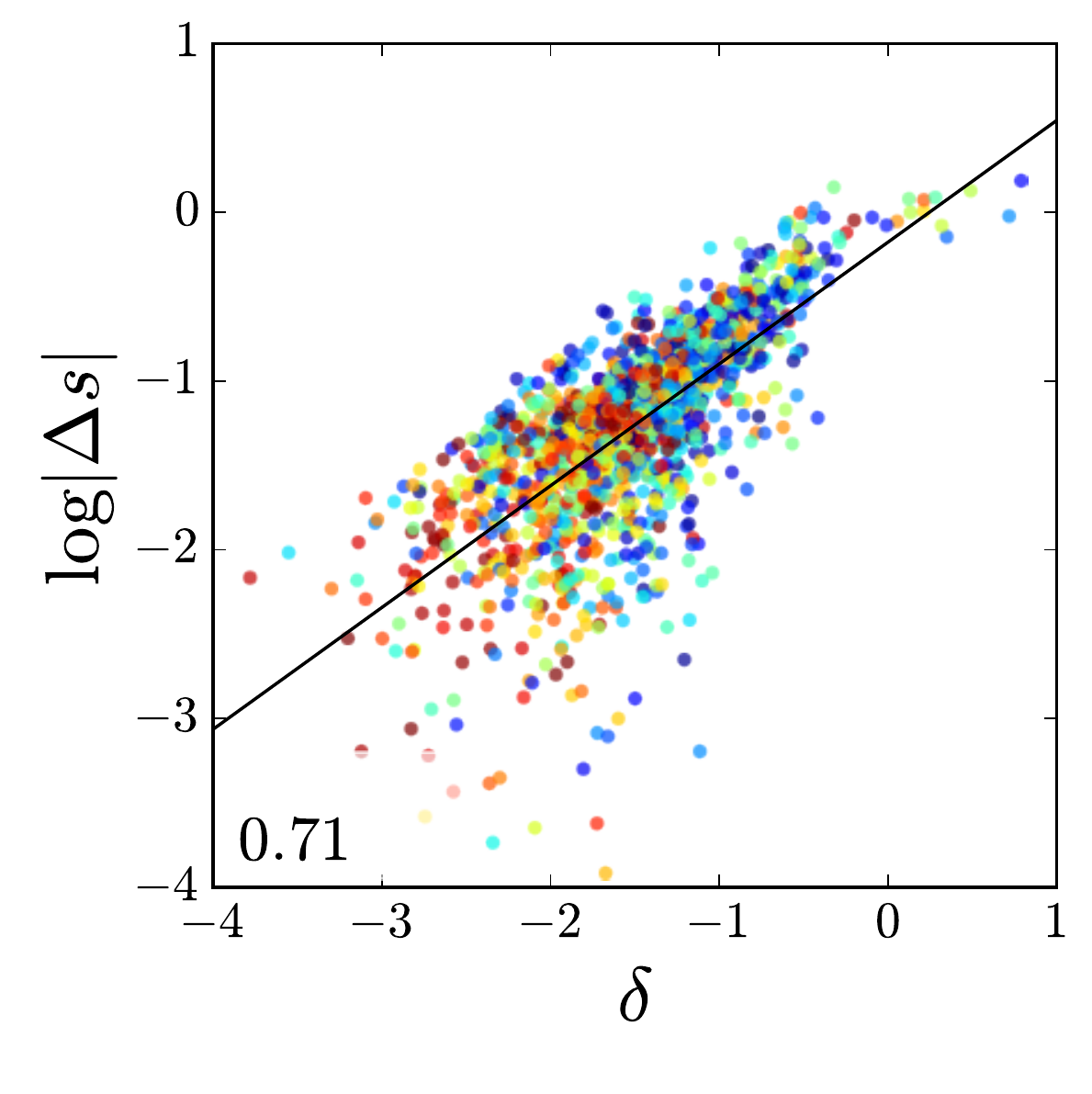}\hfill
	\includegraphics[height=0.28\linewidth,trim={1.2cm 0 4.5cm 0.6cm},clip]{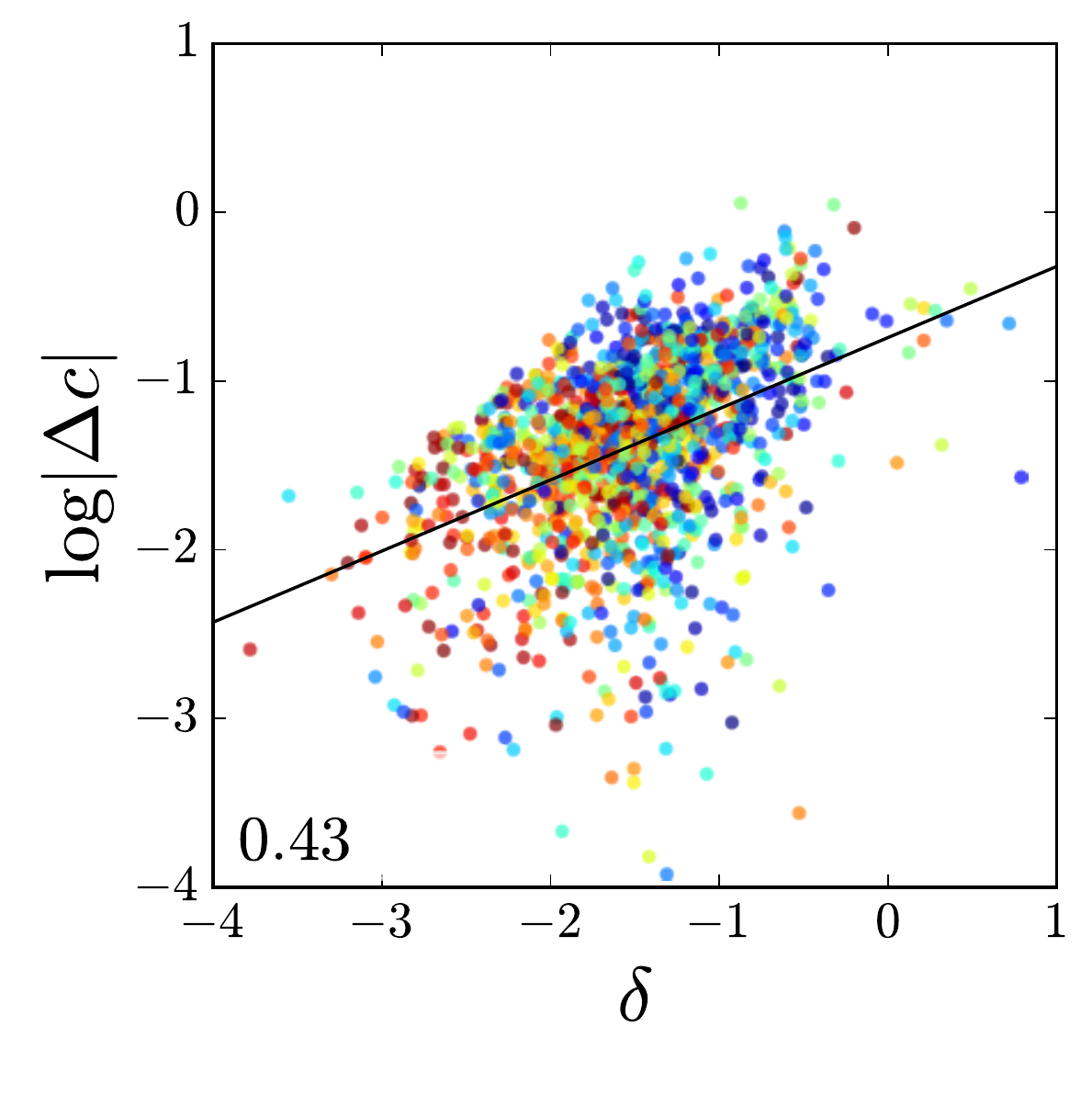}\hfill
	\includegraphics[height=0.28\linewidth,trim={0.8cm 0 0.2cm 0.6cm},clip]{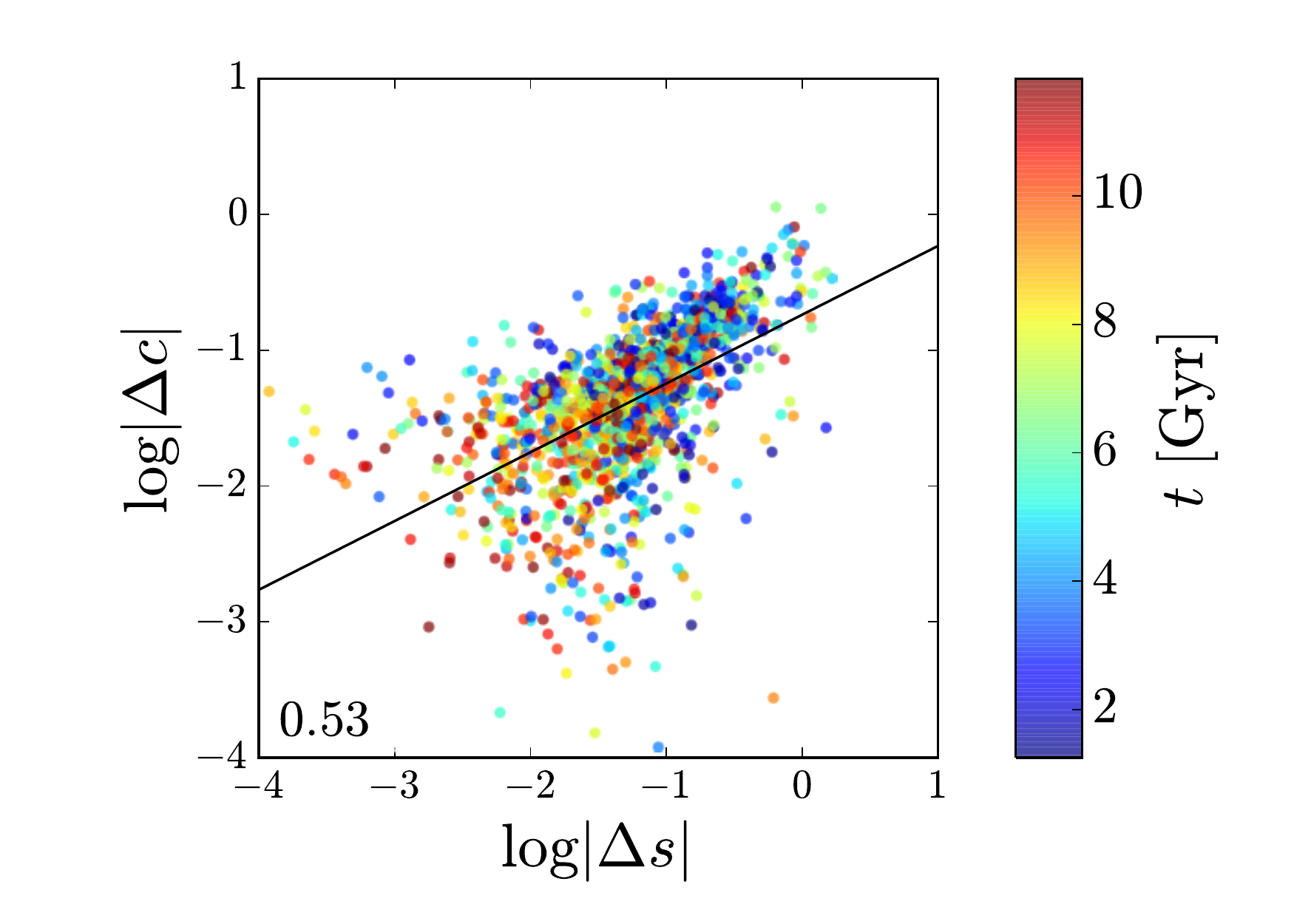}
	\vspace{-0.4cm }
	\caption{
		Correlations between the three success measures $\delta$, $\Delta s$ and $\Delta c$, colored as in Fig.~\ref{fig:correlations_all} as a function of time. The solid line shows a least-square linear fit while the Pearson correlation coefficient is quoted at the bottom left. 
	}
	\label{fig:correlations_success}
\end{figure*}

To complement Section \ref{subsection:results}, Fig.~\ref{fig:example_fprofile} shows the radial profiles of the mass ratio $f=m/M$ between the successive outputs shown in Fig.~\ref{fig:example_density}. Together with the initial density profiles, these profiles are used to predict the evolution of the mass distribution according to the model. 
Table~\ref{tab:sims} indicates the success rate of the model galaxy per galaxy and for the whole sample, differenciating the overall success rate ($p_{\rm all}$), that excluding mergers ($p_{\rm no ~mergers}$), that excluding mergers and with $|f|_{\rm RMS}\geq f_{\rm min}$ ($p_{>f_{\rm min}}$), as well as that excluding mergers but with $|f|_{\rm RMS}< f_{\rm min}$ ($p_{<f_{\rm min}}$). 

\begin{figure*}
	\centering
	\includegraphics[height=0.28\linewidth,trim={1.2cm 0 4.5cm 0.6cm},clip]{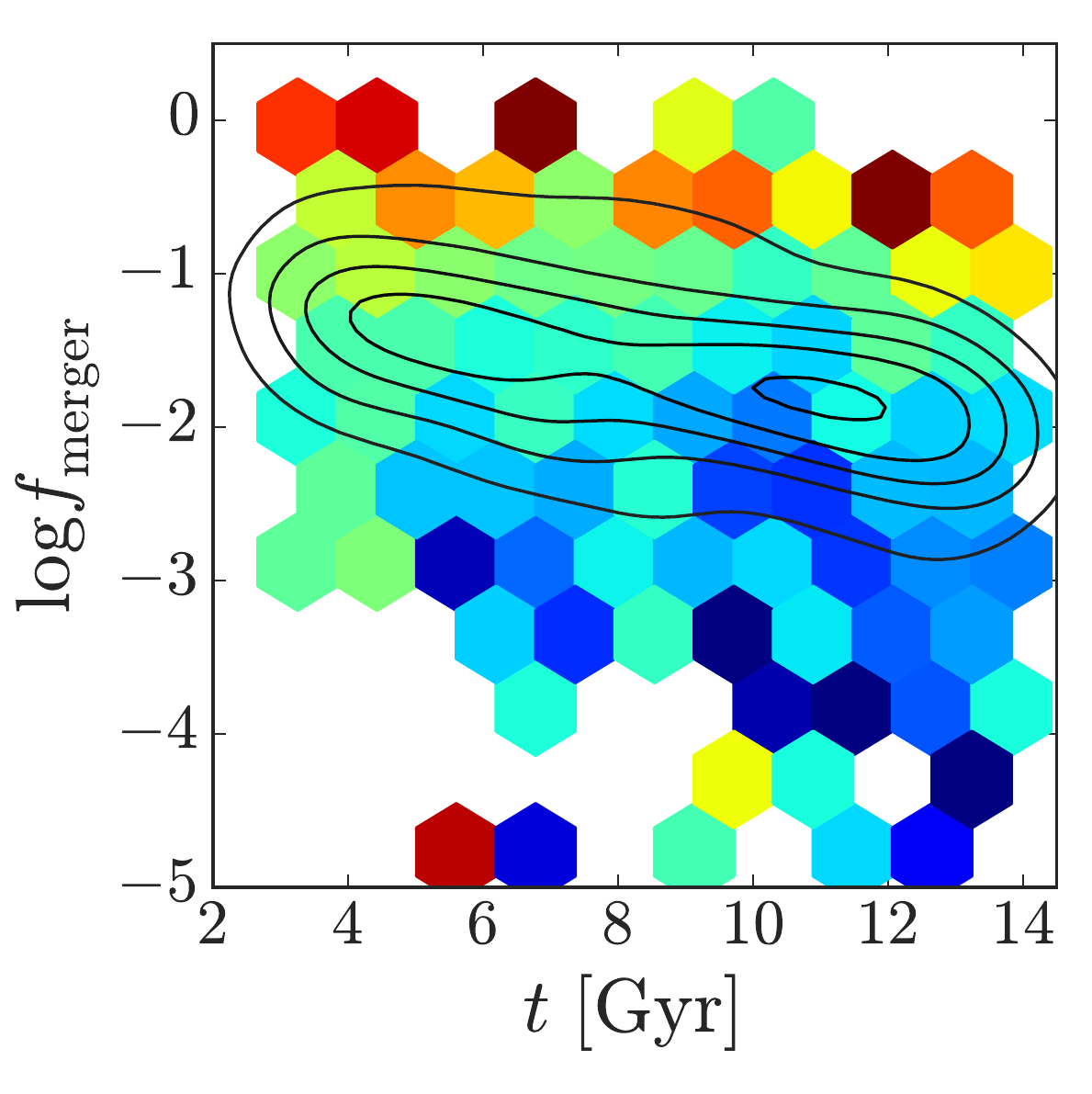}\hfill
	\includegraphics[height=0.28\linewidth,trim={1.2cm 0 4.5cm 0.6cm},clip]{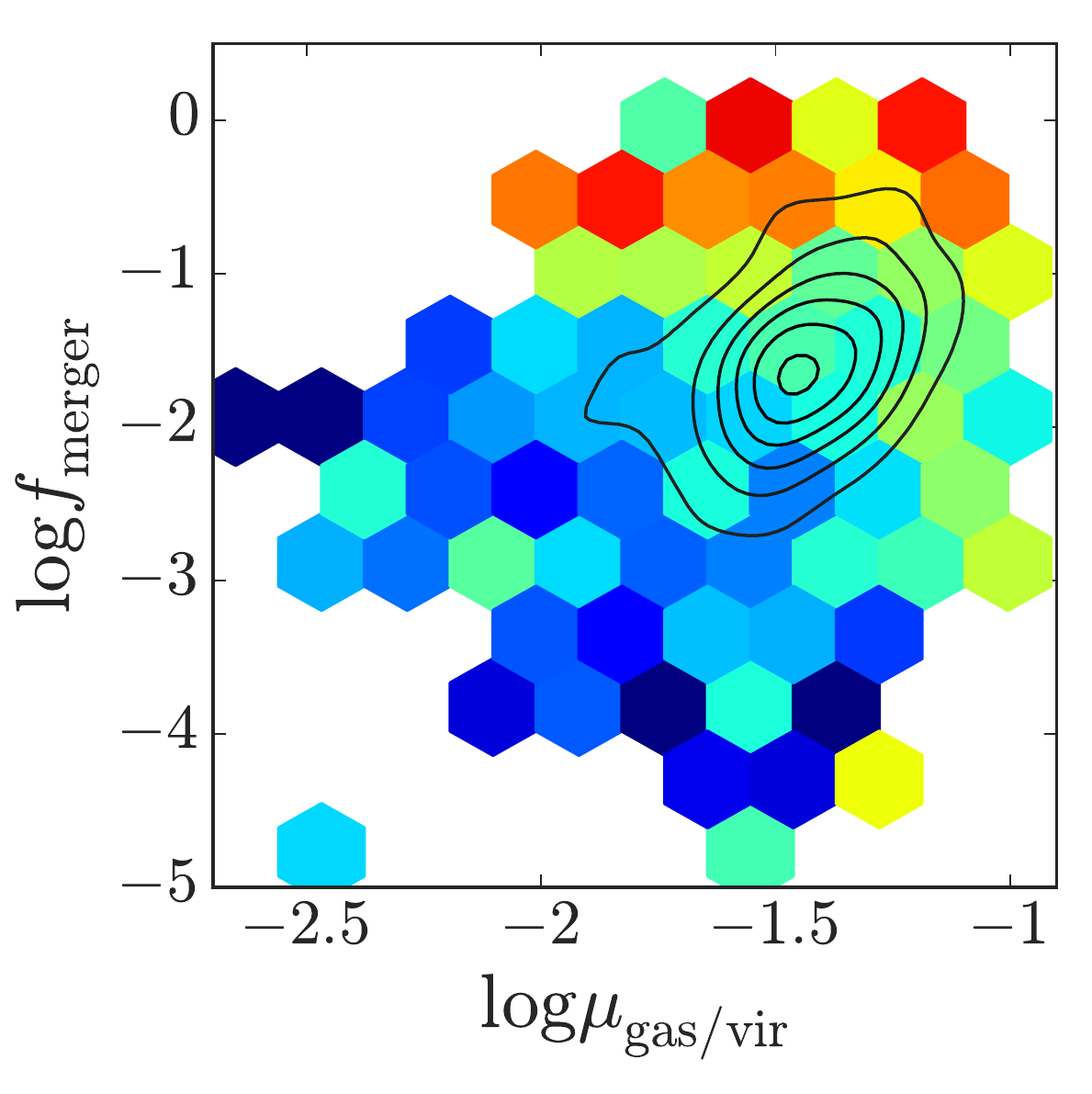}\hfill
	\includegraphics[height=0.28\linewidth,trim={0.8cm 0 0.2cm 0.6cm},clip]{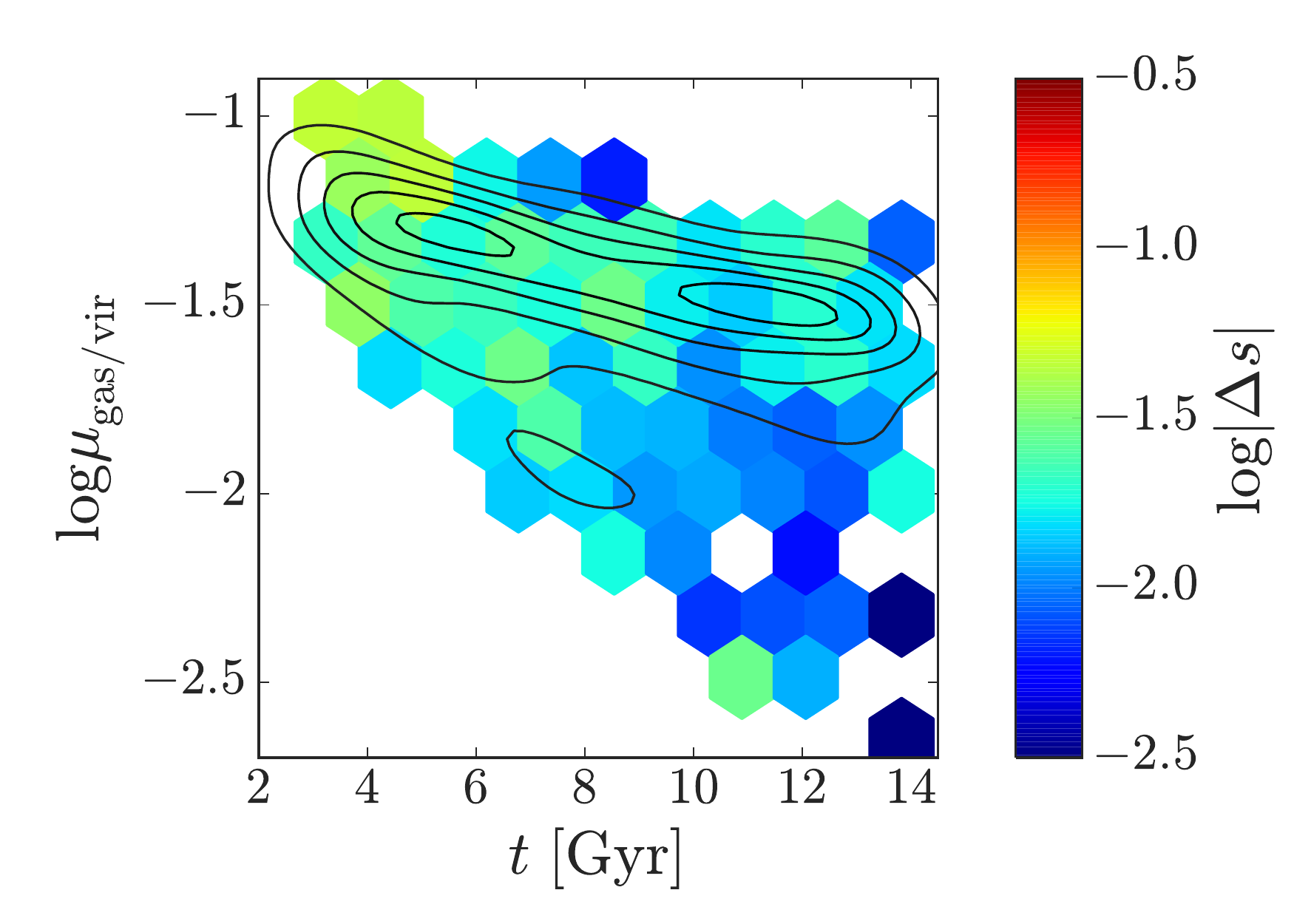}
	\\\vspace{-0.2cm}
	\includegraphics[height=0.28\linewidth,trim={.8cm 0 4.5cm 0.6cm},clip]{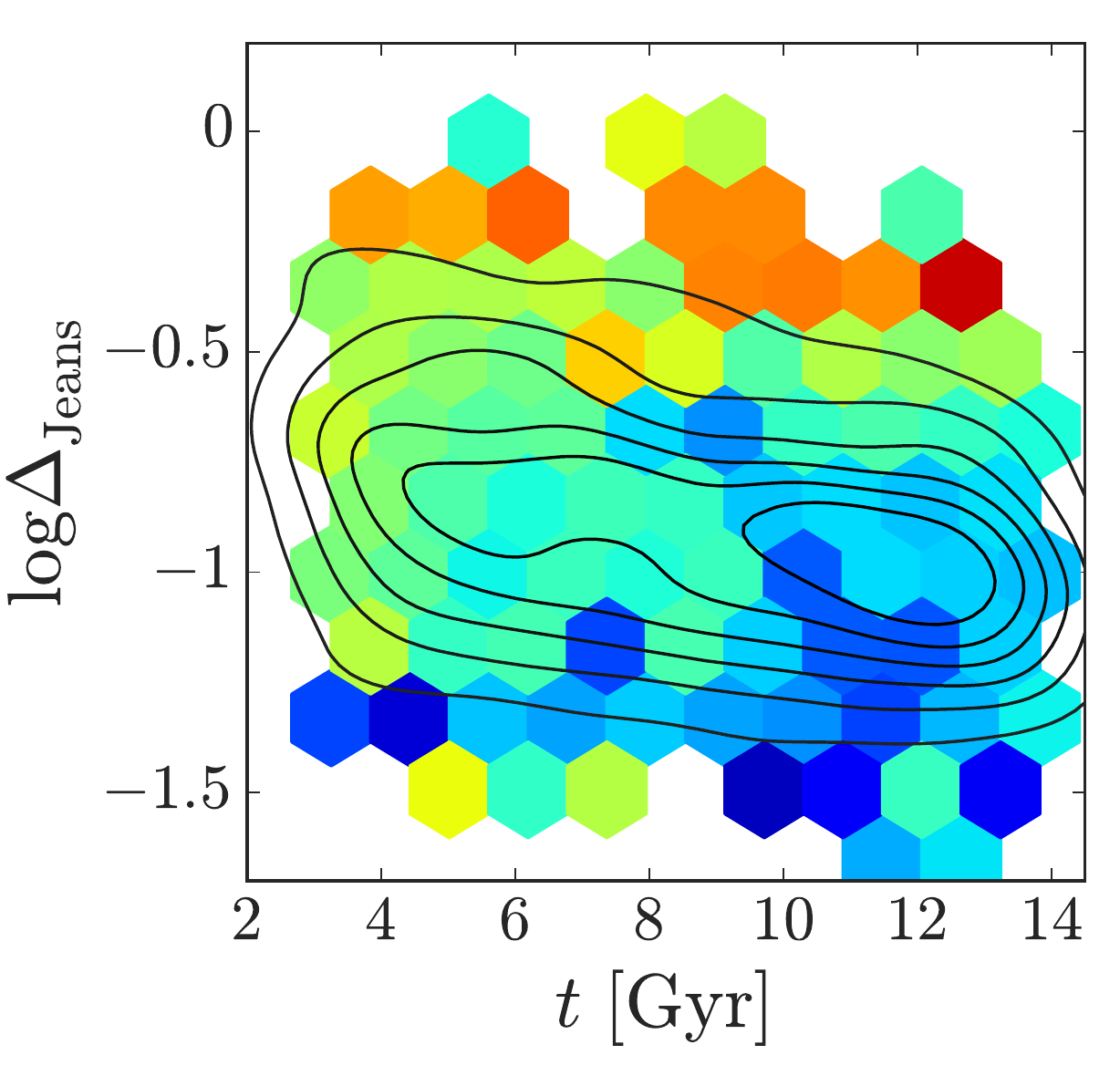}\hfill
	\includegraphics[height=0.28\linewidth,trim={.8cm 0 4.5cm 0.6cm},clip]{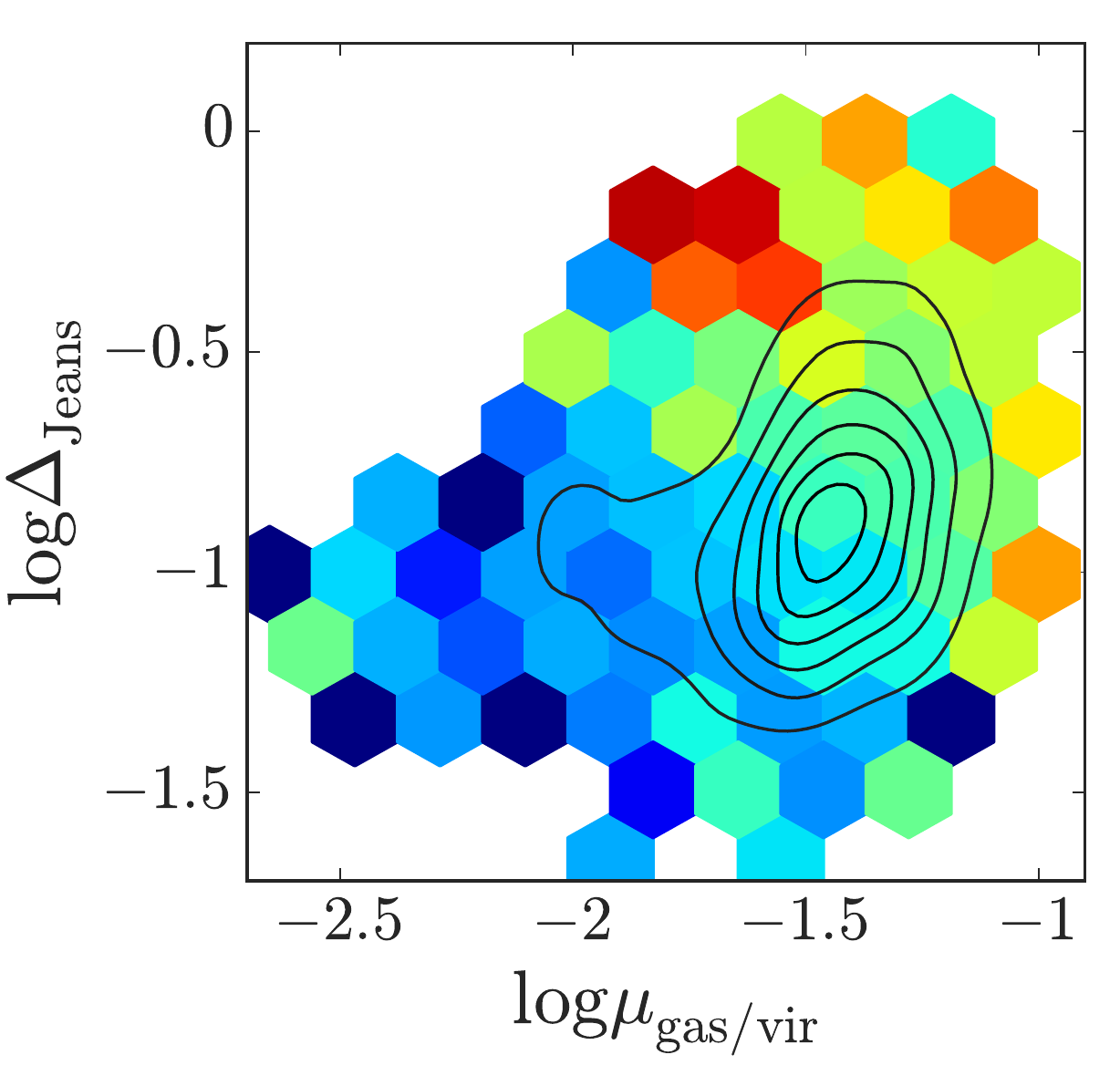}\hfill
	\includegraphics[height=0.28\linewidth,trim={0.8cm 0 0.2cm 0.6cm},clip]{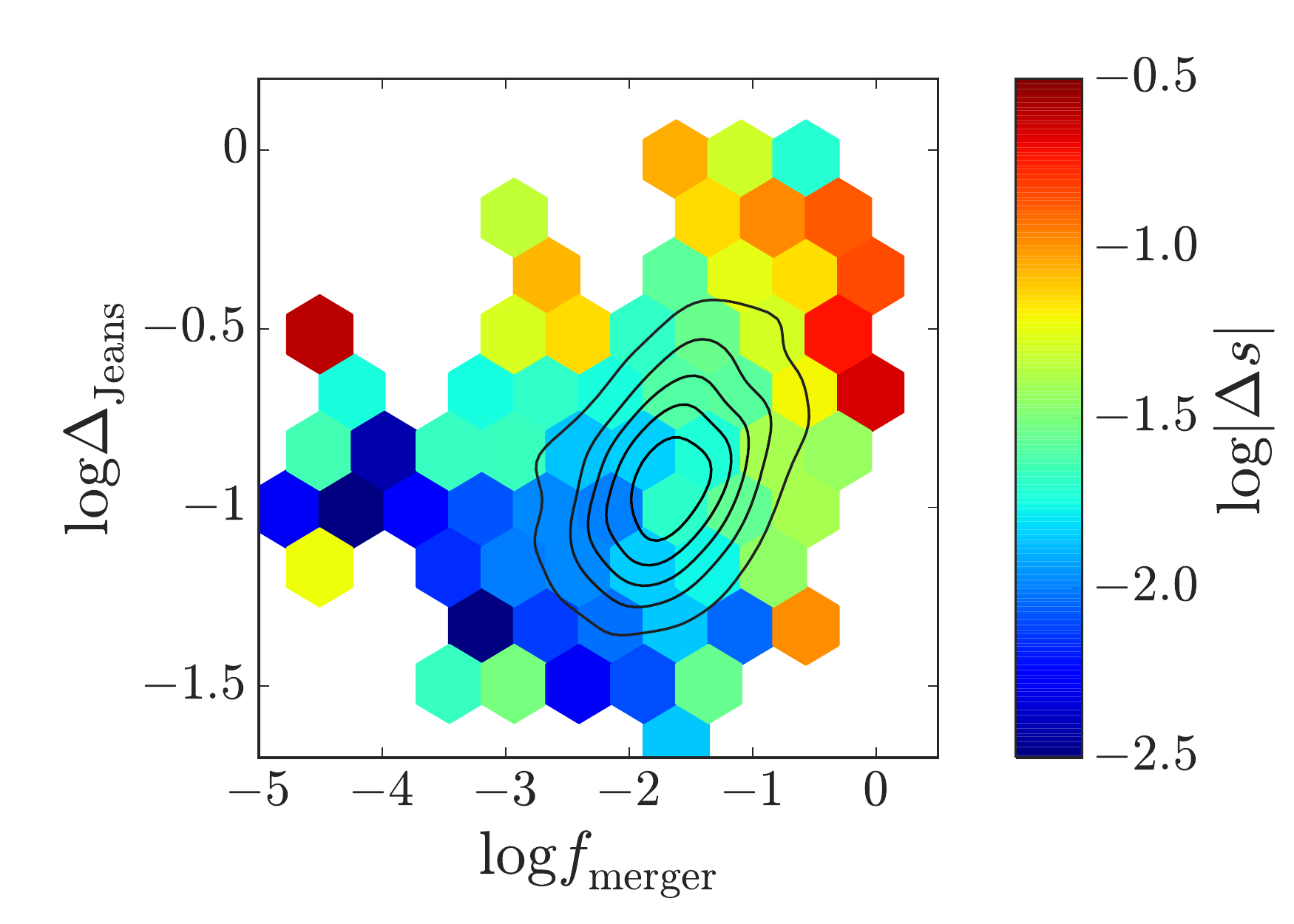}
	\\\vspace{0.2cm}
	\includegraphics[height=0.28\linewidth,trim={1.2cm 0 4.5cm 0.6cm},clip]{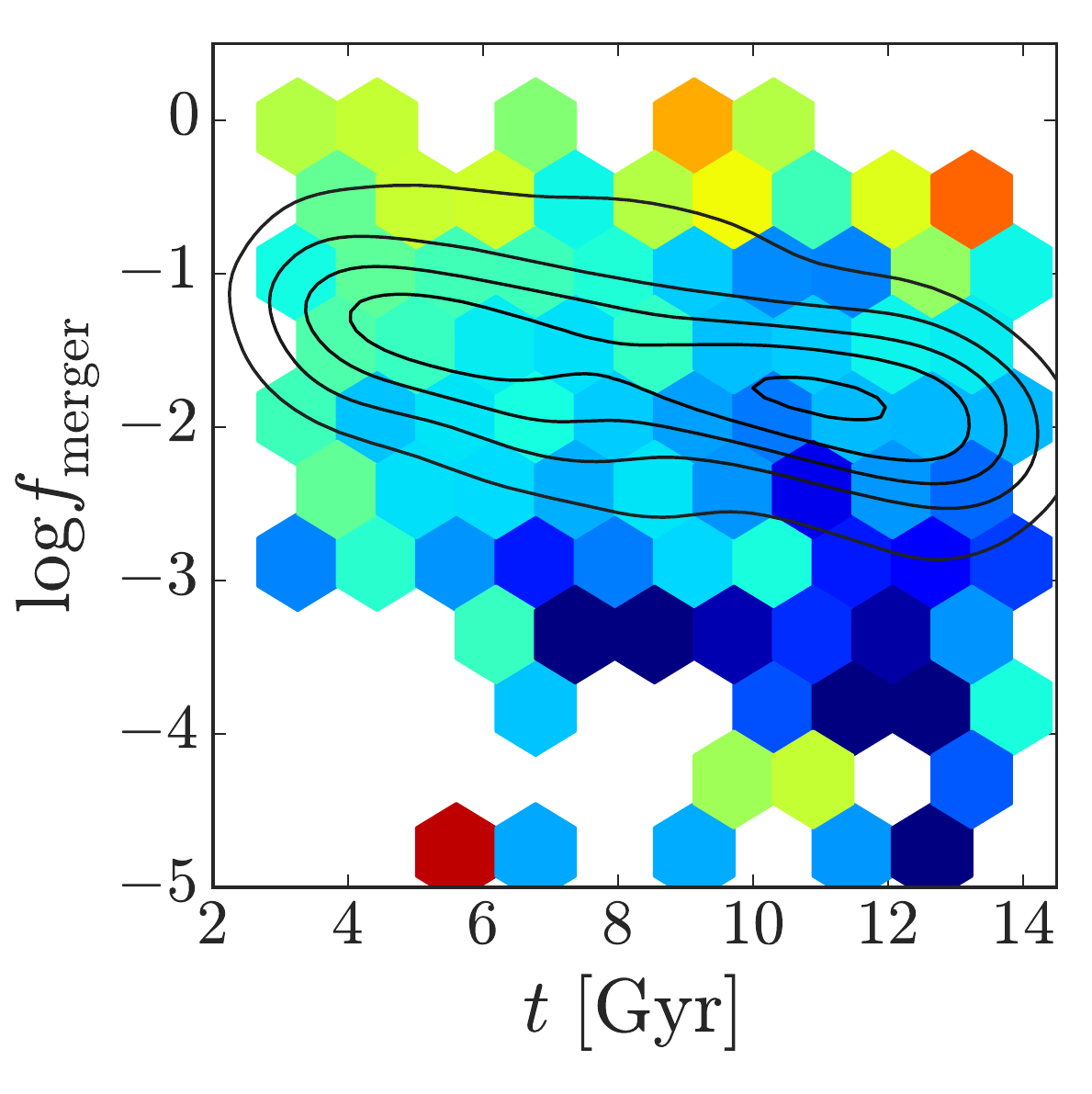}\hfill
	\includegraphics[height=0.28\linewidth,trim={1.2cm 0 4.5cm 0.6cm},clip]{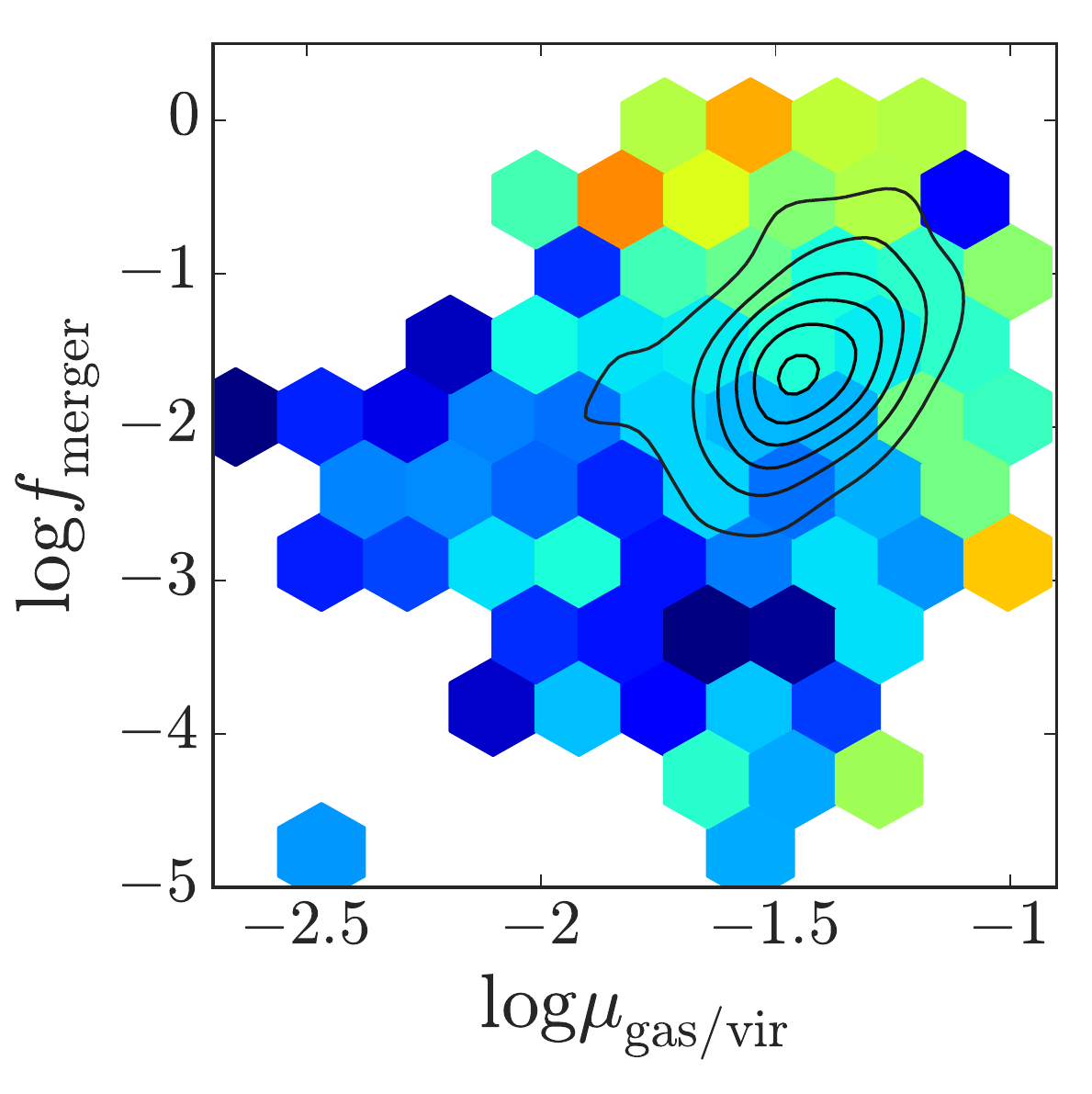}\hfill
	\includegraphics[height=0.28\linewidth,trim={0.8cm 0 0.2cm 0.6cm},clip]{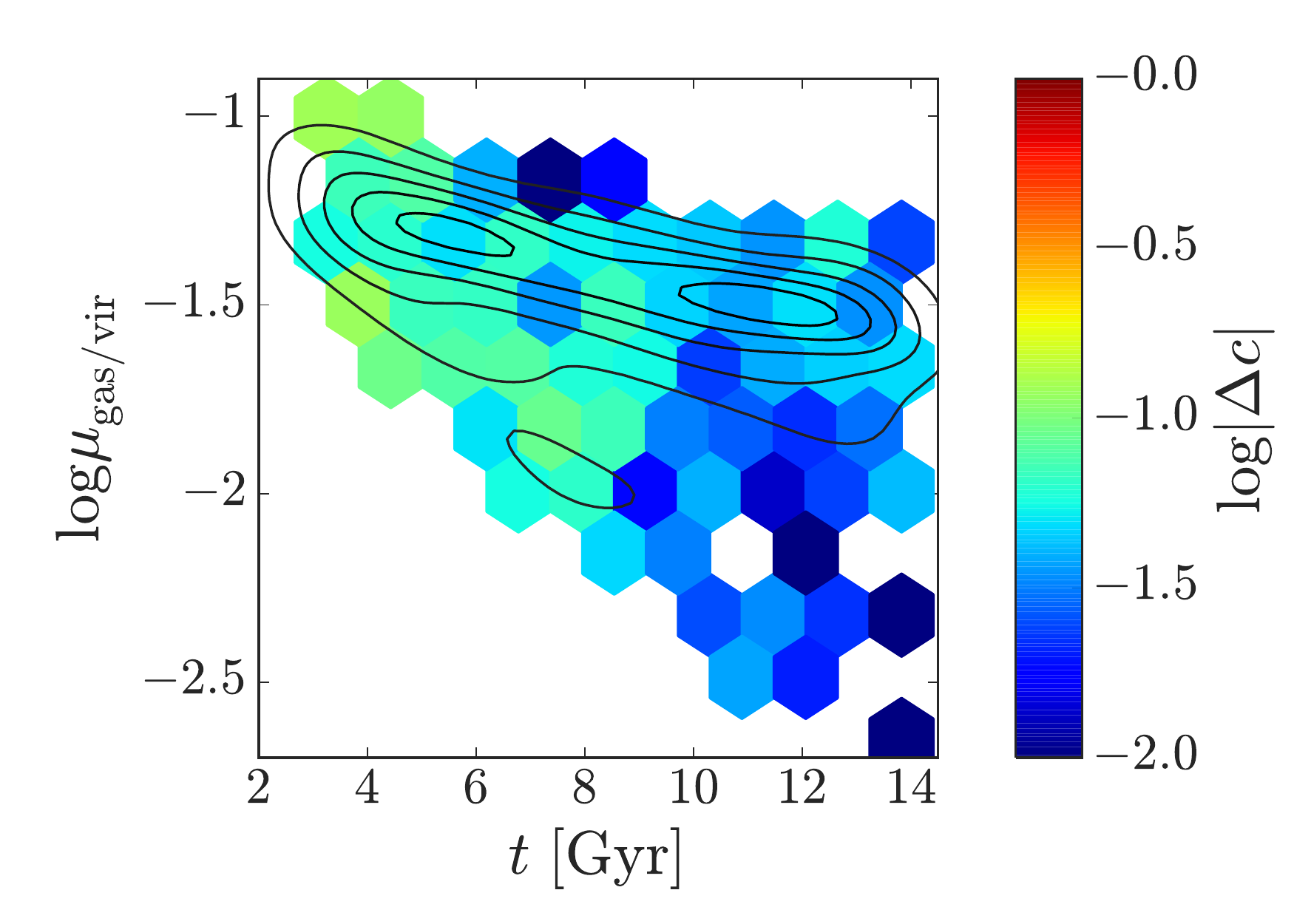}
	\\\vspace{-0.2cm}
	\includegraphics[height=0.28\linewidth,trim={.8cm 0 4.5cm 0.6cm},clip]{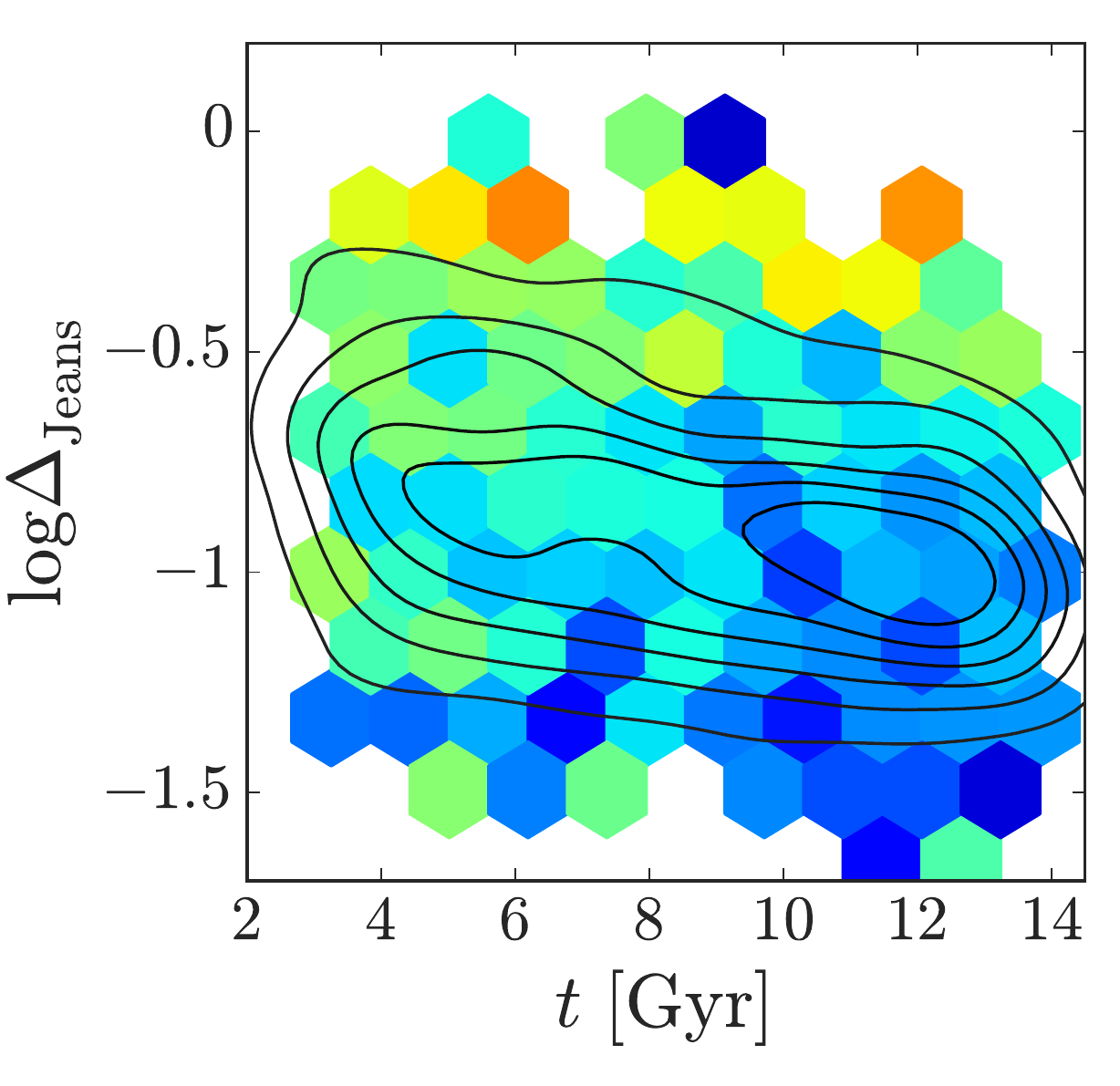}\hfill
	\includegraphics[height=0.28\linewidth,trim={.8cm 0 4.5cm 0.6cm},clip]{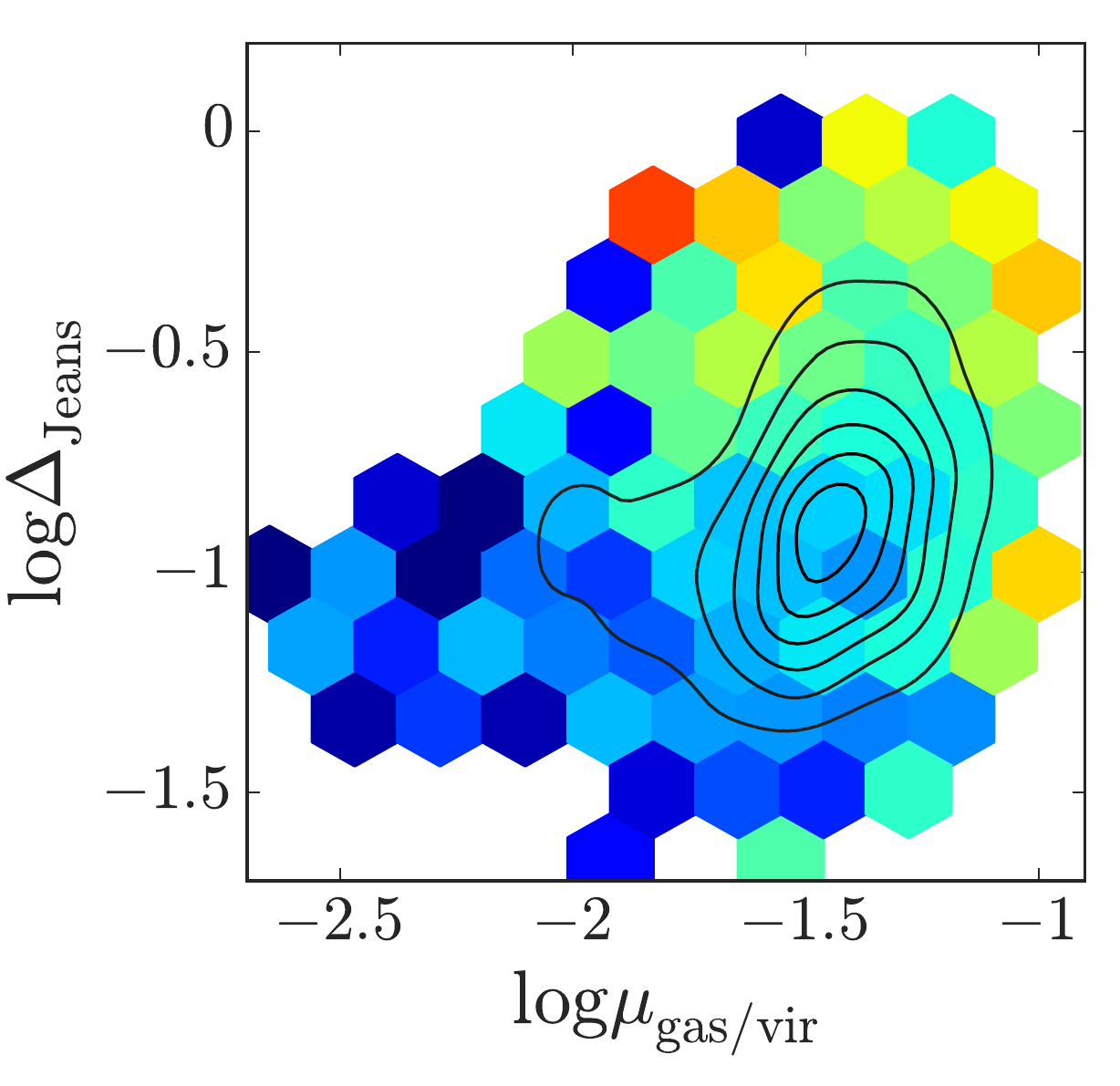}\hfill
	\includegraphics[height=0.28\linewidth,trim={0.8cm 0 0.2cm 0.6cm},clip]{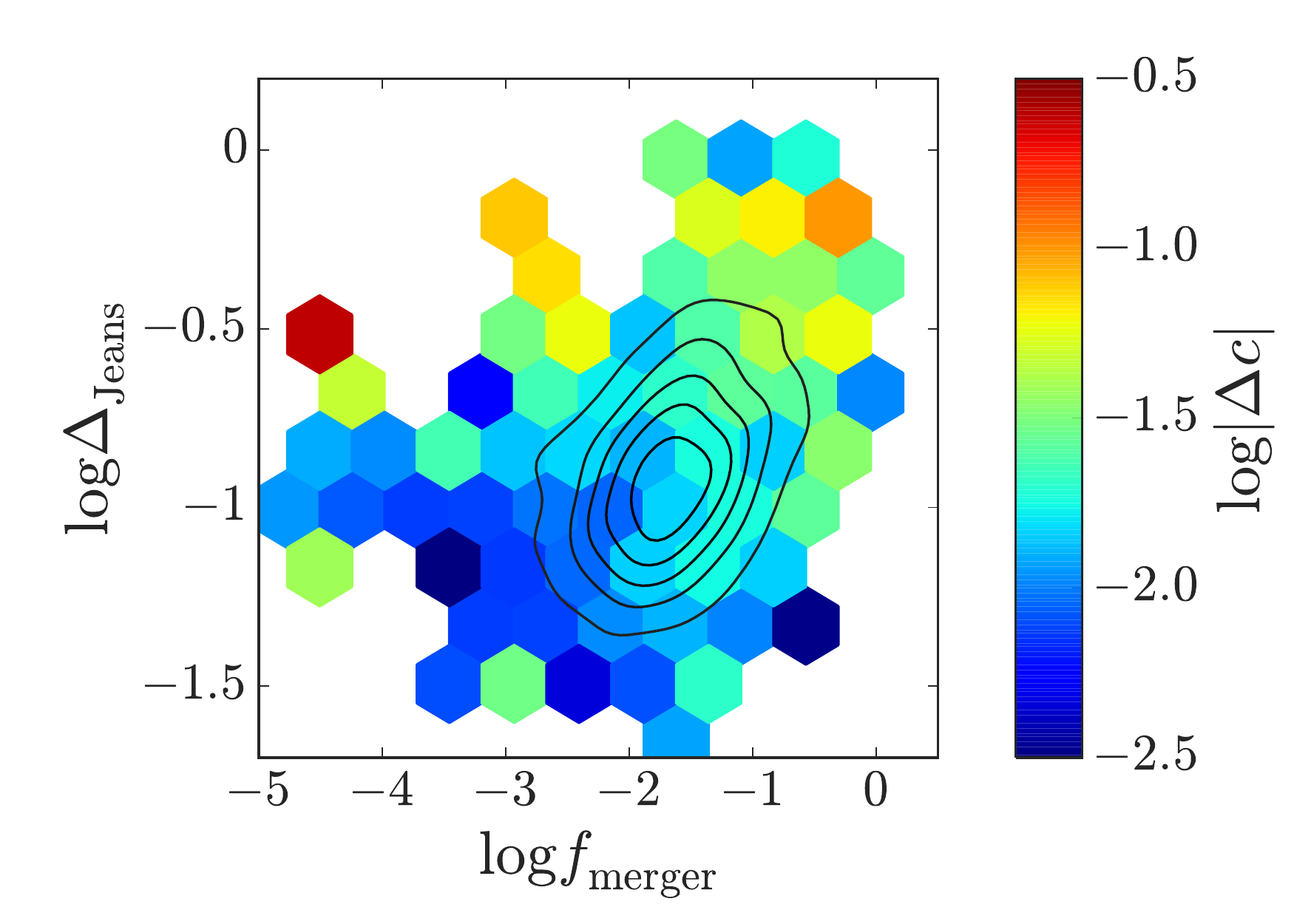}
	\vspace{-0.4cm }
	\caption{
		Distribution of the errors $\Delta s$ and $\Delta c$ on the logarithmic inner slope and effective concentration parameter (defined in Eqs.~(\ref{eq:s0}) and (\ref{eq:cmax})) as a function of time $t$, gas-to-virial mass ratio $\mu_{\rm gas/vir} = M_{\rm gas}/M_\vir$, merger indicator $f_{\rm merger}$ and deviation from equilibrium $\Delta_{\rm Jeans}$. As in Fig.~\ref{fig:correlations_hex} for $\delta$, colors correspond to the median values in each bin while contours show how the sample is distributed. 
	}
	\label{fig:correlations_hex_Dp}
\end{figure*}

Fig.~\ref{fig:correlations_all} shows the correlations between different quantities relevent to disk and halo dynamics and how the success measures $\Delta s$, $\Delta c$, and $\delta$ as well as the probability $p$ associated to $\delta<\delta_{\rm max}$ depend on them. 
The quantities we consider are the time $t$, the virial mass $M_{\rm vir}$, the stellar mass $M_{\rm star}$, the gas mass $M_{\rm gas}$, the gas-to-virial mass ratio $\mu_{\rm gas/vir}=M_{\rm gas}/M_{\rm vir}$, the stellar half-light radius $R_{1/2}$, the merger indicator $f_{\rm merger}$, and the deviation from Jeans equilibrium $\Delta_{\rm Jeans}$ defined from the relative difference shown in the lower left panel of Fig.~\ref{fig:Tjeans}.
We note that a correlation coefficient $r \gtrsim0.15$ between the success measures and any given quantity is sufficient to induce a noticeable trend of the associated probability $p$. 
The figure notably shows that the main quantities on which the model success depends are $f_{\rm merger}$, $\Delta_{\rm Jeans}$, $t$, and $\mu_{\rm gas/vir}$.
Fig~\ref{fig:correlations_success} further shows the correlations between the three success measures $\delta$, $\Delta s$ and $\Delta c$. As expected from the fact that both $\delta$ and $\Delta s$ assess the accuracy of the model to describe changes in the inner part of the average density profile while $\Delta c$ relates to the dark matter distribution at larger scales, $\delta$ is more closely related to $\Delta s$ (Pearson correlation coefficient $r=0.71$) than to $\Delta c$ ($r=0.43$). 
In Fig.~\ref{fig:correlations_hex_Dp}, we show the counterparts of Fig.~\ref{fig:correlations_hex} for $\Delta s$ and $\Delta c$ the  errors in inner logarithmic slope and concentration. The plots display similar trends as in Fig.~\ref{fig:correlations_hex}. The secondary trends as a function of the deviation from Jeans equilibrium $\Delta_{\rm Jeans}$, time $t$ and gas-to-virial mass ratio $\mu_{\rm gas/vir}$ are visible.

\label{lastpage}
\end{document}